\documentclass[final,5p,times,twocolumn]{elsarticle}

\usepackage{amssymb}
\usepackage{amsmath}

\usepackage{graphicx}
\usepackage{subfigure}
\usepackage{mathrsfs}

\usepackage[bookmarks=true,colorlinks=true,citecolor=blue,linkcolor=blue,urlcolor=magenta]{hyperref}
\usepackage[english]{babel} 
\usepackage{multirow}
\usepackage{multicol}
\usepackage{lipsum}

\usepackage{dcolumn}
\usepackage{bm}
\usepackage{breqn}
\usepackage{fancyvrb}

\usepackage{lscape}

\usepackage{longtable}
\usepackage{setspace}
\DefineVerbatimEnvironment{code}{Verbatim}{fontsize=\footnotesize}
\DefineVerbatimEnvironment{example}{Verbatim}{fontsize=\small}

\DeclareSymbolFontAlphabet{\mathcal}   {symbols}
\DeclareSymbolFontAlphabet{\mathbf}   {letters}

\usepackage{color}
\usepackage{xcolor}

\usepackage{soul}

\journal{IJMP E}

\begin{document}
\begin{frontmatter}

\author{
S. Cht. ~Mavrodiev$^{a}$ 
and 
M.A.~Deliyergiyev$^{b}$ 
}
\ead{schtmavr@yahoo.com; deliyergiyev@impcas.ac.cn}

\address{$^{a}$The Institute for Nuclear Research and Nuclear Energy, BAS, Sofia, Bulgaria}
\address{$^{b}$Department of High Energy Nuclear Physics, Institute of Modern Physics, CAS, Lanzhou, China}

\title{Modification of the Nuclear Landscape in the Inverse Problem Framework using the Generalized Bethe-Weizs\"{a}cker Mass Formula}

\author{}
\address{}

\begin{abstract}

We formalized the nuclear mass problem in the inverse problem framework. This approach allows us to infer the underlying model parameters from experimental observation, rather than to predict the observations from the model parameters. The inverse problem was formulated for the numerically generalized semi-empirical mass formula of Bethe and von Weizs\"{a}cker. It was solved in step-by-step way based on the AME2012 nuclear database. The established parameterization describes the measured nuclear masses of 2564 isotopes with a maximum deviation less than 2.6 MeV, starting from the number of protons and number of neutrons equal to 1.

The explicit form of unknown functions in the generalized mass formula was discovered in a step-by-step way using the modified least $\chi^{2}$ procedure, that realized in the algorithms which were developed by Lubomir Aleksandrov to solve nonlinear systems of equations via the Gauss-Newton method, lets us to choose between two functions with same $\chi^{2}$ the better one. In the obtained generalized model the corrections to the binding energy depend on nine proton (2, 8, 14, 20, 28, 50, 82, 108, 124) and ten neutron (2, 8, 14, 20, 28, 50, 82, 124, 152, 202) magic numbers as well on the asymptotic boundaries of their influence. The obtained results were compared with the predictions of other models.

\end{abstract}

\end{frontmatter}


\section{Introduction}

Semi-empirical mass formulas have been a staple of nuclear physics since shortly after the identification of the neutron as a constituent of the nucleus, done by Chadwick \cite{Chadwick692}, which was followed by the proposal of the composed atomic model given by D. Ivanenko \cite{Ivanenko:1932}, in 1932, and the first nuclear shell model \cite{IvanenkoGapon:1932}. The venerable formula of Bethe and von Weizs\"{a}cker, which was described in detail in \cite{Weizsacker:1935, RevModPhys.8.82}, has been extended and modified frequently over the years. 

\begin{figure*}
\centering
\subfigure[]{
\includegraphics[scale=0.185]{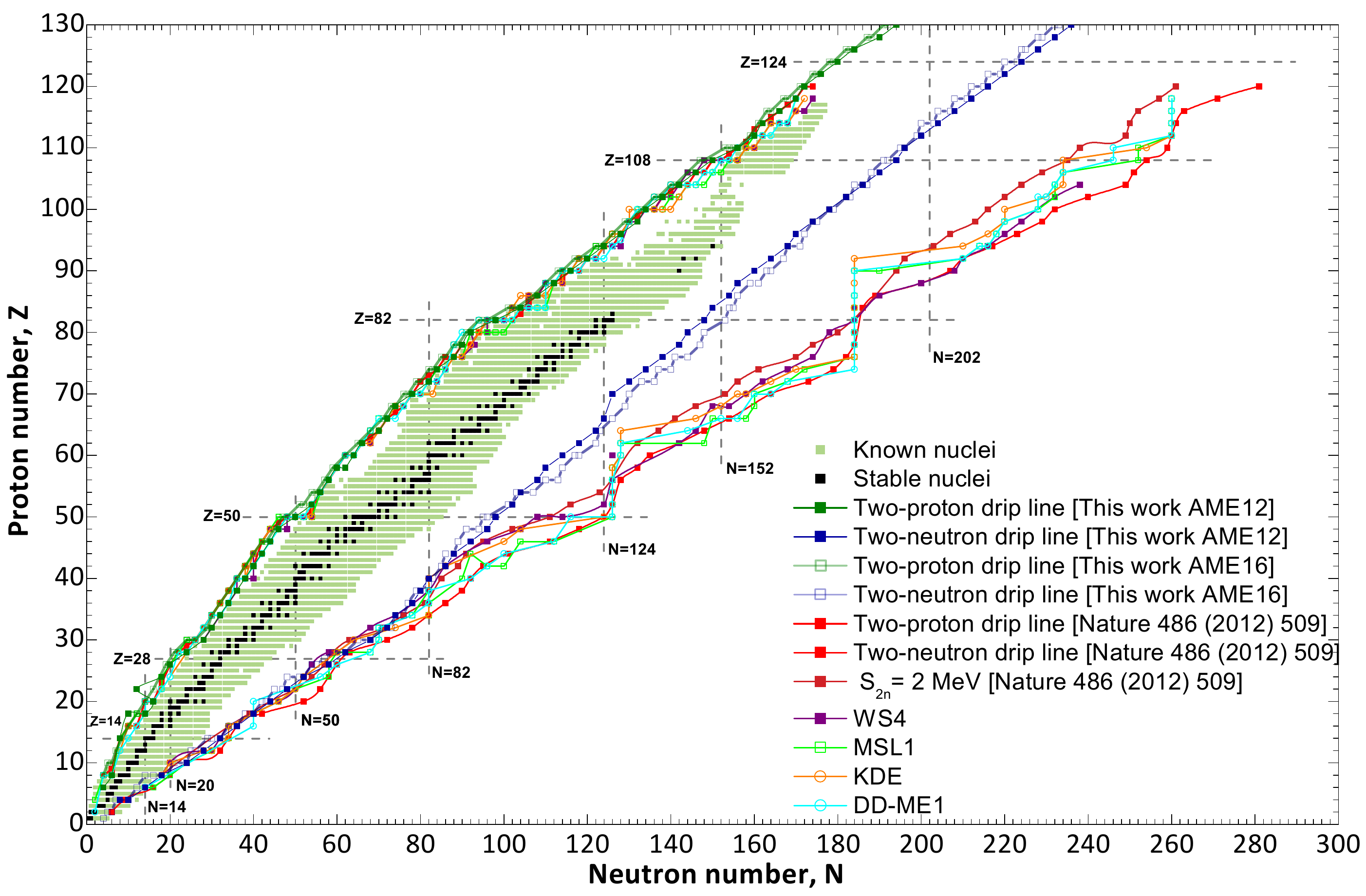}
}
\subfigure[]{
\includegraphics[scale=0.185]{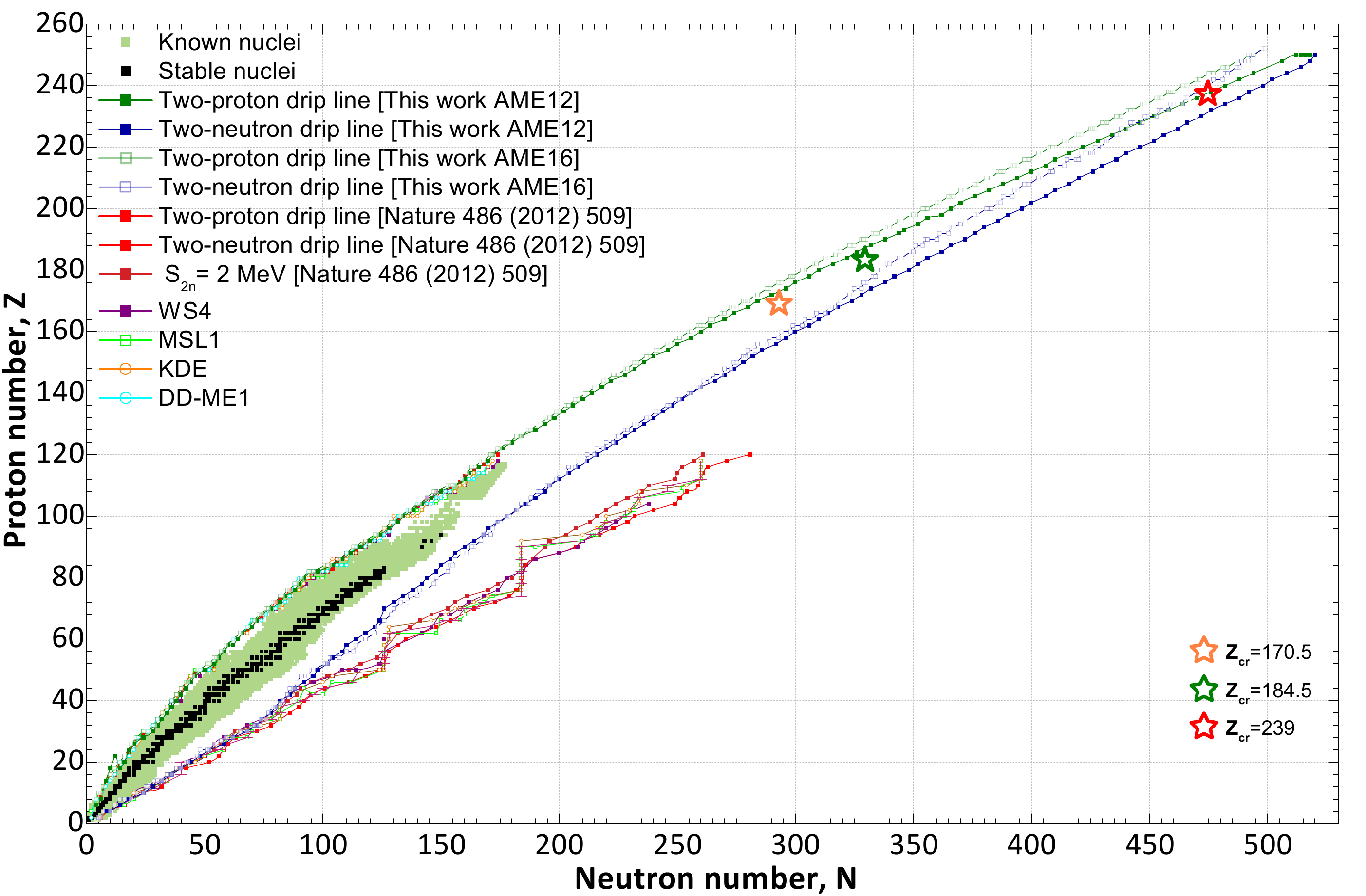}
\label{fig:DripLines_Predictions_limit}
}
\caption{
The chart of nuclei  known experimentally \cite{Ami2012_ChinPhysC} as a function of $Z$ and $N$. The stable (black squares) and radioactive (light green squares). Mean drip lines (red) are shown together with the $S_{2n}=2$ MeV line (brown).  The landscape of bound even-even nuclei as obtained from the microscopic density functional theory (DFT) calculations with two Skyrme interactions KDE \cite{PhysRevC.85.024305}, MSL1 \cite{Zhang2013234} and one relativistic interaction, DD-ME1 (cyan circles) \cite{PhysRevC.89.054320}. The prediction from Weizs\"{a}cker-Skyrme mass formula with WS4 \cite{Wang2014215} is also included for comparison (purple squares). The data is extracted from Ref. \cite{Erler_nature:2012, Wang:2014mra} and references therein. Predictions based on the AME2012 database for the two-neutron (olive) and two-proton drip (royal) lines are also shown. The two-neutron (light olive) and two-proton drip (light royal) lines obtained from the blind analysis of the AME2016 database using solution from the AME2012. In the left pad we show the limit of the predicted drip-lines in this work. Stars on the left panel indicate the critical charge values for a spherical nucleus (lowest states) \cite{Popov2001}.
}
\label{fig:DripLines_Predictions}
\end{figure*}

The principal domain of low-energy nuclear physics is the table of the nuclides. 
There are several hundred stable nuclei, 288, of the several thousand nuclides, or isotopes, that inhabit the nuclear landscape are either stable or practically stable (that is, have half-lives longer than the expected life of the Solar System) \cite{Erler_nature:2012}. Exactly these 288 nuclei form the so-called valley of stability. However, the total number of nuclides is unknown, one may note the white gap between known nuclei and the two-neutron drip line, see Fig.\ref{fig:DripLines_Predictions}, with theoretical estimates suggesting approximately seven thousand in total \cite{Erler_nature:2012}. 
The recent dramatic expansion of experimental nuclear physics facilities, allowing the measurement of nuclear properties very far from that valley, has triggered, among others, a renewed interest in nuclear astrophysics. Only in 2011, 100 new nuclides were discovered \cite{Thoennessen_nature:2011}. They joined the approximately 3,000 stable and radioactive nuclides that either occur naturally on Earth or are synthesized in the laboratory \cite{ENSDF:2006, Thoennessen:2004, Wapstra2003129, Audi2003337}. We hope that many new unstable nuclei (`rare isotopes') will be created and studied in new and planned experimental facilities around the world. By moving away from this valley, by adding nucleons, nuclear physics enters the vast territory of short-lived radioactive nuclei, which disintegrate by emitting $\beta$- and $\alpha$-particles or split into smaller parts through spontaneous fission.

As known nuclear existence ends at the drip lines, where there is no longer enough binding energy to prevent the last nucleons from escaping the nucleus. The placement of this line for the heavier elements is based on uncertain hypothetical predictions. However, it is not known how uncertain it is or how many protons and neutrons can be bound in a nucleus. The proton-rich border of the nuclear territory has been experimentally delineated up to protactinium \cite{ENSDF:2006} (proton number, $Z=91$), as indicated in Fig.\ref{fig:DripLines_Predictions}. The neutron-rich boundary is known only up to oxygen ($Z=8$) because of the long distance separating the valley of stability from the neutron drip line \cite{Thoennessen:2004}. The super heavy nucleus with $Z=120$ and $A=299$ \cite{PhysRevC.74.044602, Oganessian201562} marks the current limit of nuclear charge and mass. The borders of the super heavy region are unknown and difficult to predict because competition between Coulomb and shell effects can cause voids and form exotic topologies (see Section 4 of Ref.\cite{Nazarewicz2002165}).

Although the masses of many very exotic nuclei are now amenable to measurement \cite{PhysRevC.74.044602, Oganessian201562, 0954-3899-34-4-R01, PhysRevC.76.011601}, many more nuclei of great interest in astrophysics remain beyond the capabilities of modern mass measurements \cite{Erler_nature:2012}. This has conferred increased urgency on the development of reliable extrapolations of measured masses \cite{Moller1995185}, one of the classic applications of mass formulas. 
However, there is significance in the numerical values of the coefficients of the semi-empirical mass formula, influencing, for instance, the deduction of the binding energy per nucleon of infinite nuclear matter, or the symmetry energy and surface tension of nuclear matter. These are not easily accessed by the  macroscopic-microscopic mass formulas \cite{NILSSON19691, RevModPhys.44.320,PhysRevC.5.1050} that have been elaborated in order to reproduce the irregularity of the masses as functions of $A$ and $Z$, partly due to shell closings and proton and neutron number parity. 
Nevertheless a number of models such as 
the finite-range droplet model (FRDM) \cite{Moller1995185, MYERS1974186, Moller1988213}, 
the liquid drop model (LDM), an improved LDM by inclusion of schematic microscopic shell corrections \cite{Mendoza-Temis:2008ztv}, 
the Relativistic Mean Field theory (RMF) \cite{Serot1992}, also known as the relativistic Hartree (RH) \cite{Walecka:1974qa}, received wide attention due to its many successes in describing lots of nuclear phenomena \cite{doi:10.1142/9789812816634_0026, Zhou:2009sp, Niu:2013npa} as well as successful applications in astrophysics \cite{Baohua:2008tu,Niu:2009cw, Niu:2008kx, Meng2011, Xu:2012yf, Niu:2012rc},
the Duflo-Zuker model  (DZ, DZ10, DZ31) \cite{PhysRevC.52.R23, PhysRevC.59.R2347, PhysMexicRevS54.129}, 
the Garvey-Kelson model (GK) \cite{PhysRevLett.16.197, RevModPhys.41.S1, Janecke1988265, Cheng:2014uha}, 
an infinite nuclear matter (INM) model \cite{0305-4616-13-6-009, Nayak1999213}, 
the Liran-Zeldes model (LZ) \cite{Liran1976431}, 
the Koura-Tachibana-Uno-Yamada model (KTUY) \cite{Koura200047, doi:10.1143/PTP.113.305}, 
the extended Thomas-Fermi plus Strutinsky integral model (ETFSI) \cite{doi:10.1063/1.1361389}, 
and the Weizs\"{a}cker-Skyrme model (WS, WS3, WS4) \cite{Wang:2011hxa, Wang2014215} have been developed to predict masses of exotic nuclei to model the genesis of the chemical elements \cite{PhysRev.73.803, Rolf1988}.

The rapid development of computer technology in the new century allow us to achieve a great progress in a fully microscopic framework. Based on Hartree-Fock-Bogoliubov (HFB) theory \cite{GORIELY2001311, PhysRevC.66.024326, Samyn2002142, 0954-3899-31-11-R01} with a Skyrme \cite{Stoitsov:2004pe} or Gogny \cite{Niksic:2014dra} force, a series of microscopic mass models have been proposed with accuracy comparable with that of traditional macroscopic-microscopic mass models \cite{Goriely:2009zz, Goriely:2010bm}. The microscopic density functional theory (DFT) model has achieved remarkable success in describing the data on known nuclei with two Skyrme interactions KDE \cite{PhysRevC.85.024305}, MSL1 \cite{Zhang2013234} and one relativistic interaction, DD-ME1 \cite{PhysRevC.89.054320}.

However, as one may note from the recent literature there is still lack of knowledge about the masses of new nuclides both in the super-heavy element region and the regions close to the proton and neutron drip lines. There are still large deviations among the mass predictions of different models. Moreover, the model predictions do not completely agree between each other, even in the region close to known masses. 
The accuracy and predictive power of these nuclear mass models have been tested in the Refs.\cite{Athanassopoulos2004222, RevModPhys.75.1021, Geng:2005yu, Mendoza-Temis:2008agm, doi:10.1142/S0218301312500383, 1402-4896-2013-T154-014001}). Further improvements of the accuracy of nuclear mass models have been done with the help of integration the image reconstruction techniques based on the Fourier transform \cite{Morales:2010zza} to the nuclear models, significantly reducing the rms deviation to the known masses with the CLEAN algorithm \cite{1974AS.Suppl417, 1980AA89.377C,  doi:10.1146/annurev.aa.24.090186.001015}. Later on, the radial basis function (RBF) approach was developed to improve the mass predictions of several theoretical models \cite{Wang:2011hxa, Niu:2013hda}. Compared with the CLEAN reconstruction, the RBF approach more effectively reduces the rms deviations with respect to the masses first appearing in AME03 \cite{Wang:2011hxa}.

All these works represent an on-going challenge for low-energy nuclear theory, which is to describe the structure and reactions of all nuclei, whether measured or not.

The main goal of our studies was to determine how well the existing data, and only data, determines the mapping from the proton and neutron numbers to the mass of the nuclear ground state. Another is to find presumed regularities by analysis of observed nuclei masses \cite{AleksandrovGadjokov1971}. In addition is to provide reliable predictive model that can be used to forecast mass values away from the valley of stability, where no possibility exist for direct measurements of the nuclei masses and one have to apply indirect methods.
The results suggest that with further development this approach may provide a valuable complement to conventional global models. 

A set of experimental nuclear masses from AME2012, the most recent evaluation database, that was published in December 2012 in \cite{Ami2012_ChinPhysC, NUBASE2012_ChinPhysC}, constitutes the raw material for this work.  Only measured nuclei are included into our consideration. The masses extrapolated from systematics and marked with the symbol $\#$ in the error column are not taken into account here. Therefore, we use only 2564 experimental nuclear masses, including the Hydrogen atom, to provide a deep understanding of the mutual influence of terms in the semi-empirical formula previously investigated in \cite{Kirson200829}. Those experimental values play a crucial role in our study of the separated influence of the Wigner term, the curvature energy and different powers of the relative neutron excess $I=(N-Z)/A$. The improvement of the generalized liquid drop model formula (GLDM) previously proposed in \cite{Royer1985477}, whose coefficients were recently determined once again in \cite{Chowdhury:2004jr} and \cite{PhysRevC.73.067302}, is also the focus of this paper. 

Later the obtained solution will be used in the blind analysis of the AME2016 \cite{Ami2016_ChinPhysC} and the DGFRS separator \cite{Oganessian201562} data. 

In present work we demonstrate the applicability of the inverse problem approach for solution of such nuclear physics problems. 
The desired result  is archived by very complicated procedures for processing the experimental data.
First, we formalize the nuclear mass problem in the framework of the inverse problem. Second we propose the generalized form of the Bethe-Weizs\"{a}ker (BW) mass formula, which helps us to discover the latent regularities in the nuclear masses from AME2012.  And afterwards we provide a solution of the formalized inverse problem that was obtained with the help of the Alexandrov dynamic auto-regularization method of the Gauss-Newton type for ill-posed problems (Dubna REGN -- Regularized Gauss-Newton iteration method) \cite{Alexandrov:1970,Alexandrov:1973, Alexandrov:1983, Aleksandrov197146, Alexandrov:1982,  AleksandrovPriv1997, Alexandrov66:1977}, which is a constructive development of the Tikhonov regularization method \cite{Tikhonov:1963, Tikhonov:1983, Tikhonov:1986, Tikhonov:1995} for ill-posed problems.

The formalism of the applied approach is given in Sec.~\ref{Theory_and_Method}. The basis of the classic BW mass formula is sketched in Sec.~\ref{Nuclear Binding Energy} of this paper. 
The numerical generalization of the BW mass formula is described in Sec.~\ref{Parametrization_of_the_BWmass_formulae}. 
The main conclusions of the paper are drawn in Sec.~\ref{Results}, which also includes a discussion of the principal results, the resulting rms deviations of the predicted mass from the measured once and the parameter values. 
Predicted binding energies, nuclear and atomic masses, and mass excess for the recently discovered nuclei at the DGFRS separator based on fusion reactions of $^{48}\rm{Ca}$ with  $^{238}\rm{U}$-$^{249}\rm{Cf}$ target nuclei \cite{Oganessian201562} are shown in \ref{Predicted binding energies for super heave nuclei}.

\section{Theory and Method}
\label{Theory_and_Method}
This section begins with the concept of latent regularities in the experimental data. An attempt is made to provide a general mathematical formulation of this problem and to connect its occurrence to an effective research strategy. It explains regularization technique within the Aleksandrov method. Furthermore, it describes how to construct solution of non-regular problems, based on the iteration processes for the solution of nonlinear equations. The example of such a process is given. 

The aforementioned goals stimulated us to try to clarify the features and to find hidden regularities of the well known semi-empirical mass formula of Bethe-Weizs\"{a}ker, based exclusively on experimental data. This idea motivate us to apply the inverse problem approach to solve this task, which will be briefly discussed in this and the next section.

\subsection{The problem of analyzing the latent regularities}
\label{AleksandrovMethod_LatentRegul}

The conventional approach to physical data analysis consists of constructing different physical models and comparing the theoretical predictions computed for these models with the observed data. Numerical modeling of physical data for given model parameters is usually called a $forward$ problem. The forward problem solution makes it possible to predict physical  observation for specific physical areas. 

Usually we approximate real physics by a more or less simple model and try to determine the model parameters from the data. One may call this problem an $inverse$ problem. The success of physical interpretation depends on our ability to approximate real physical dependencies by reasonable models, and to solve the corresponding inverse problems effectively. 

The concept  of an inverse problem was formulated by Hadamard at the beginning of the 20th century \cite{Hadamard:1902}. For a long time it was assumed that such problems have nothing to do with physics and techniques. However, the importance of such type of problems was fully recognized in the 40's especially from the application point of view, this boost an intensive developments of the inverse problem method. At present time a wide variety of nonlinear inverse problems exist in science and engineering, and many examples can be found in the monographs and  surveys by Tikhonov and Arsenin \cite{Tikhonov:1995, TikhonovArsenin:1977},  Hofmann  \cite{BHofmann:1986}, Banks and  Kunisch \cite{Banks:1989}, Engl \cite{Engl:1993}, Groetsch \cite{Groetsch:1993}, and Vasin-Ageev \cite{Vasin:1995}. Such problems are based on the comparison of the theoretical and experimental data by solving the system of the nonlinear operator equations on the field of real numbers $\bm{R}$  of following type: 
\begin{equation}
\begin{split}
&f_{j}(x_{1}, x_{2},\dots, x_{n};p_{1}, p_{2},\dots, p_{\nu}) = y_{j}, \\
&j=1,2,\dots,m; ~~1 \leq n \leq  m;
\label{eq:fxy}
\end{split}
\end{equation}
where $x\in D_{f} \subseteq \bm{R}^{n}$ and $y \in D_{f} \subseteq \bm{R}^{m}$, $\bm{R}^{n}$ and $\bm{R}^{m}$ are the $n$ and $m$ dimensional real Cartesian canonical space correspondingly, $D_{f}$ is the convex unbounded domain, the given nonlinear function
\begin{equation}
f_{j}(x_{1}, x_{2},\dots, x_{n};p_{1}, p_{2},\dots, p_{\nu}) \in C^{2}(D_{f}), ~~j=1,2,\dots,m,
\label{eq:GivenNonlinearFunc}
\end{equation}
where  
\begin{equation}
\begin{split}
&D_{f}=(x_{1}^{\prime}, x_{1}^{\prime\prime})\times \dots \times (x_{n}^{\prime}, x_{n}^{\prime\prime}),\\
&x_{i}^{\prime}, x_{i}^{\prime\prime} \in \bm{R}^{1}; p_{1}, p_{2}, \dots,p_{\nu} \in \bm{R}^{1}, ~~\nu \geq 0
\label{eq:GivenSpaceForNonlinearFunc}
\end{split}
\end{equation}
are the given parameters. 

The solution of the system of Eqs.\eqref{eq:fxy} can be found with respect to the unknowns $x_{i}\in(x_{i}^{\prime}, x_{i}^{\prime\prime})$, where $i=1,2,\dots, n$. Taking into account that $\bm{R}^{n}$ and $\bm{R}^{m}$ are the $n$ and $m$ dimensional real coordinate space, that corresponds to the $x$ and $y$ respectively and using the following notation 
\begin{equation}
\begin{split}
&x=(x_{1},x_{2},\dots,x_{n})^{T} \in D_{f} \subseteq \bm{R}^{n}\\ 
&y=(y_{1},y_{2},\dots,y_{m})^{T} \in D_{f} \subseteq \bm{R}^{m}.
\label{eq:XYvariables}
\end{split}
\end{equation}
The system of Eqs.\eqref{eq:fxy} can be written in a vector form as:
\begin{equation}
F(x) = y.
\label{eq:fxy_VectorForm}
\end{equation}
Operator $F$ in Eq.\eqref{eq:fxy_VectorForm} is a forward modeling nonlinear operator, that transforms any model $x$ into the corresponding data $y$. The inverse problem is formulated as the solution of the operator Eq.\eqref{eq:fxy_VectorForm}. Therefore, Eqs.\eqref{eq:fxy_VectorForm} connect the unknown parameters of the model with some given quantities (variables) describing the model, in our case atomic mass number, $A$, and proton mass number, $Z$. These quantities take the form of the so-called input data. Generally, the input data as well as the unknown parameters are the elements of certain metric spaces, in particular, Banach or Hilbert spaces, with the operator of the model acting from the solution space to the data space. 

Suppose that observed function, which describes some property of some physical phenomenon may be denoted by $f(t)$. Note, that in our case the role of this function will be played by the BW mass formula, we come to this in later sections. Let us assume that $f(t)$ is determined on the closed segment $S = [t_{1}, t_{2}]$, 
$-\infty \textless t_{1} \textless t_{2} \textless \infty$. For sake of simplicity, the independent variable $t$ and the function $f(t)$ will be considered scalar quantities.
Let a set $L$ of scalar functions $g_{i}(i = 1, 2,\dots, n)$, depending on the parameters $p_{ij}(i=1,2,\dots,n;j=1,2,\dots,s_{i})$ and on the independent variable $x\in S$, be given:
\begin{equation}
L=\{g_{1}(p_{11},\dots, p_{1n_{1}};t),\dots,g_{n}(p_{n 1},\dots, p_{n s_{n}};t)\}.
\label{eq:L_scalarFunc}
\end{equation}
Further, a certain degree of smoothness of the function $g_{i}$ with respect to the parameters $p_{ij}$ is assumed, namely, that continuous second-order derivatives exist.

Based on the previous knowledge of the observed functions or on some other conceptions (theory, analogy, phenomenology, etc.), the following basic hypothesis is made: the observed function $f(t)$ may be approximated by a linear combination of the elements of the set $L$ as:
\begin{equation}
f(t)=\sum_{i=1}^{n}{\rm{a}}_{i} g_{i}(p_{i1},\dots, p_{is_{i}};t)= 
{\rm{a}}_{0}+\sum_{i=1}^{n}{\rm{a}}_{i}{\rm{e}}^{-h_{i}t},
\label{eq:Comb_Elements_Lf}
\end{equation}
where ${\rm{a}}_{i}\geq 0$ and the number of exponents in the general case is unknown. Due to high sensitivity of the solution to the infinitesimal changes of values $f(t)$ one has to use special steady solutions of the systems Eq.\eqref{eq:Comb_Elements_Lf}.

Let us demonstrate the high sensitivity of the solution of this system. Suppose that $f(u)$ will define the transformation of the right hand side of Eq.\eqref{eq:Comb_Elements_Lf}, that transforms the points:
\begin{equation}
u=({\rm{a}}_{0}, {\rm{a}}_{1},\ldots, {\rm{a}}_{n}, h_{1},\dotsc, h_{n} ) \in R_{2n+1}
\label{eq:TransformFunction}
\end{equation}
into the points $\upsilon=(y_{1},y_{2},\dotsc,y_{m}) \in R_{n_{i}}$. Let  $\Omega$ be the union of the all subspaces
\begin{equation}
{\rm{a}}_{0}=0, ~~{\rm{a}}_{i}=0, ~~ h_{p}=h_{q}, ~~~i,p,q=1,2,\ldots,n, ~~~p\neq q,
\label{eq:H_Subspaces}
\end{equation}
and $U=\{ u/u\in R_{2n+1}, {\rm{det}}J(u)\neq 0 \}$ at $2n+1=m$, where $J(u)$ is the Jacobi matrix of the function $f(u)$.

Thus we have $U\cap \Omega=\O$. One may easily check, that once $u\in f(U)$, then the non-linear problem Eq.\eqref{eq:Comb_Elements_Lf} is correct \cite{Tikhonov:1963, Tikhonov:1995}.

In the case where the point $u$ is located close enough to the manifold $\Omega$ one may note a sensitivity feature of the solution Eq.\eqref{eq:Comb_Elements_Lf} with respect to small deviations of the point $\upsilon$.

Now we back to the made hypothesis, according to it the set $L$ is related to the function $f(t)$. In this case $L$ will be called ``characteristic function set" and denoted as $L_f$. In addition, we can make the following definition \cite{AleksandrovGadjokov1971}.

{\bf{Definition~1}}: The problem of finding the set $L_f$, the parameters $p_{ij}(i=1,2,\dots,n;j=1,2,\dots,s_{i})$ and the amplitudes ${\rm{a}}_{i}(i=1,2,\dots,n)$ is defined as the full problem of analysis of the latent regularities.

Let us analyze the series of observations of the function $\{f(t_{k})\}$\footnote{By the series of observation we will consider the list of the observed nuclei masses, that is given by the AME2012 database}, where $k = 1,2,\dots,m$, and a set $L_f$ with a known number $n$ of presumed relationships be given.

Based on relation Eq.\eqref{eq:Comb_Elements_Lf}, the unknown quantities $p_{ij}$ and ${\rm{a}}_{i}$ can be found by solving the following non-linear (in general, over-determined) system
\begin{equation}
f(t_{k})={\rm{a}}_{0}+\sum_{i=1}^{n}{\rm{a}}_{i}{\rm{e}}^{-h_{i}t_{k}}, ~~k=1,2,\ldots, m,
\label{eq:NonLinearSystem}
\end{equation}

The following vector denotations are introduced:
\begin{equation}
\begin{split}
&x=( {\rm{a}}_{1},\dots,{\rm{a}}_{n},p_{11},\dots, p_{1s_{1}},\dots, p_{n 1},\dots, p_{n s_{n}} ) \in \bm{R}^{l(n)}\\ 
&l(n)=n+\sum_{i=1}^{n}s_{i},\\
&F(x)=[F_{1}(x),\dots,F_{m}(x)]\in \bm{R}^{m},\\
\end{split}
\label{eq:VectorDenotation}
\end{equation}
where
\begin{equation}
F(x_{k})=\left({\rm{a}}_{0}+\sum_{i=1}^{n}{\rm{a}}_{i}{\rm{e}}^{-h_{i}x_{k}} \right) -f(x_{k}), ~~k=1,2,\ldots, m.
\label{eq:VectorDenotation}
\end{equation}

The non-linear system Eqs.\eqref{eq:NonLinearSystem} may be expressed in the vector form as written in Eq.\eqref{eq:fxy_VectorForm}.

Let us assume that the operator $F$ transforms the limited convex subset $X \subseteq \bm{R}^{l(n)}$ into $\bm{R}^{m}$. The Jacobian matrix of the operator $F$ with respect to $x$ is denoted as $F^{\prime}$. When the different relationships have no common parameter the Jacobi matrix can be written in following form \cite{AleksandrovGadjokov1971}:
\begin{equation}
F^{\prime}(x)=\\
\begin{bmatrix}
	g_{1}(t_{1}),..,g_{n}(t_{1}) &{\rm{a}}_{1}\frac{\partial g_{1}(t_{1})}{\partial p_{11}} 
	&...~{\rm{a}}_{1}\frac{\partial g_{1}(t_{1})}{\partial p_{1s_1}} 
	&...~{\rm{a}}_{n}\frac{\partial g_{n}(t_{1})}{\partial p_{n s_{n}}}\\
	g_{1}(t_{2}),..,g_{n}(t_{2}) &{\rm{a}}_{1}\frac{\partial g_{1}(t_{2})}{\partial p_{11}} 
	&...~{\rm{a}}_{1}\frac{\partial g_{1}(t_{2})}{\partial p_{1s_1}} 
	&...~{\rm{a}}_{n}\frac{\partial g_{n}(t_{2})}{\partial p_{n s_{n}}}\\
	\dots\dots\dots\dots
    &\dots\dots\dots
    &\dots\dots\dots
    &\dots\dots\dots\\
	g_{1}(t_{m}),..,g_{n}(t_{m}) &{\rm{a}}_{1}\frac{\partial g_{1}(t_{m})}{\partial p_{11}} 
	&...~{\rm{a}}_{1}\frac{\partial g_{1}(t_{m})}{\partial p_{1s_1}} 
	&...~{\rm{a}}_{n}\frac{\partial g_{n}(t_{m})}{\partial p_{n s_{n}}}
\end{bmatrix}
\label{eq:JacobiMatrix_of_F}
\end{equation}
If two or more relationships exist which have a common parameter (or parameters) the modifications in matrix $F^{\prime}(x)$ are not essential for further considerations.

The operator $H(x) = F^{\prime T}(x)F^{\prime}(x)$ will be considered regular. Let us consider its condition number \cite{GeorgeForsythe:1967}:
\begin{equation}
{\rm{cond}}H(x)=\Arrowvert  H^{-1}(x) \Arrowvert, ~~ \Arrowvert  H(x) \Arrowvert =\frac{\mu_{1}}{\mu_2}\geq 1,
\label{eq:condition}
\end{equation}
where $\mu_{1}$ and $\mu_{2}$ are its greatest and smallest singular numbers \cite{GeorgeForsythe:1967}, respectively. The operator $H(x)$ is only then irregular when $\mu_2=0$. In this case, if $\mu_1 \neq 0 $, the condition number in Eq.\eqref{eq:condition} will be considered as infinitely large. 

Let us define a real number $r$, such as $1\textless r \textless \infty$. Then

{\bf{Definition~2}}: If the non-equality 
\begin{equation}
{\rm{cond}}H(x)~\textless~ r
\label{eq:Nonequality}
\end{equation}
is fulfilled for all $x\in X$, the solution of Eq.\eqref{eq:fxy_VectorForm} is defined as ``the regular problem with the modulus $r$".

It can be easily verified that regularity of the problem Eq.\eqref{eq:fxy_VectorForm} depends on the closeness of the relationships or on their smallness with regard to the modulus, and also on the smallness of the amplitudes ${\rm{a}}_{i}$ or on the smallness with regard to the modulus of the derivatives of the relationships.

If one will assume that in Eq.\eqref{eq:fxy_VectorForm} $\mu_{1}(x)\neq 0$ for all $x\in X$. Then, if $r$ is fixed, 
there will always exist a sufficiently small real number $\delta \textgreater 0$, which makes the Eq.\eqref{eq:fxy_VectorForm} nonregular, if at least one of the non-equalities is fulfilled: 
\begin{equation}
\begin{cases} 
&\sum_{k=1}^{m}\left( g_{i_{1}}(t_{k})-g_{i_{2}}(t_{k}) \right)^{2}\leq\delta; i_{1},i_{2}=1,2,\dots,n; i_{1}\neq i_{2}\\
&{\rm{a}}_{i}~\leq~\delta\\
&\vert g_{i}(t_{k}) \vert ~\leq~\delta;~~k=1,2,\dots,m\\
&\vert \frac{\partial g_{i}(t_{k})}{\partial p_{ij}}\vert~\leq~\delta;~~k=1,2,\dots,m \\
\end{cases}
\label{eq:System_Of_Nonequalities}
\end{equation}

It should be particularly stressed that perturbations of the observed function $\{f(t_k)\}$ (and the presence of perturbations is very characteristic for the problems concerning latent regularities) may also infringe on the regularity of problem Eq.\eqref{eq:fxy_VectorForm}, if at least one of the non-equalities Eq.\eqref{eq:System_Of_Nonequalities} is fulfilled with a sufficiently small value of $\delta$.

When the problem Eq.\eqref{eq:fxy_VectorForm} is non-regular, general calculation algorithms ensuring its solution can be constructed based on the regularized iteration processes for the solution of non-linear equations \cite{Alexandrov5137:1970, Alexandrov:1970, Alexandrov:1973}.

\subsection{Aleksandrov method for Ill-posed nonlinear systems of equations}
\label{AleksandrovMethod_REGN}

When it is not feasible to give a definite indication of the type and number of regularities (i.e. to define exactly the set $L_f$), the full analysis of latent regularities becomes much more difficult. However, even in this case it is possible to find an approximate solution of problem Eq.\eqref{eq:fxy_VectorForm} based on special methods. 
Techniques known as regularization or an auto-regularized\footnote{The subgroup of the regularized Newton-Kantorovich type processes (called ``automatically controlled" or ``auto-regularized"), for which the regularization in each iteration (with number $n$) of the process is performed with respect to value of ``defect" norm of equation ($\vert\vert F(x_{n-1}) \vert\vert$), which was obtained basing on the previous iteration (with number $n-1$).} iteration processes \cite{Tikhonov:1995, TikhonovArsenin:1977, Alexandrov:1970, Alexandrov:1973, Alexandrov:1983, Aleksandrov197146, Alexandrov:1982,  AleksandrovPriv1997} have been developed to deal with this ill-posedness, to get stable approximations of solutions of Eq.\eqref{eq:fxy_VectorForm}. Thus on one hand the classic solution to ill-posed problems is to regularize the unknowns, e.g. by penalizing deviations from a prior model. It consists in approximating a solution of Eq.\eqref{eq:fxy} or Eq.\eqref{eq:fxy_VectorForm} by minimizing the functional
\begin{equation}
x\mapsto \Arrowvert  f(x)-y^{\delta}\Arrowvert^{2} +  \alpha\Arrowvert  x - x_{0}\Arrowvert^{2},
\label{eq:TikhonovRegularization}
\end{equation}
where $x_{0}$ typically unifies all available $a~priori$ information on the solution, or the initial input data (approximation), $\alpha$ is a positive number. On another hand one may use the iterative approaches which are attractive alternatives to regularization, since for solution of the ill-posed problems it is usually necessary to approximate the initial, often infinite-dimensional, problem by a finite-dimensional problem, for which numerical algorithms and computer programs have been devised, for instance, the steepest descent method \cite{ScherzerOmar:1996}, the Landweber iteration \cite{Hank:1995}, and the Landweber-Kaczmarz method, also called algebraic reconstruction techniques (ART) \cite{Herman:1980}. However, in the current work we apply the REGN method \cite{Alexandrov:1970,Alexandrov:1973, Alexandrov:1983, Aleksandrov197146, Alexandrov:1982,  AleksandrovPriv1997, Alexandrov66:1977}, which we found an appropriate choice for our inverse problem that will be formulated in the next section.

The ill-posedness degree provides indications on how badly the deviations from the experimental data are echoed on the quality of the solution. For severely ill-posed problems, the effect of that noise is highly damaging to the solution and has a major impact on the numerical methods. That is why savvy practitioners use a safe approximation of it. A relevant computed solution cannot be achieved without sophisticated regularization strategies. Regarding these issues, we refer to \cite{Alexandrov:1982, Tikhonov:1986, PerChristian:1991, RainerKress:1999} for how to numerically handle severely ill-posed problems, in general, and to \cite{0266-5611-22-4-012, FrankWubbeling:1993}, for the Cauchy problem, specifically. 

The solution of the inverse problem, Eq.\eqref{eq:fxy_VectorForm}, consists in determining such a model $x$ (predicted model) that generates the  predicted data $y^{\delta}$, that fit well the observed data $y$, this solution is generally obtained by deducing system parameters from observations of the system behavior. Such kind of problems, typically involve the estimation of certain quantities based on indirect measurements of these quantities. The estimation process is often ill-posed in the sense that noise in the data may give rise to significant errors in the estimate. In other words, the problem Eq.\eqref{eq:fxy_VectorForm} is ill-posed \cite{Tikhonov:1986} if its solution does not depend continuously on the right hand side $y$, which is often obtained by measurement and hence contains errors:
\begin{equation}
\Arrowvert  y^{\delta}-y \Arrowvert ~\leq\delta.
\label{eq:ySigma}
\end{equation}
Here we denote the measured perturbed data by $y^{\delta}$, $\delta$ is the experimental uncertainty (noise level), $\delta\textgreater0$. The $y$ in Eqs.\eqref{eq:ySigma} and \eqref{eq:fxy_VectorForm} one may interpret as a data set, which can also be characterized as a function of the observation point (in the case of continuous observations), or as a vector (in the case of discrete observations). 

One question, which arises when solving an inverse task, is whether the results obtained are proven. This question can be answered by comparing the results obtained by independent methods. To illustrate such a comparison, the binding energies and atomic masses were used as an example, see next section.

Alexandrov in his works \cite{Alexandrov:1970, Alexandrov:1973, Alexandrov:1983, Aleksandrov197146, Alexandrov:1982,  AleksandrovPriv1997, Alexandrov66:1977} propose to search for the solution of Eq.\eqref{eq:fxy}, or Eq.\eqref{eq:fxy_VectorForm},  for two formulations
\begin{equation}
y\in \bm{R}_{f}\equiv FD_{f},\\
\label{eq:Formulations1}
\end{equation}
\begin{equation}
y\in \bm{R}^{m} \setminus \bm{R}_{f},
\label{eq:Formulations2}
\end{equation}
which are covered by transition to the  ``$\varphi$-transformed" \cite{Alexandrov:1973}  problem
\begin{equation}
F(x) \equiv \varphi(x) (f(x)-y)=0,
\label{eq:PhiTransformedProblem}
\end{equation}
where 
\begin{equation}
\varphi(x):\bm{R}_{\varphi} (  \equiv {\bm{R}}_{f}\setminus \{ y \} ) \rightarrow D_{f}
\label{eq:varphi_x}
\end{equation}
is the given linear operator function. Since $\varphi(x)$ as the linear operator transform $\bm{R}_{\varphi}$ to $D_{f}$, and as the nonlinear function of the $x$ transform $D_{f}$ into the space of the confined real matrices, ${\mathcal{L}}(\bm{R}^{n},\bm{R}^{m})$. The space of the confined real matrices is characterized by the norm
\begin{equation}
\Arrowvert A \Arrowvert_{\infty} =\max_{i}  \left(  g_{i}   \sum_{i=1}^{n}  \frac{\vert a_{ij}\vert}{g_{i}} \right),
\label{eq:RealMatrixNorm}
\end{equation}
where $A=\{ a_{ij}\}  \in {\mathcal{L}}(\bm{R}^{n},\bm{R}^{m})$.

The transformation function, $\varphi(x)$ in Eq.\eqref{eq:PhiTransformedProblem} in the basic case (second Gauss transformation) can be written as:
\begin{equation}
\varphi(x)=(f^{\prime}(x))^{T}=\Bigg\{ \frac{ \partial f_{j}(x)}{ \partial x_{i}}  \Bigg\}^{T}_{{i=1,2,\dots,n;}{j=1,2,\dots,m;}} 
\label{eq:PhiTransfFunction}
\end{equation}

Therefore, the Eq.\eqref{eq:PhiTransformedProblem} for both formulations, Eqs.\eqref{eq:Formulations1}-\ref{eq:Formulations2} in accordance to the Alexandrov proposal \cite{Alexandrov:1970,Alexandrov:1973, Alexandrov:1983, Aleksandrov197146, Alexandrov:1982,  AleksandrovPriv1997, Alexandrov66:1977} reduced to search for the solution of the ``$f^{\prime T}(x)W^{1/2}$-transformed" \cite{Alexandrov66:1977} and the $\alpha$-regularized \cite{Tikhonov:1986} problem
\begin{equation}
F(x_{k}) \equiv f^{\prime T}(x_{k})W^{1/2}(f(x_{k})-y)+\alpha x_{k}=0, ~~\alpha\textgreater 0,
\label{eq:RegularizedChi2}
\end{equation}

where 
\begin{equation}
W={\rm{diag}}(\omega_{1},\omega_{2},\dots, \omega_{m}), ~~\omega_{m}\textgreater 0,
\label{eq:WeightMatrix}
\end{equation}
is the weighting matrix,  $\omega_{m}\in (0, {\infty})$ are the given constants, and $\alpha: D_{f}\rightarrow D_{f}$ is the given nonlinear operator; $F$ is a continuous and Frechet differential nonlinear operator.

In the case, when the $\bm{R}^{n}$ is the Banach space, then it is normalized by the vector norm
\begin{equation}
\Arrowvert x \Arrowvert_{\infty} =\max_{i}(g_{i} \vert x_{i}\vert).
\label{eq:VectorNorm}
\end{equation}
where $g_{i}~\textgreater~0 ~(i=1,2,\dots, n)$ are the given weights. When the $\bm{R}^{n}$ is the Euclidean space then it is characterized by the Euclidean norm:
\begin{equation}
\Arrowvert x \Arrowvert_{2} = \left(  \sum_{i=1}^{n} x_{i}^{2} \right)^{1/2},
\label{eq:EuclideanNorm}
\end{equation}

Further, by applying the iteratively regularized Gauss-Newton method within double regularization and Eq.\eqref{eq:RegularizedChi2} we can build the following regularized process:
\begin{equation}
\begin{split}
R_{\varepsilon_{k}}: x_{0}, x_{k+1} & = x_{k}-\left[ f^{\prime T}(x_{k})W^{1/2}f^{\prime}(x_{k})+(\alpha + \varepsilon_{k}) I_{\bm{R}^{n}} \right]^{-1} F(x_{k})
\label{eq:GaussNewton}
\end{split}
\end{equation}
which can be used to find an approximate solution of Eq.\eqref{eq:fxy_VectorForm}. 
Here $\alpha$ is the Tikhonov regularization parameter, $\varepsilon_{k}$ is the Aleksandrov feedback regularization parameter. The both regularization parameters are adjustable from iteration to iteration. $x_{0}$ and $\varepsilon_{0}$ are the initial input data (approximation), in other words is an initial guess which may incorporate $a~priori$ knowledge on an exact solution, in our case it is the generalized BW mass formula. The $x_{k}$ denotes the iterative solution, $k=1,2,\dots,k^{\ast}\textless \infty$,  where $k^{\ast}$ is the pseudo solution $x_{k^{\ast}}$ (i.e. the solution in the best iteration).
The process, $R_{\varepsilon_{k}}$ is converge according to the theorem stated in Ref.\cite{Alexandrov:1970}. 

Therefore, as one may note the key idea of any Newton type method consists in repeatedly linearizing the operator Eq.\eqref{eq:fxy_VectorForm} around some approximate solution $x_{k}$, and then solving the linearized problem Eq.\eqref{eq:GaussNewton} for $x_{k+1}$. However, usually these linearized problems are also ill-posed if the nonlinear problem is ill-posed and, therefore, they have to be regularized.

The linear Eq.\eqref{eq:GaussNewton} can be solved with the help of one of the three methods \cite{Alexandrov:1970}: Cholesky Decomposition \cite{GeneGolub:1996, WilliamHPress}, Gauss-Jordan Elimination \cite{Steven:102307} or Singular Value Decomposition \cite{WilliamHPress_C:1992}. Aleksandrov in turn propose \cite{Alexandrov:1973, Alexandrov:1983, Aleksandrov197146,  Alexandrov66:1977} the following regularizator for the process Eq.\eqref{eq:GaussNewton}:
\begin{equation}
\begin{split}
&R_{\varepsilon_{k}}:\varepsilon_{k}=\frac{\alpha_{2}}{2} \left[ \left( \tau_{k}^{2} + 4 N_{0}(\varrho_{k})^{p_{r}} \right)^{1/2} - \tau_{k}\right].
\label{eq:AleksandrovReg}
\end{split}
\end{equation}
Here, $\varrho_{k}$ and $\chi^{2}_{k}$ are both the iteration process criteria, namely the ``target criteria", which can be written in the following form:
\begin{equation}
\begin{split}
\varrho_{k}& = \Arrowvert F(x_{k})\Arrowvert_{\infty} ~~~{\rm{or}}~~~\varrho_{2k} = \Arrowvert F(x_{k})\Arrowvert_{2},\\
\chi^{2}_{k}& = \Arrowvert  W^{1/2}(f(x_{k})-y) \Arrowvert_{2}^{2}.
\label{eq:AleksandrovVar}
\end{split}
\end{equation}
Based on these criteria one may say that obtained solution of the inverse problem is a fake or true. The $F(x_{k})$ for $\varrho_{k}$ defined from Eq.\eqref{eq:RegularizedChi2}. The functionals $\tau_{k}$ and $N_{0}$ in Eq.\eqref{eq:AleksandrovReg} are the iteration process behavior criteria, that defined as:
\begin{equation}
\begin{split}
\tau_{k} &=\Arrowvert f^{\prime T}(x_{k})W^{1/2}f^{\prime}(x_{k})  +\alpha I_{\bm{R}^{n}} \Arrowvert_{\infty}, \\
N_{0} & = \frac{\alpha_{1}}{\varrho_{0}} \left(\tau_{0}\varepsilon_{0} + \varepsilon_{0}^{2}\right) ,  ~~\varepsilon_{0}\geq 0.
\label{eq:IteraProcBehaviorCriteria}
\end{split}
\end{equation}
where $\varrho_{0}\neq 0$, or $N_{0}$ is a given positive number.  
The behavior criteria provides information about the degree of degeneracy of the Jacobi matrix on each step of iteration, and thus sheds light on the stability of the implemented process Eq.\eqref{eq:GaussNewton}. The $\alpha_{1}$, $\alpha_{2}$ and $p_{r}$ are the internal coefficients of the auto-regularized process. 
By default they are equal to unity, however in the more general case $\alpha_{1},\alpha_{2}\in [0,1]$ and $p_{r}\geq 1$.

It was proven in the Ref.\cite{Alexandrov:1970,AlexandrovLCH:1976} that under condition
\begin{equation}
\begin{split}
\Arrowvert  \left(\varphi(x_{k}) -  \varphi(x_{k^{\ast}}) \right)  (f(x_{k})-y) \Arrowvert ~\leq~ Q\Arrowvert \varphi(x_{k}) (f(x_{k})-y) + \alpha x_{k} \Arrowvert,
\label{eq:ConvergeCreteria}
\end{split}
\end{equation}
where $Q\in[0,1)$ the process Eq.\eqref{eq:GaussNewton} quadratically converge and Eq.\eqref{eq:RegularizedChi2} has a solution.
In the case when the condition Eq.\eqref{eq:ConvergeCreteria} is not satisfied, then the process Eq.\eqref{eq:GaussNewton} converge linearly.

With analogy to the Gauss-Newton processes, Eq.\eqref{eq:GaussNewton}, one may consider the auto-regularized Newton-Kantorovich and Schroder \cite{Schroder:1870} type of processes. For this purpose one may need to construct expressions similar to Eq.\eqref{eq:AleksandrovReg}. For instance, one may consider the following Newton-Kantorovich  process
\begin{equation}
\begin{split}
&R_{NK}:\\
&x_{0}, x_{k+1} = x_{k}-v_{k}\left[ f^{\prime T}(x_{k})f^{\prime}(x_{k}) + f^{\prime\prime}(x_{k}(f(x_{k}-y) + \varepsilon_{k}I_{\bm{R}^{n}}\right] F(x_{k}), 
\label{eq:NewtonKantorovich}
\end{split}
\end{equation}
for which the regularization parameter of Eq.\eqref{eq:GaussNewton} is taken in the form of:
\begin{equation}
\varepsilon_{k}=\frac{1}{2} \left(  (\lambda_{k})^{2}  + 4 N\varrho_{k} - \lambda_{k} \right), 
\label{eq:AleksandrovAutoReg}
\end{equation}
where $\lambda_{k}=\lambda(x_{k})$ is the norm of the linear symmetric operator which operates on the $D_{f}\rightarrow D_{f}$
\begin{equation}
\lambda_{k}=\Arrowvert f^{\prime T}(x_{k})f^{\prime}(x_{k})+f^{\prime\prime}(x_{k})(f(x_{k})-y) \Arrowvert,
\label{eq:labdamin}
\end{equation}
and
\begin{equation}
\begin{split}
N_{0} & = \frac{1} {\varrho_{0}}  \left( \lambda_{\rm{min}}^{0}\varepsilon_{0} + \varepsilon_{0}^{2} \right),\\
\varrho_{k}& = \Arrowvert F(x_{k}) \Arrowvert_{2},
\label{eq:AleksandrovAutoRegVar}
\end{split}
\end{equation}
where $f^{\prime}(x)$, $f^{\prime\prime}(x)$ is the first and the second Frechet derivative correspondingly, $\varrho_{0}\neq 0$, or $N_{0}$ is a given positive number.  

In an analysis of the experimental data, the standard deviation, $\sigma_{m}$, of the measured quantities $y_{j}(j=1,2,\dots,m)$ are often well defined in advance. 
If so, in Eq.\eqref{eq:RegularizedChi2}, one may use the weighting matrix as:
\begin{equation}
W={\rm{diag}}(\sigma_{1}^{-1},\sigma_{2}^{-1},\dots, \sigma_{m}^{-1}).
\label{eq:WeightMatrixAutoRegEuclNorm}
\end{equation}
When the mathematical formulation of the problem is clear, but the experimental uncertainties
\begin{equation}
((\Delta y)_{1},(\Delta y)_{2},\dots, (\Delta y)_{m})^{T}
\label{eq:ExpUncertaities}
\end{equation}
are very rough or unknown, one may use the robust methods to determine the weights \cite{AleksandrovMavrodiev:1976, Aleksandrov1980520} or the LCH-weighting procedure\footnote{LCH is the modified least $\chi^{2}$ procedure, which investigates the possibility of the existence of the minimal $\chi^{2}$ from the parameters that were excluded from the considered model due to strong correlation.} \cite{AleksandrovMavrodiev:1976,  AlexandrovLCH:1976,  Aleksandrov2004519, LCH:1999, Mishev:2733070}.
The meaning of this procedure is based on the preliminary solution of the problem Eq.\eqref{eq:RegularizedChi2} with the help of the weight matrix $W=I_{\bm{R}^{n}}={\rm{diag}}(1,1,\dots, 1)$. This gives us the intermediate pseudo solution $x^{\ast}$ of the system Eq.\eqref{eq:RegularizedChi2} (or Eq.\eqref{eq:GaussNewton}) and then we solve it again, with the help of the weight matrix $W$ \cite{AleksandrovMavrodiev:1976, Aleksandrov1980520}:
\begin{equation}
W={\rm{diag}} \left( \frac{1}{ \vert f_{1}(x^{\ast})-y_{1}\vert }, \frac{1}{\vert f_{2}(x^{\ast})-y_{2}\vert } \dots, \frac{1}{\vert f_{m}(x^{\ast})-y_{m}\vert } \right).
\label{eq:W2}
\end{equation}
To be sure, that the true solution is converge one may repeat this process several times.

The essential difference between Aleksandrov method and similar methods \cite{Bjork:1996, Nguyen_UMJ:2005, ARGYROS20151318, YANG20151339} 
based on especially effective ideology of regularization of the inverse problem, is that at each step (iteration) one may control both the decision and the uncertainty of the solution, which is important. Aleksandrov regularization stabilizes the solution of our ill-posed operator equations by minimizing the weighted sum of a strictly convex regularization term. The regularization term stabilizes the solution of the problem at the expense of biasing the solution. Note, that the regularizator proposed by Aleksandrov to solve nonlinear ill-posed problems with the help of the iterative approach is the constructive development of the Tikhonov-Glasko regularizator  \cite{Tikhonov196593}:
\begin{equation}
\alpha_{k}=\left( \frac{1}{2} \right)^{k-1} x_{k}
\label{eq:TikhonovGlasko}
\end{equation}
and the regularization parameter proposed by Ramm \cite{Ramm:2000}
\begin{equation}
\alpha_{k}=\left( \frac{1}{\rm{e}} \right)^{k-1} x_{k}.
\label{eq:Ramm}
\end{equation}

Therefore, the purpose of such functions as in Eqs.\eqref{eq:AleksandrovReg}, \eqref{eq:TikhonovGlasko} and \eqref{eq:Ramm}, to regularize the solution by filtering out the contributions of the noise in $y$.%

The processing of the experimental data of different nature often leads to the automatized solution of one-type systems, connected with the analysis of hidden (latent) regularities, i.e. hidden exponents, hidden Gauss- or Breit-Wigner functions and so on \cite{ AleksandrovGadjokov1971, Alexandrov:19701285, AleksandrovHRegul:1972}. The number of the hidden regularities involved in each problem is unknown in the general case. In the next section we show how these regularities been discovered in the description of the nuclear binding energies.

Note, that with the help of the justification of the ``enhanced Newton’s method" developed by Collatz Ref.\cite{LotharCollatz:1964}, Alexandrov manage to obtain the general theory of convergence and attraction, which cover the iterative regularized Newton-Kantorovich and Gauss-Newton processes. Prior to work \cite{Alexandrov:1973} the  general theory of convergence and attraction for the ordinary methods of the Newton-Kantorovich and Gauss-Newton type did not exist.

\section{The Bethe-Weizs\"{a}cker mass formula and binding energy of the nucleus}
\label{Nuclear Binding Energy}

During the first few decades of nuclear physics, the term `nuclear forces' was often used as being synonymous with nuclear physics as a whole \cite{Rosenfeld:1948, Eder:1965}. Nevertheless, even today with the onset of quantum chromodynamics, in any first approach towards a nuclear structure problem, one assumes the nucleons to be elementary particles.

In the theory of the liquid drop model proposed by George Gamow \cite{GGamow:1931}, the binding energy per nucleon is given by formula
\begin{equation}
\begin{split}
E_{B}(A,Z)&=\alpha_{vol}+\alpha_{surf}\frac{1}{A^{1/3}}+\alpha_{comb}\frac{Z(Z-1)}{A^{4/3}}\\
&+\alpha_{sym}\frac{(N-Z)^{2}}{A^{2}}+\alpha_{pairing}\frac{\delta(A,Z)}{A}+\alpha_{\rm{Wigner}}\frac{1}{A},
\label{eq:LiquidDrop_BW}
\end{split}
\end{equation}
when $\alpha_{\rm{Wigner}} \rightarrow 0$, Eq.~\eqref{eq:LiquidDrop_BW} reduces to the usual BW formula. 

Due to historical reasons the BW mass formula, Eq.\eqref{eq:LiquidDrop_BW}, provides the baseline fit, relative to which all refinements can be judged. That is why we decided to use this formula in our formulation of the inverse problem, which will be discussed later in the paper. 

One may note, that the leading linear behavior of the binding energy as a function of the number of nucleons  is a manifestation of the saturation of nuclear forces, see Fig.\ref{fig:BindingEnergyA}. Another manifestation of saturation, following Bethe \cite{RevModPhys.8.82}, the proportionality of both the total binding energy and the nuclear volume at first approximation to the number of nucleons. Nuclei have diffuse surfaces, so the notion of volume is somewhat fuzzy, but an indicator is provided by the measured root-mean squared (rms) radius, $\langle r^{2} \rangle^{1/2}$, which sets a linear scale of the nuclear size. The volume term, $\alpha_{vol}$, in  Eq.\eqref{eq:LiquidDrop_BW} is expected to be positive, the other three $\alpha$'s negative. 

Another an obvious feature in the BW is the effect of the repulsive Coulomb interaction between protons, that information is encoded in the third term of Eq.\eqref{eq:LiquidDrop_BW}. On the basis of simple electrostatics and dimensional arguments, the total Coulomb energy of the nucleus is expected to be proportional to the square of its charge and inversely proportional to its length scale, $E_{C}\propto Z^{2}e^{2}/A^{1/3}$ (called also the proton form-factor correction to the Coulomb energy in \cite{Moller1995185}), where $e^{2} = 1.44$ MeV fm in convenient units. 

The surface energy term, second term of Eq.\eqref{eq:LiquidDrop_BW}, takes into account the deficit of binding energy of the nucleons at the nuclear surface and corresponds to semi-infinite nuclear matter. 

The symmetry term, the fourth term of Eq.\eqref{eq:LiquidDrop_BW}, has another motivation in the BW mass formula, it produces a rough balance between the number of neutrons $N$ and number of protons $Z$ among the $A = N + Z$ nucleons. In the absence of this term, the semi-empirical mass formula would suggest maximum binding for a nucleus consisting only of neutrons.  The symmetry energy has two contributing factors $\textendash$ a kinetic energy term and a potential energy term \cite{Angeli2004185}, that have the same form and proportional to $(N-Z)^{2}/A$. Since the isoscalar part of the interaction does not distinguish between protons and neutrons, a qualitative estimate of the potential energy contribution is obtained by considering the isovector part of the nucleon-nucleon interaction. It has the same form as the kinetic energy, which is usually estimated in a Fermi gas model and proportional to $(N-Z)^{2}/A$. Note the symmetry term, recalling binding energy connection to the isospin $I$, which was introduced by Myers and Swiatecki \cite{WDMyers:1966}. It should be noted that a more sophisticated treatment of the nuclear interaction (as in the seniority shell model or the Wigner supermultiplet theory \cite{Talmni:1993}) would produce symmetry energies of the form $I(I+1)$ or $I(I+4)$, respectively, where the nuclear isospin $I=\vert N-Z\vert/2$.

Despite the widespread use of formula, Eq.\eqref{eq:LiquidDrop_BW}, in nuclear physics it was found that it is severely inadequate for light-mass nuclei, especially away from the line of stability \cite{Heyde:1999}, after the discovery of neutron rich light nuclei. The unusual stability of nuclei with preferred nucleon numbers, commonly referred to as magic numbers, can be clearly delineated by comparing the experimental binding energies with predictions of the BW mass formula. However, this comparison does not indicate the recently observed features like the disappearance of some traditional magic numbers \cite{GuillemaudMueller198437, Motobayashi19959, PhysRevLett.83.496, Keller:1994ZurPhysA,  PhysRevLett.84.5493} and extra stability for some newly observed nuclei.

These issues motivate the recent phenomenological searches for a more generalized BW formula \cite{PhysRevC.65.037301, ADHIKARI:2004, doi:35IWNTheory, doi:IJNonLinearAn_EnSyst}, which can explain the gross features of the shapes of the binding energy versus neutron number curves of all the elements from $Li$ to $Bi$. Nevertheless in order to understand and appreciate the relevance and reliability of the semi-empirical mass formula, it is necessary to study not only the effect of adding specific terms to the original BW formula, but also the interplay between the different terms. At this point, we would like to reduce the excessive inflation of discussion about different terms of the BW, the interested reader is referred to this Ref.~\cite{Kirson200829} and references therein to familiarize oneself with rich history proposals of additional terms for inclusion in the formula. 
\begin{table}[ht]
	\begin{center}
		\begin{tabular}{ccccc}
			\hline \hline
			\multicolumn{1}{c}{$\alpha_{vol}$}
			&\multicolumn{1}{c}{$\alpha_{surf}$}
			&\multicolumn{1}{c}{$\alpha_{comb}$}
			&\multicolumn{1}{c}{$\alpha_{sym}$}
			&\multicolumn{1}{c}{$\alpha_{pairing}$}\\
			\hline
			\scriptsize{16.00126\cite{Moller1995185}} 
			&\scriptsize{21.18466\cite{Moller1995185}}  
			&\scriptsize{-} 
			&\scriptsize{-}  
			&\scriptsize{-}\\
			\scriptsize{16.24\cite{Myers1996141}} 
			&\scriptsize{18.63\cite{Myers1996141}}  
			&\scriptsize{-} 
			&\scriptsize{-}  
			&\scriptsize{-}\\
			\scriptsize{15.36(3)\cite{Kirson200829}} 
			&\scriptsize{16.42(8)\cite{Kirson200829}} 
			&\scriptsize{0.691(2)\cite{Kirson200829}} 
			&\scriptsize{22.53(7)\cite{Kirson200829}}  
			&\scriptsize{-}\\  
			\scriptsize{15.77(3)\cite{Samanta:2004et}} 
			&\scriptsize{18.341\cite{Samanta:2004et}}   
			&\scriptsize{0.711\cite{Samanta:2004et}}  
			&\scriptsize{23.211\cite{Samanta:2004et}}
			&\scriptsize{11.996\cite{Chowdhury:2004jr}}\\ 
			\scriptsize{15.75\cite{Rohlf:1994}} 
			&\scriptsize{17.8\cite{NIX19651, Rohlf:1994}}  
			&\scriptsize{0.711\cite{Rohlf:1994}} 
			&\scriptsize{23.7\cite{Rohlf:1994}}  
			&\scriptsize{11.18\cite{Rohlf:1994}}\\
			\scriptsize{15.835\cite{Wapstra:1958}} 
			&\scriptsize{18.3299\cite{Wapstra:1958}}  
			&\scriptsize{0.1785\cite{Wapstra:1958}} 
			&\scriptsize{23.203\cite{Wapstra:1958}}  
			&\scriptsize{$\pm 11.2^{\ast}$\cite{Wapstra:1958}}\\ 
			\scriptsize{-} 
			&\scriptsize{14 \cite{PhysRev.89.1102}}  
			&\scriptsize{-} 
			&\scriptsize{-}  
			&\scriptsize{-}\\
			\scriptsize{-} 
			&\scriptsize{17.9439\cite{WDMyers:1966}}  
			&\scriptsize{-} 
			&\scriptsize{-}  
			&\scriptsize{-}\\
			\scriptsize{-} 
			&\scriptsize{20.69\cite{PhysRevC.33.2039}}  
			&\scriptsize{-} 
			&\scriptsize{-}  
			&\scriptsize{-}\\
			\scriptsize{15.4920\cite{PhysRevC.67.044316}} 
			&\scriptsize{16.9707\cite{PhysRevC.67.044316}}  
			&\scriptsize{-} 
			&\scriptsize{-}  
			&\scriptsize{-}\\
			\scriptsize{15.3982 \cite{Royer20131}} 
			&\scriptsize{17.3401 \cite{Royer20131}}  
			&\scriptsize{0.6 \cite{Royer20131}} 
			&\scriptsize{-}  
			&\scriptsize{-}\\
			\hline
			\scriptsize{-15.794 \cite{GORIELY2001311}} 
			&\scriptsize{17.77 \cite{GORIELY2001311}}  
			&\scriptsize{-} 
			&\scriptsize{27.95 \cite{GORIELY2001311}}  
			&\scriptsize{-}\\
			\scriptsize{-15.805 \cite{Samyn2002142}} 
			&\scriptsize{17.5 \cite{Samyn2002142}}  
			&\scriptsize{-} 
			&\scriptsize{27.81 \cite{Samyn2002142}}  
			&\scriptsize{-}\\
			\scriptsize{-15.794  \cite{PhysRevC.66.024326}} 
			&\scriptsize{17.5  \cite{PhysRevC.66.024326}}  
			&\scriptsize{-} 
			&\scriptsize{28.00  \cite{PhysRevC.66.024326}}  
			&\scriptsize{-}\\
			\scriptsize{-16.247 \cite{MYERS1974186}} 
			&\scriptsize{22.92(22) \cite{MYERS1974186}}  
			&\scriptsize{-} 
			&\scriptsize{32.73(35) \cite{MYERS1974186}}  
			&\scriptsize{-}\\
			\hline
			\hline
			\multicolumn{1}{c}{$\langle \alpha_{vol} \rangle$}
			&\multicolumn{1}{c}{$\langle \alpha_{surf} \rangle$}
			&\multicolumn{1}{c}{$\langle \alpha_{comb} \rangle$}
			&\multicolumn{1}{c}{$\langle \alpha_{sym} \rangle$}
			&\multicolumn{1}{c}{$\langle \alpha_{Wigner} \rangle$}\\
			\hline
			\scriptsize{19.12231} 
			&\scriptsize{18.1929}  
			&\scriptsize{0.518735} 
			&\scriptsize{12.54048}  
			&\scriptsize{28.9866}  \\
			\hline \hline
		\end{tabular}
		\caption{
List of the BW mass formula coefficients (in MeV) obtained using different fitting procedures and the corresponding rms deviations (in keV). The sign ``-" in the $\alpha_{pairing}$ column corresponds to even-even nuclei, and the ``+" to odd-odd nuclei. Macroscopic parameters of the different mass formulas based on Hartree-Fock-Bogoliubov theories \cite{Samyn2002142, GORIELY2001311, PhysRevC.66.024326} and the FRDM \cite{Moller1995185, Moller1988213, MYERS1974186} are shown in the center of the table and are underlined by a horizontal line. 
The quantities in parentheses for the FRDM denote the values used in the microscopic part of the model. ${\ast}$  denotes the most accurate value of the pairing term obtained from the $\beta$ decay energies of isodiapheres as a function of the mass number. The less accurate value is 33.5 MeV \cite{Wapstra:1958}. Results of the current work shown in the bottom line. 
		}
		\label{tab:BW_CoeffFits}
	\end{center}
\end{table}

The existence of the heaviest atomic nuclei bound against immediate disintegration depends on the detailed arrangement of protons, $Z$, and neutrons, $N$, in shells that provide a second effect on nuclear stability, which is analogous to the way electrons fill atomic orbitals; see, for instance, \cite{0954-3899-34-4-R01,  Cwiok:2005, GRYZINSKI1973131, Gryzinski1976180} and references therein. Some of the effective interactions used in the shell model approach \cite{PhysRev.75.1969, PhysRev.75.1766.2} can be traced back to the nucleon-nucleon Brueckner $G$-matrix \cite{PhysRev.95.217, PhysRev.96.508, PhysRev.97.1344}. Closed shells are energetically favorable, and this modulates the smooth trend of the binding energy predicted by the liquid drop model. It has proven very successful, that few valence nucleons (outside closed shells) can describe the properties of nuclei \cite{Talmni:1993}, 
including energy levels, magnetic and quadrupole moments, electromagnetic transition probabilities, beta-decay rates and reaction cross-sections. Moreover, this approach has also been used as the theoretical basis for several algebraic nuclear models \cite{DaGeng:1997, Cseh2015213}. Therefore, the shell model became the standard model for describing the systematics observed in the spectra and transition intensities of $p-$ \cite{COHEN19651}, $sd-$ \cite{Wildenthal19845, Brown1988191, BrownAnnRev:1988} and lower $fp$-shell \cite{French:1969, RICHTER1991325, PhysRevC.50.225, PhysRevC.55.187} nuclei. Since the size of the model space increases rapidly with the number of valence nucleons and/or orbits, full major shell calculations were limited to nuclei with $A \textless 49$ \cite{PhysRevC.50.225, PhysRevC.55.187}.

Therefore continuous improvements are still needed to determine of the nuclear masses and their structure. 

\section{Parametrization of the Bethe-Weizs\"{a}cker mass formula}
\label{Parametrization_of_the_BWmass_formulae}

So far, due to absence of  rigorous theory of the nucleus the role of the various correction terms in the BW formula remain unclear \cite{PhysRevC.65.037301, ADHIKARI:2004, Kirson200829}. As far as their empirical basis is concerned, the introduction of correction terms into the equation for the binding energy, despite the addition of new parameters, does not lead to any appreciable improvement in the agreement with experiment. Nevertheless, the introduction of purely empirical corrections to the parity and the effects of the shells considerably improves the agreement with experiment \cite{Cameron:1957, KUMMEL1966129, Strutinskii:1966}.

The present paper attempts to provide a general mathematical formulation of this problem and to relate its occurrence to a new research strategy. A general calculational approach to the analysis of latent regularities is suggested, based on auto-regularized iteration processes for solving non-linear problems, see Sec.\ref{AleksandrovMethod_REGN}.

The aim of our study was to investigate the BW mass formula correction terms correlation for the whole set of nucleus masses that have been measured \cite{Ami2012_ChinPhysC} and to provide information on latent regularities that may affect the nuclei formation. Recent fits to the AME2012 atomic mass evaluation \cite{Wapstra2003129, Ami2012_ChinPhysC} are given in Table~\ref{tab:BW_CoeffFits}. One may consider the AME database as a good playground for developing and testing different nuclear models.

In this section, we analyze fluctuations of the nuclei and atomic masses as a many-body quantum system; we rewrite the BW mass formula Eq.\eqref{eq:LiquidDrop_BW} in the functional analytical framework of an operator equation. In particular, we discuss the existence and origin of a chaotic part in the binding energy and its possible implications on the accuracy of theoretical mass predictions in atomic nuclei. A careful analysis of the previous attempts of calculation nucleus masses reveals that all models can reproduce experimental/empirical trends on the average 
\cite{
Moller1995185, 
Wapstra2003129, 
Chowdhury:2004jr, 
Kirson200829,
WDMyers:1966,
Myers1996141, 
Samanta:2004et, 
Rohlf:1994, 
NIX19651, 
Wapstra:1958,
PhysRev.89.1102, 
PhysRevC.33.2039,
PhysRevC.67.044316,
Royer20131}. 
In order to overcome this issue we consider a parameterized nonlinear dynamical system of equations for determining nuclei and atomic masses from the experimental bound-state energies, which can be written using the vector notation of Eq.\eqref{eq:fxy_VectorForm} as: 
\begin{equation}
FE_{B,j}^{\rm{Th}}(A,Z,\{a_{i}\}) =E_{B,j}^{\rm{Expt}}(A,Z),
\label{eq:BE_j}
\end{equation}
\begin{equation}
FM_{a.m.,j}^{\rm{Th}}(A,Z,\{a_{i}\}) = M_{a.m.,j}^{\rm{Expt}}(A,Z),
\label{eq:Mam_j}
\end{equation}
\begin{equation}
FM_{n.m.,j}^{\rm{Th}}(A,Z,\{a_{i}\}) = M_{n.m.,j}^{\rm{Expt}}(A,Z),
\label{eq:Mnm_j}
\end{equation}
\begin{equation}
F\Delta m_{j}^{\rm{Th}}(A,Z,\{a_{i}\}) = \Delta m_{j}^{\rm{Expt}}(A,Z),
\label{eq:MExess_j}
\end{equation}
where $F$ is the rectangular $d\times m$ $\{d=1,\dots,{\mathcal{N}}_{\rm{data}}; m=1,\dots,{\mathcal{N}}_{\rm{param}}\}$ Jacobian matrix composed of the Frechet derivatives with respect to the model parameters. Each system of Eqs.\eqref{eq:BE_j} - \eqref{eq:MExess_j} contains 2564 equations, which correspond to the number of the experimental data-points. Therefore, these systems are overdetermined because the number of considered equations exceeds the number of parameters used in the fit. The right hand side of these equations represented by the vector of the observed experimental data, $j$ is the component of this vector, also it indicates the nonlinear equation in the system.

The solution of the overdetermined system of Eqs.\eqref{eq:BE_j} for the binding energy and its model is given by the real values of the parameters $a_{i}$, where $i=1,\dots, {\mathcal{N}}_{\rm{param}}$ is the number of unknown parameters. 
Due to ill-posedness of the formulated problem to obtain a true solution we apply the iterative approach, which is based on the auto-regularized iteration process of Gauss-Newton type that was described in the previous sections (Sec.\ref{Theory_and_Method}). For more details we recommend to read the following references \cite{Alexandrov:1970, Aleksandrov197146, Alexandrov:19701285}.

The main idea of the any regularization algorithm is to consider, instead of one ill-posed inverse problem Eq.\eqref{eq:BE_j}, a family of well-posed problems,
\begin{equation}
F_{\varepsilon}E_{B,j}^{\rm{Th}}(A,Z,\{a_{i}\}) =E_{B,j}^{\rm{Expt}}(A,Z),
\label{eq:BE_WellposedFamily}
\end{equation}
which approximate the original inverse problem in some sense. The scalar parameter $\varepsilon > 0$, which is defined using Eq.\eqref{eq:AleksandrovReg}, is called a regularization parameter. The considered regularization algorithm Eq.\eqref{eq:BE_WellposedFamily} allows one to impose the following requirement on the system (Eqs.\eqref{eq:BE_j}--\eqref{eq:MExess_j})
\begin{equation}
E_{B,j}^{\rm{Th}}(A,Z,\{a_{i}\}) \rightarrow  E_{B,j}^{\rm{true}}(A,Z),~{\rm{if}}~ \varepsilon \rightarrow 0,
\label{eq:BE_true}
\end{equation}
where $E_{B,j}^{\rm{Th}}(A,Z,\{a_{i}\})$ is the solution of the inverse problem  Eq.\eqref{eq:BE_WellposedFamily}, and $E_{B,j}^{\rm{true}}(A,Z)$ is the true solution of the original problem Eq.\eqref{eq:BE_j}. Thus, we replace the solution of one ill-posed inverse problem by the solutions of the family of well-posed problems, assuming that these solutions, $E_{B,j}^{\rm{Th}}(A,Z,\{a_{i}\})$, tend asymptotically to the true solution, as $\varepsilon$ tends to zero. 

In other words, the applied regularization algorithm, as in any other,  is based on the approximation of the noncontinuous inverse operator $F^{-1}$ by the family of continuous inverse operators $F^{-1}_{\varepsilon} E_{B,j}^{\rm{Expt}}(A,Z)$  that depend on the regularization parameter $\varepsilon$. The regularization must be such that, as $\varepsilon$ vanishes, the operators in the family should approach the exact inverse operator $F^{-1}$.

The regularized solution of the ill-posed inverse problem, Eq.\eqref{eq:BE_j}, is provided by minimization of the corresponding parametric functional
\begin{equation}
\begin{split}
&{\mathcal{J}}_{\varepsilon}(E_{B}^{\rm{Th}}(A,Z,\{a_{i}\})) =\\
&{\rm{min}}~\Bigg\{  \Arrowvert F E_{B}^{\rm{Th}}(A,Z,\{a_{i}\}) - E_{B}^{\rm{Expt}}(A,Z)\Arrowvert^{2} + (\alpha+\varepsilon)  \Arrowvert {\mathcal{W}}  E_{B}^{\rm{Th}}(A,Z,\{a_{i}\})\Arrowvert^{2} \Bigg\},
\label{eq:Parametric_functional}
\end{split}
\end{equation}
where the $j$ and $k$ indices are omitted for brevity, the Tikhonov and Aleksandrov regularization parameters are denoted as, $\alpha$ and ${\varepsilon}$ correspondingly. ${\mathcal{W}}\in \bm{R}^{p\times {\mathcal{N}}_{\rm{param}}}$ is a positively determined linear continuous operator in the space of model parameters that imposes smoothness. We assume that ${\rm{ker}}F \cap {\rm{ker}}{\mathcal{W}} = \{ 0 \}$, which guarantees the uniqueness of the minimizer, and the rank of the matrix ${\mathcal{W}}$ is ${\mathcal{N}}_{\rm{param}}$.

The solution of the penalized least-squares problem, Eq.\eqref{eq:Parametric_functional}, is given by Eq.\eqref{eq:GaussNewton}, which can be rewritten as
\begin{equation}
\begin{split}
&E_{B}(A,Z,\{a_{i}\})_{k+1} = E_{B}(A,Z,\{a_{i}\})_{k}\\
&- \left[ F^{\prime T}(E_{B}(A,Z,\{a_{i}\})_{k} ) F^{\prime}(E_{B}(A,Z,\{a_{i}\})_{k})+(\alpha + \varepsilon_{k}) I_{\bm{R}^{n}} \right]^{-1} \\
&\times \left[ F^{\prime T}(E_{B}(A,Z,\{a_{i}\})_{k}) \left(F(E_{B}(A,Z,\{a_{i}\})_{k})- E_{B}^{\rm{Expt}}(A,Z)_{k} \right) \right.\\
&\left. + \alpha E_{B}(A,Z,\{a_{i}\})_{k}^{}\right].
\label{eq:SolutionPenalizedLeastSquaresPr}
\end{split}
\end{equation}

We assume that the misfit between the observed data and the true data is less than the given level of the experimental uncertainties. We will just comment here that in this statement we just follow the exposition given by Tikhonov and Arsenin \cite{Tikhonov:1986} and Lavrenti'ev \cite{Lavrentiev:1986}. For more details, please see Ref.\cite{Zhdanov:2002}. So, in this situation, it is natural to search for a solution in the set of the models such that
\begin{equation}
\begin{split}
\Arrowvert F E_{B,j}^{\rm{Th}}(A,Z,\{a_{i}\}) - E_{B,j}^{\rm{Expt}}(A,Z)\Arrowvert^{2}~\le~\delta
\label{eq:CondOfMinimization}
\end{split}
\end{equation}
with a given experimental uncertainty $\delta \textgreater 0$. 

Note, that here we just described the basic idea of the applied method focusing only on the binding energy, the same method have been used for the rest of the systems, Eqs.\eqref{eq:Mam_j} -\eqref{eq:MExess_j}. In order to proceed further one have to choose the initial model for description of the binding energy. The BW mass formula was chosen for this role, the motivation for this choice has been already explained in the previous sections. It provides the baseline fit for all the rest models 
\cite{
Moller1995185, 
Wapstra2003129, 
Chowdhury:2004jr, 
Kirson200829,
WDMyers:1966,
Myers1996141, 
Samanta:2004et, 
Rohlf:1994, 
NIX19651, 
Wapstra:1958,
PhysRev.89.1102, 
PhysRevC.33.2039,
PhysRevC.67.044316,
Royer20131}. 

The generalization form of the BW mass formula  in the inverse problem framework, Eq.\eqref{eq:LiquidDrop_BW}, for the binding energy per nucleon  can be written as:
\begin{equation}
\begin{split}
& E_{B}(A,Z,\{a_{i}\}) = \alpha_{vol}(A,Z,\{a_{i}\}_{1})-\alpha_{surf}(A,Z,\{a_{i}\}_{2}) \frac{1}{A^{p_{1}(A,Z, \{a_{i}\}_{6})}}\\
&-\alpha_{comb}(A,Z,\{a_{i}\}_{3})\frac{Z(Z-1)}{A^{p_{2}(A, Z, \{a_{i}\}_{7})}} -\alpha_{sym}(A,Z, \{a_{i}\}_{4} )\frac{(N-Z)^{2}}{A^{p_{3}(A,Z, \{a_{i}\}_{8} )}}\\
&+\alpha_{Wigner}(A,Z,\{a_{i}\}_{5})\frac{\delta(A,Z)}{A^{p_{4}(A, Z, \{a_{i}\}_{1}) }}+K_{MN}(A,Z,\{a_{i}\}_{10}).\\
\end{split}
\label{eq:BWparametrization_Wigner_CorMN}
\end{equation}
The $\delta(A,Z)$ is denoted as:
\begin{equation}
\delta(A,Z) = 
\begin{cases} 
+1, & \mbox{for even}~N,Z\\ 
-1, & \mbox{for odd}~N,Z\\
0, & \mbox{for odd}~(Z+N)\rightarrow \mbox{for odd}~A.\\
\end{cases}
\label{eq:pairingTerm}
\end{equation}
\begin{figure*}
\centering
\subfigure[]{
\includegraphics[scale=0.295]{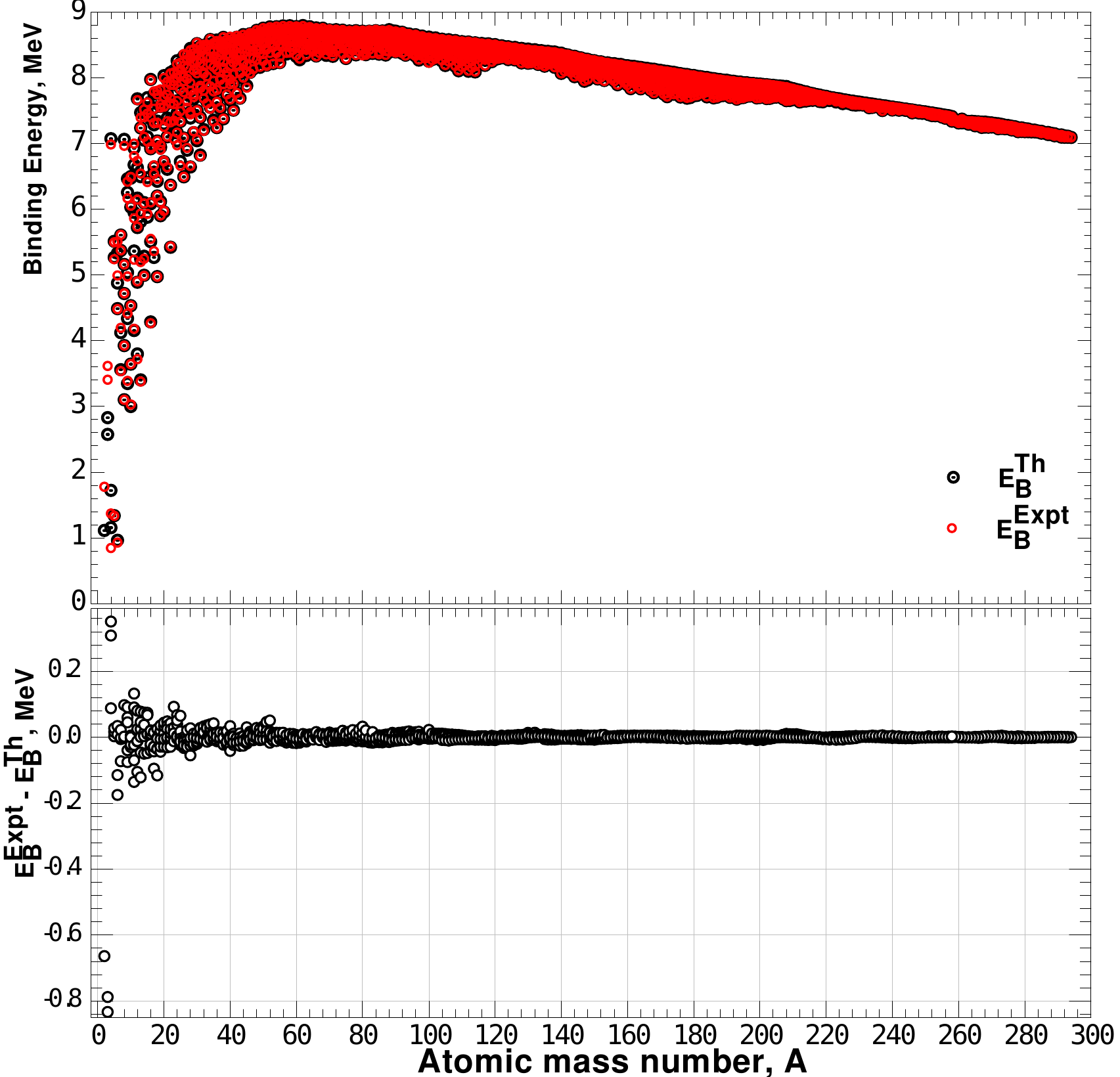}
}
\subfigure[]{
\includegraphics[scale=0.295]{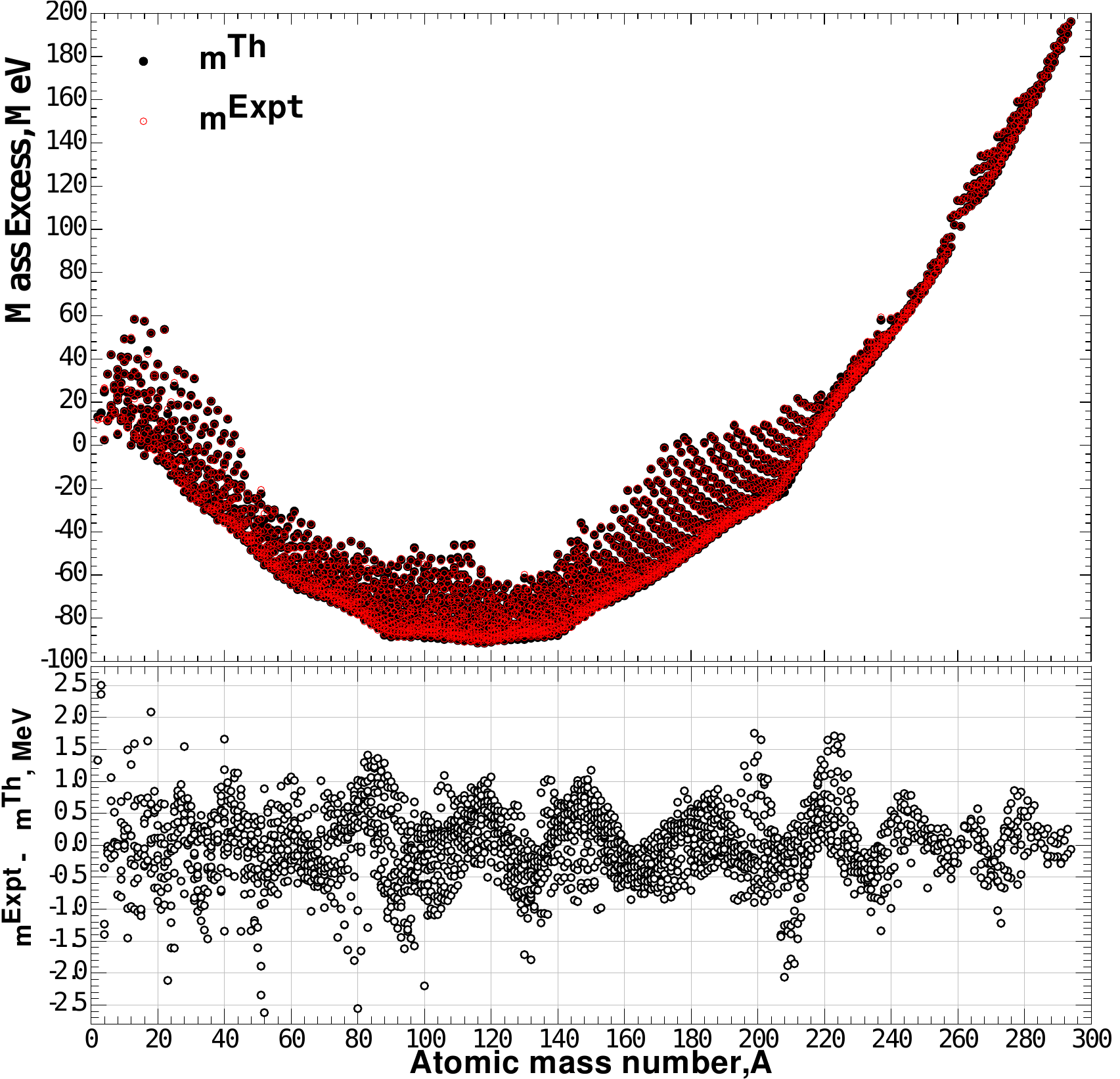}
}
\caption{
Left -- binding energy per nucleon as a function of atomic mass number, $A$. Right -- mass excess as a function of atomic mass number, $A$. There are 2564 data-point representing isotopes within experimentally known atomic masses (red circles), results of the our calculation (black circles). The difference between the experimental data and the resulting true solution are shown on bottom pads respectively.
}
\label{fig:BindingEnergyA}
\end{figure*}

Note, 
to perform the generalization of the BW mass formula we assume, that all the BW constant terms that are obtained as results of the different fitting procedure (see Table~\ref{tab:BW_CoeffFits}) in the inverse problem framework are unknown functions of $A$, $Z$, and $N$, namely structures, and the set of the unknown parameters $\{a_{i}\}$. In addition, we involve the Wigner term, and use the correction term instead of the pairing term, which is sensitive to the magicity of the given atomic mass number. Since the BW mass formula has no shell effect incorporated and thus when the shell effect in a nucleus quenches, by including this term, the theoretical mass should come closer to the experimental ones. Also, the original Coulomb has been implemented in order to have the possibility to describe the light nuclei. Therefore, we assume that all structures corresponding to the appropriate type of energy, namely, volume, surface, Coulomb, asymmetry and pairing, as well as the power factors $A^{-1/3}$, $A^{-4/3}$, $A^{-2}$, $A^{-3/2}$, are all determined with the help of the function of $A$, $Z$  and unknown number of the parameters $a_{i}$, where $i=1,\dotsc,{\mathcal{N}}_{\rm{param}}$.

The atomic masses that are related to the nuclear mass $M_{n.m}(A,Z,\{a_{i}\})$ described by
\begin{equation}
M_{a.m}(A,Z,\{a_{i}\})=Zm_{p}+Nm_{n}-AE_{B}(A,Z,\{a_{i}\}),
\label{eq:AtomicMassFunc}
\end{equation}
$M_{a.m}(A,Z,\{a_{i}\})$ may thus be connected to the experimental atomic masses given in \cite{Ami2012_ChinPhysC} since:
\begin{equation}
M_{n.m}(A,Z, \{a_{i}\})=M_{a.m}(A,Z,\{a_{i}\})-Zm_{e}+B_{e}(Z).
\label{eq:Mnm}
\end{equation}
The cloud of electrons that defines atomic dynamical \cite{PhysRevLett.24.45}, chemical and mechanical properties generates its own binding field, driven by the strong force that exists between all of its constituent nucleons (neutrons and protons). As such, nuclei have a much less well-defined `center'. Thus the $B_{e}(Z)$ in Eq.\eqref{eq:Mnm} is the binding energy of all removed electrons that is fitted by a power law accordingly to \cite{PhysRev.83.397, RevModPhys.75.1021} as
\begin{equation}
B_{e}(Z)=a_{el}Z^{a_{240}}+b_{el}Z^{a_{241}},
\label{eq:Electrons_BindEn}
\end{equation}
with $a_{el}=1.44381\times10^{-5}$ MeV and $b_{el}=1.55468\times10^{-12}$ MeV. 

The mass excess of a nuclide is
\begin{equation}
\Delta m(A,Z,\{a_{i}\})=M_{a.m}(A,Z,\{a_{i}\})-Au,
\label{eq:Mnm_Strashemir}
\end{equation}
where 
$m_{H} = 938.782303(0.084)$ MeV, 
$m_{n} = 939.56538(4.56)$ MeV, 
$m_{p} = 938.272046(21)$ MeV, 
$m_{el} = 0.510998928(11)$ MeV \cite{Ami2012_ChinPhysC, PhysRevD.86.010001} and
$u = 931.494061$ MeV \cite{Ami2012_ChinPhysC, PhysRevD.86.010001}. The $1-\sigma$ uncertainties in the last digits are given in parentheses after the values.

In order to solve all systems of Eqs.\eqref{eq:BE_j}-\eqref{eq:MExess_j} we adopt the following procedure. First, we searched for the appropriate solution for the binding energy for the given $A$ and $Z$. The obtained solution was used as an approximation (`fake solution') to find the atomic masses, the solution for atomic masses was in turn used to find nuclear mass. Then, the fit is performed again, allowing the binding energy to vary for the given $A$ and $Z$. The resultant binding energy is then taken to be a new seed, and this procedure is repeated iteratively until convergence is reached for all systems. Indeed the system converge due to applied iterative process that was discussed in Sec.\ref{AleksandrovMethod_REGN}.
\begin{figure*}[ht]
	\centering
	\subfigure[]{
		\includegraphics[scale=0.31]{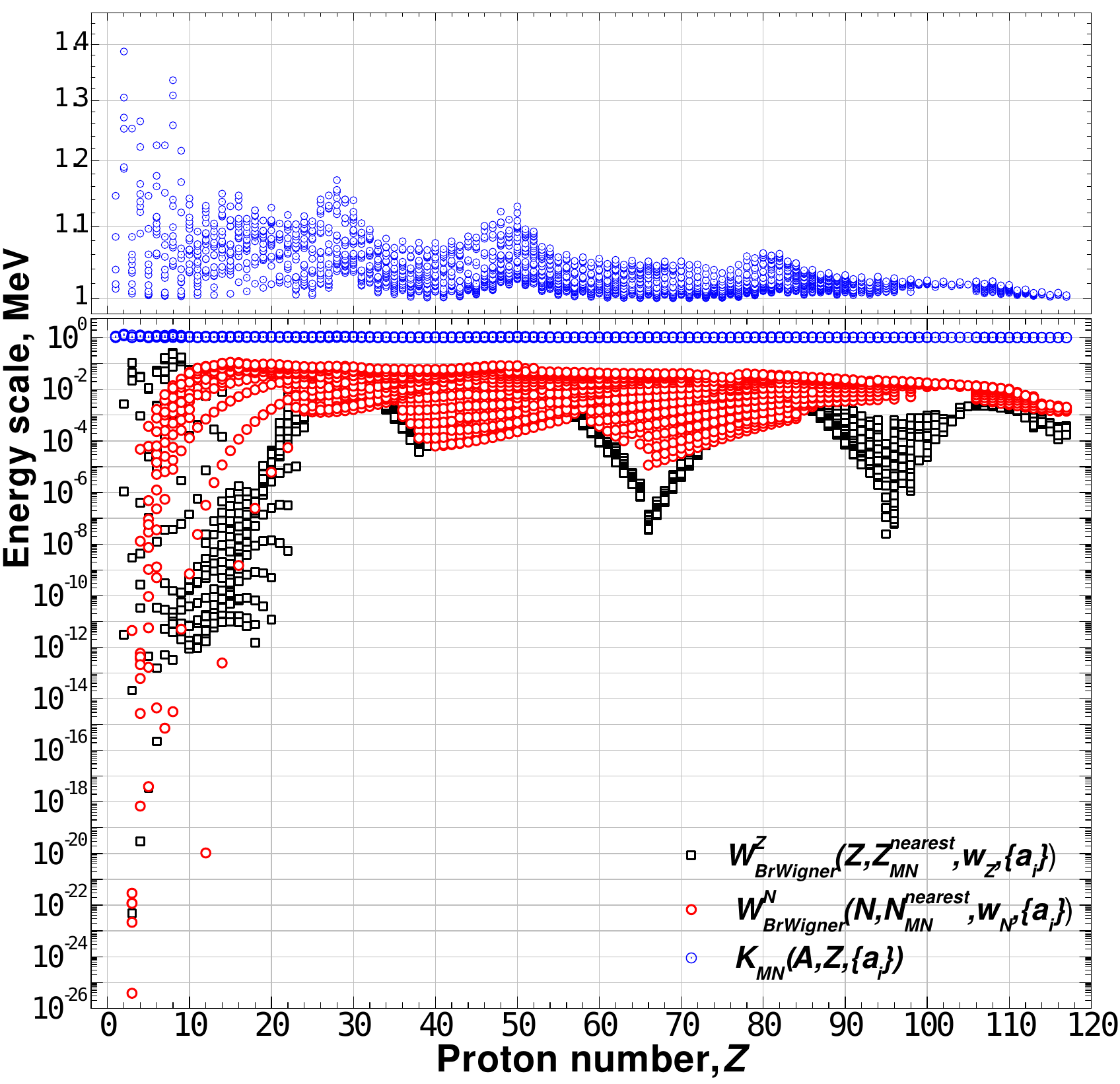}
	}
	\subfigure[]{
		\includegraphics[scale=0.31]{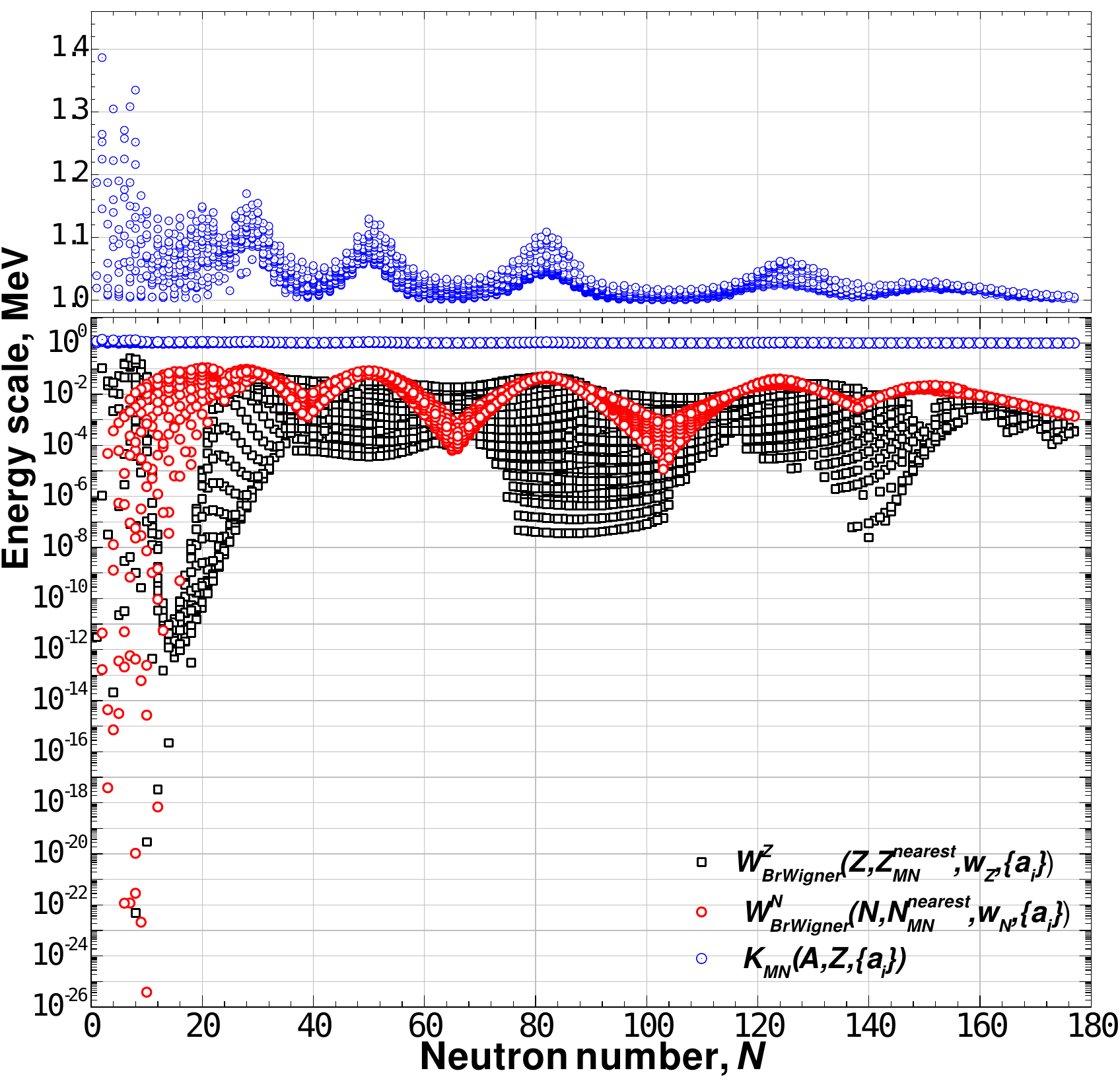}
	}
	\caption{
		Distribution of the functions $W_{\rm{BrWigner}}^{Z}$ and $W_{\rm{BrWigner}}^{N}$, see Eq.\eqref{eq:func_CorGam}, for proton and atomic mass numbers shown together with the shell correction energy function, see Eq.\eqref{eq:CorMN_func}. Zoomed view of the shell correction function shown in the top pad.
	}
	\label{fig:BetheWeizsackerCorMNA}
\end{figure*}

In Figs.\ref{fig:BindingEnergyA}, we plot the result binding energy and the mass excess. 
In the bottom pads, we show the difference between true solution and the experimental data for each of the isotopes. 
There is no significant deviation for the binding energy, while the maximum absolute deviation for the mass excess is less than 2.6 MeV.

By solving the over determined nonlinear systems with conditions from Eqs.\eqref{eq:BE_true} and \eqref{eq:CondOfMinimization}, using the different models  for unknown functions $\alpha_{vol}(A, Z, \{a_{i}\})$,  $\alpha_{surf}(A, Z, \{a_{i}\})$,  $\alpha_{comb}(A, Z, \{a_{i}\})$, 
$\alpha_{sym}(A, Z, \{a_{i}\})$, $\alpha_{Wigner}(A, Z, \{a_{i}\})$ in the step-by-step way we obtained their explicit form, which can be expressed as follows:
\begin{equation}
\begin{split}
&\alpha_{vol}(A, Z, \{a_{i}\}_{1})={\rm{exp}} \left(a_{1} + P\left(\upsilon, \{a_{i}\}_{1}  \right) \right),\\
&\alpha_{surf}(A, Z, \{a_{i}\}_{2})={\rm{exp}}  \left(a_{2}+P\left(\upsilon, \{a_{i}\}_{2} \right)\right),\\
&\alpha_{comb}(A, Z, \{a_{i}\}_{3})={\rm{exp}} \left(a_{3}+P\left(\upsilon, \{a_{i}\}_{3} \right)\right),\\
&\alpha_{sym}(A, Z, \{a_{i}\}_{4})={\rm{exp}}\left(a_{4} +P\left(\upsilon, \{a_{i}\}_{4} \right) \right),\\
&\alpha_{Wigner}(A, Z, \{a_{i}\}_{5})={\rm{exp}}\left(a_{5} +P\left(\upsilon, \{a_{i}\}_{5}\right) \right) +\omega_{W}(A,Z,\{a_{i}\} ).\\
\end{split}
\label{eq:BW_coeff}
\end{equation}
The explicit form of functions in Eq.\eqref{eq:BW_coeff} has been obtained by implementation of the LCH procedure \cite{Aleksandrov2004519, LCH:1999} in the iteratively regularized method \cite{Alexandrov:1970} for inverse problems of discovering the explicit form of the unknown functions, which is realized in the Dubna REGN program \cite{Alexandrov:1973, Alexandrov:1983, Alexandrov:1982}, and allows us to choose the better one of two functions with the same $\chi^{2}$.

The parameterized powers that were implemented in Eq.\eqref{eq:BWparametrization_Wigner_CorMN},
have been obtained using the same LCH procedure, that was used for the structure functions, and can be written as:
\begin{equation}
\begin{split}
& p_{1}(Z,A, \{a_{i}\}_{6})={\rm{exp}} \left(a_{6}+P(\upsilon, \{a_{i}\}_{6} )\right),\\
& p_{2}(Z,A, \{a_{i}\}_{7})={\rm{exp}} \left(a_{7}+P(\upsilon, \{a_{i}\}_{7} )\right),\\
& p_{3}(Z,A, \{a_{i}\}_{8})={\rm{exp}} \left(a_{8}+P(\upsilon, \{a_{i}\}_{8} )\right),\\
& p_{4}(Z,A, \{a_{i}\}_{9})={\rm{exp}} \left(a_{9}+P(\upsilon, \{a_{i}\}_{9} )\right).
\end{split}
\label{eq:BW_Powers_coeff}
\end{equation}

The function $P(\upsilon, \{a_{i}\})$  in Eqs.\eqref{eq:BW_coeff} is defined using the exponential distributions as:
\begin{equation}
\begin{split}
&P(\upsilon, \{a_{i}\}_{j})= {\rm{exp}}\left(   -\left(\sum_{k=1}^{4} c_{k}(\upsilon, \{a_{i}\}_{j}) \right)^{2}  \right),
\end{split}
\label{eq:func_CorPow}
\end{equation}
it determines corrections to the structure terms, and also implies corrections to the power factors with respect to the different isotopes. The functions $c_{k}(\upsilon, \{a_{i}\}_{j})$ defined through linear series of the variables $\upsilon$ and parameters $a_{i}$ as
\begin{equation}
\begin{split}
&c_{1}(\upsilon, \{a_{i}\}_{j} )=a_{i_1}\upsilon_{1}+a_{i_2}\upsilon_{2}+a_{i_3}\upsilon_{3}+a_{i_4}\upsilon_{4},\\
&c_{2}(\upsilon, \{a_{i}\}_{j})=a_{i_5}(\upsilon_{1})^{2}+a_{i_6}(\upsilon_{2})^{2}+a_{i_7}(\upsilon_{3})^{2}+a_{i_8}(\upsilon_{4})^{2},\\
&c_{3}(\upsilon, \{a_{i}\}_{j})=a_{i_9}(\upsilon_{1})^{3}+a_{i_{10}}(\upsilon_{2})^{3}+a_{i_{11}}(\upsilon_{3})^{3}+a_{i_{12}}(\upsilon_{4})^{3},\\
&c_{4}(\upsilon, \{a_{i}\}_{j})=a_{i_{13}}\upsilon_{6}+a_{i_{14}}\upsilon_{5}.
\end{split}
\label{eq:func_CorGam_coeff}
\end{equation}
Here $\{a_{i}\}_{j}$ represents the subset of the free parameters from the whole set of parameters that is used for the particular structure function. The consecutive index $j$ defines a mapping for the chosen subset in the full set of free parameters used in this approach. For $j=1,\dots,9$ the mapping is defined is the following way\footnote{For an explicit example, let us examine the case when the consecutive index $j=2$ in Eq.\eqref{eq:func_CorPow}. Therefore, we have $c_{k}(\upsilon, \{a_{i}\}_{2})$, whose sequence of the parameters, taken into account Eq.\eqref{eq:ParamMappring_j19}, can be written in the following way $a_{i_{1}}\rightarrow a_{24}, a_{i_2}\rightarrow a_{25},\cdots, a_{i_{14}}\rightarrow a_{37}$. Thus, for each of the functions $P(\upsilon, \{a_{i}\})$ we assign the 14 parameters from the whole set $\{a_{i}\}$.
}:
\begin{equation}
\begin{split}
& \{a_{i}\}_{1}\rightarrow a_{10},a_{11},\dots,a_{23},\\
& \{a_{i}\}_{2}\rightarrow a_{24},a_{25},\dots,a_{37},\\
& \cdots\\
& \{a_{i}\}_{9}\rightarrow a_{122},a_{123},\dots,a_{135}.
\end{split}
\label{eq:ParamMappring_j19}
\end{equation}
\begin{figure*}[ht]
\centering
\subfigure[]{
\includegraphics[scale=0.272]{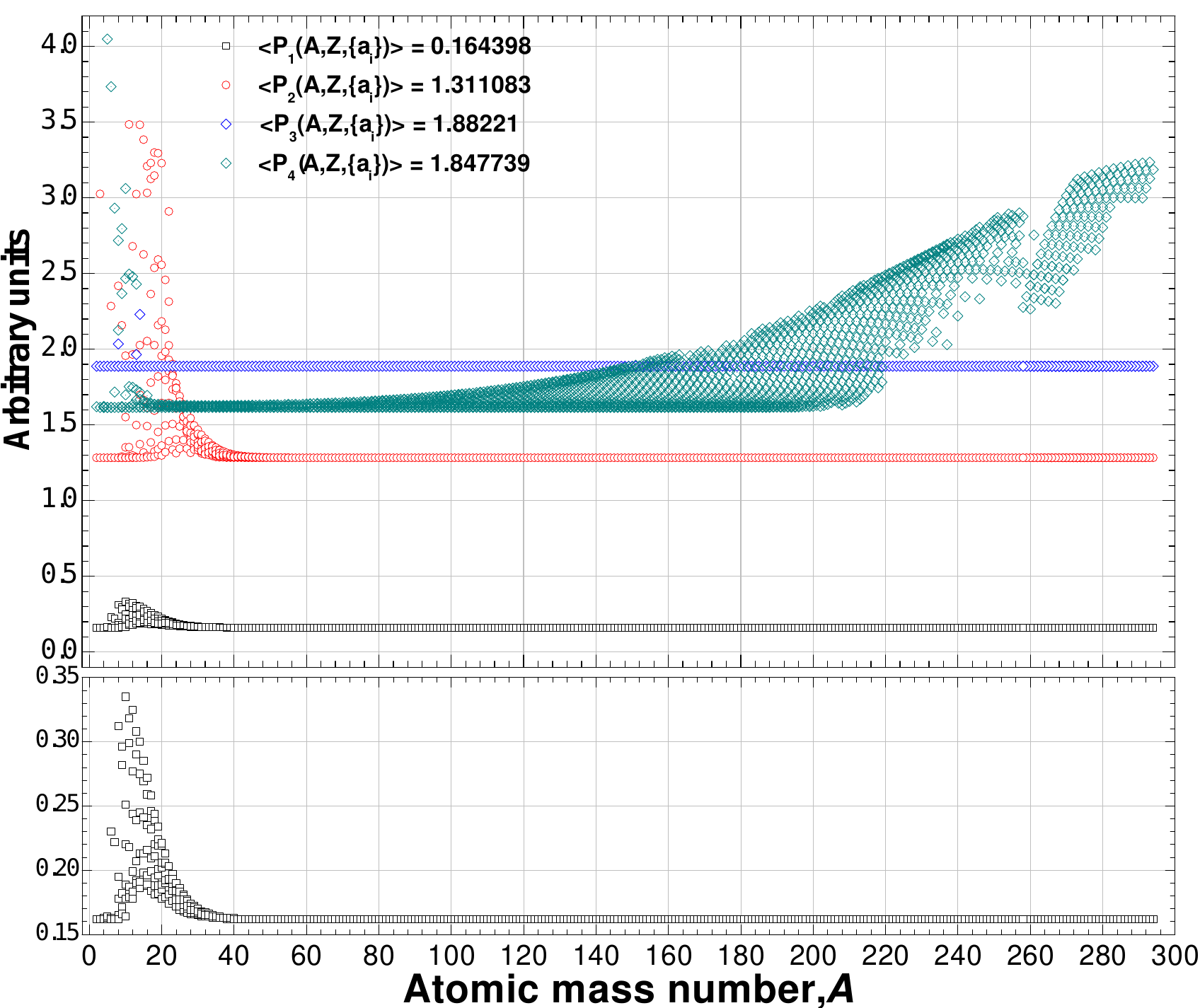}
\label{fig:PowerFactors_A_2564}
}

\subfigure[]{
\includegraphics[scale=0.272]{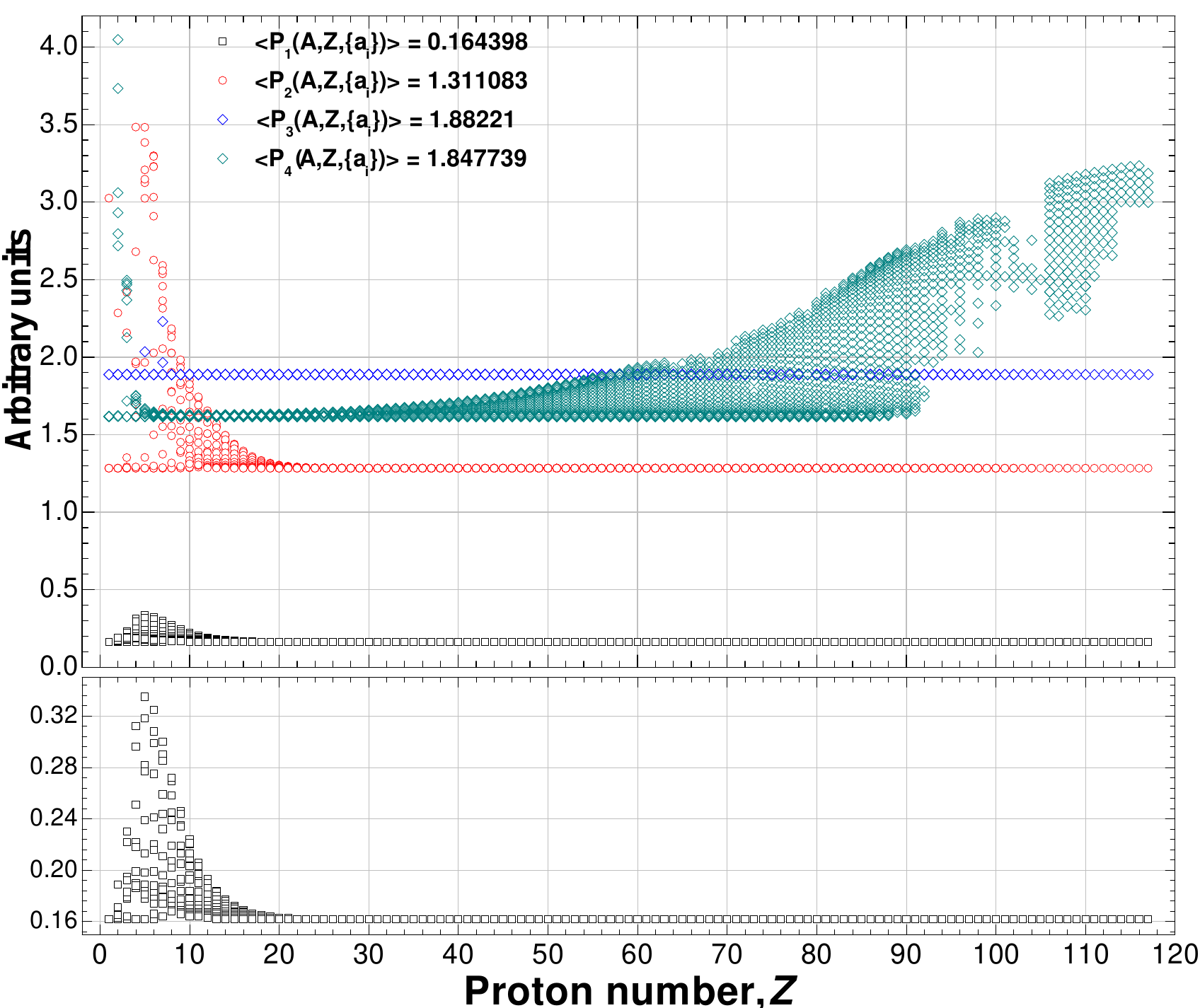}
\label{fig:PowerFactors_Z_2564}
}
\subfigure[]{
\includegraphics[scale=0.272]{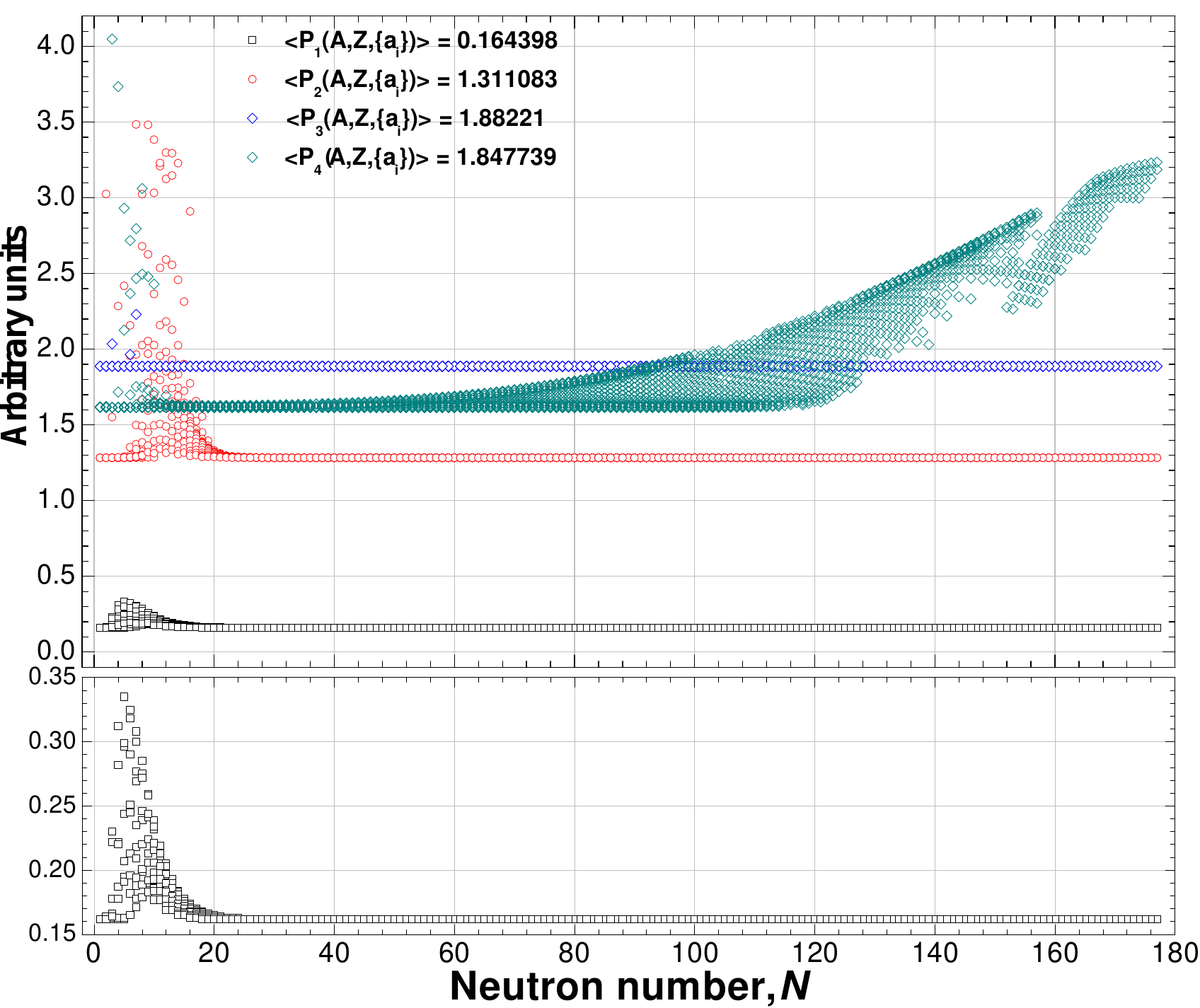}
\label{fig:PowerFactors_N_2564}
}
\caption{
Evolution of the power factors (see Eq.\eqref{eq:BW_Powers_coeff}) as function of the proton, neutron and atomic mass numbers respectively. On the bottom pad we show the zoomed distribution of the $p_{1}(A,Z,a_{i})$ power factor.
}
\label{fig:PowerFactors_AZN}
\end{figure*}

\noindent {\bf{Energy corrections based on magic numbers}}\\
The presented approach enables calculations of the nuclear properties in large model spaces. The concrete subjects of investigation are the nucleus masses and the structure behavior of the BW mass formula, whose bound states are interpreted as highly stable elements $\textendash$ nuclei with closures shells, magic numbers. In order to obtain such kind of dependencies from magic numbers and boundaries between them, which define their influence, different inverse problems were formulated using different values of proton and neutron magic numbers. The LCH analysis of the solutions of these additional problems helps us to establish the explicit form of these corrections, which depend on nine proton and ten neutron magic numbers. 
However, the blind analysis of the AME2016 data generates the slightly modified list of magic numbers, which in turn affected the correction energies, see Fig.\ref{fig:MagNumberList} and Table~\ref{tab:ZN_magnumbers}.

The function that determines the correction energy with respect to the magic numbers is: 
\begin{equation}
\begin{split}
&K_{MN}(Z,A,  \{a_{i}\}_{10} )\\
&=W_{\rm{BrWigner}}\left(Z, Z_{MN}^{\rm{nearest}}, w_{Z}, N, N_{MN}^{\rm{nearest}}, w_{N}, \{a_{i}\}_{10} \right)\\
&+\omega_{MN}(A,Z, \{a_{i}\}),\\
\end{split}
\label{eq:CorMN_func}
\end{equation}
where $w_{Z}$ and $w_{N}$ are the mean values of the two closest proton magic numbers and the two neutron magic numbers respectively.  For this function we use subset of $\{a_{i}\}_{10}$ which now includes 15 parameters, starting from $a_{136}$ parameter in the sequence. This correction function provides an indicator about the deviation in the level structure of nuclei away from closed shells. In addition, it includes the pairing effects, which can provide reliable information about magicity of the different proton and neutron numbers. The maximum values of the correction energy corresponds to shell closure at certain nucleon number, as will be shown later in the paper.

The $\omega_{W}$ function in the Eq.\eqref{eq:BW_coeff} is defined as:
\begin{equation}
\begin{split}
&\omega_{W}(A,Z,\{a_{i}\})=
{\rm{exp}}\left( a_{237}\frac{\upsilon_{7}}{A}+a_{238}\frac{\upsilon_{8}}{Z}+a_{239}\frac{\upsilon_{9}}{N+1}\right).
\end{split}
\label{eq:omegaW_Functions}
\end{equation}

The $\omega_{MN}$ function in Eq.\eqref{eq:CorMN_func} defined in the similar way as $\omega_{W}$, the only difference it depends on the different subset of the parameters $a_{i}$ as:
\begin{equation}
\begin{split}
&\omega_{MN}(A,Z, \{a_{i}\})
={\rm{exp}}   \left( a_{234}\frac{\upsilon_{7}}{A}+a_{235}\frac{\upsilon_{8}}{Z}+a_{236}\frac{\upsilon_{9}}{N+1} \right),\\
\end{split}
\label{eq:omegaMN_Functions}
\end{equation}
where the variables $\upsilon_{7}$, $\upsilon_{8}$ and $\upsilon_{9}$ that used in these functions will be defined later in the paper.

The Breit-Wigner function, $W_{\rm{BrWigner}}$, in Eq.\eqref{eq:CorMN_func} was constructed in the following form:
\begin{equation}
\begin{split}
&W_{\rm{BrWigner}}(Z,Z_{MN}^{\rm{nearest}},w_{Z}, N, N_{MN}^{\rm{nearest}}, w_{N}, \{a_{i}\}_{10} )\\
&=W_{\rm{BrWigner}}^{Z}(Z,Z_{MN}^{\rm{nearest}},w_{Z}, \{a_{i}\})+W_{\rm{BrWigner}}^{N}(N,N_{MN}^{\rm{nearest}},w_{N}, \{a_{i}\}),\\
\end{split}
\label{eq:BrWirnerDistribution}
\end{equation}
where
\begin{equation}
\begin{split}
&W_{\rm{BrWigner}}^{Z}(Z,Z_{MN}^{\rm{nearest}},w_{Z}, \{a_{i}\})={\mathcal{A}}_{Z}\frac{  {\rm{exp}} \left(   \frac{- \left(Z-Z_{MN}^{\rm{nearest}}\right)^{2} } {2Q(\upsilon, \{a_{i}\}_{12} ) } \right)} {\left(Z-Z_{MN}^{\rm{nearest}}\right)^{2}+Q(\upsilon, \{a_{i}\}_{12} )},\\
&W_{\rm{BrWigner}}^{N}(N,N_{MN}^{\rm{nearest}},w_{N}, \{a_{i}\})={\mathcal{A}}_{N}\frac{ {\rm{exp} }\left(   \frac{- \left(N-N_{MN}^{\rm{nearest}}\right)^{2} } {2Q(\upsilon, \{a_{i}\}_{13}) } \right)  }{\left(N-N_{MN}^{\rm{nearest}}\right)^{2}+Q(\upsilon, \{a_{i}\}_{13} )},\\
\end{split}
\label{eq:func_CorGam}
\end{equation}
Here $Z$ is the proton number, $N$ is neutron number, $w_{Z}$ and $w_{N}$ are half-width of the two closest magic numbers for the given proton and neutron numbers respectively. One may note, that these widths are representing the arguments of the generalized BW mass function, which is defined as the product of the two distributions, the Gaussian and the Breit-Wigner. Given that prescription, one may conclude that the nearest proton or neutron number is the location parameter, specifying the location of the peak of the binding energy distribution.
\begin{figure*}
\centering
\subfigure[]{
\includegraphics[scale=0.31]{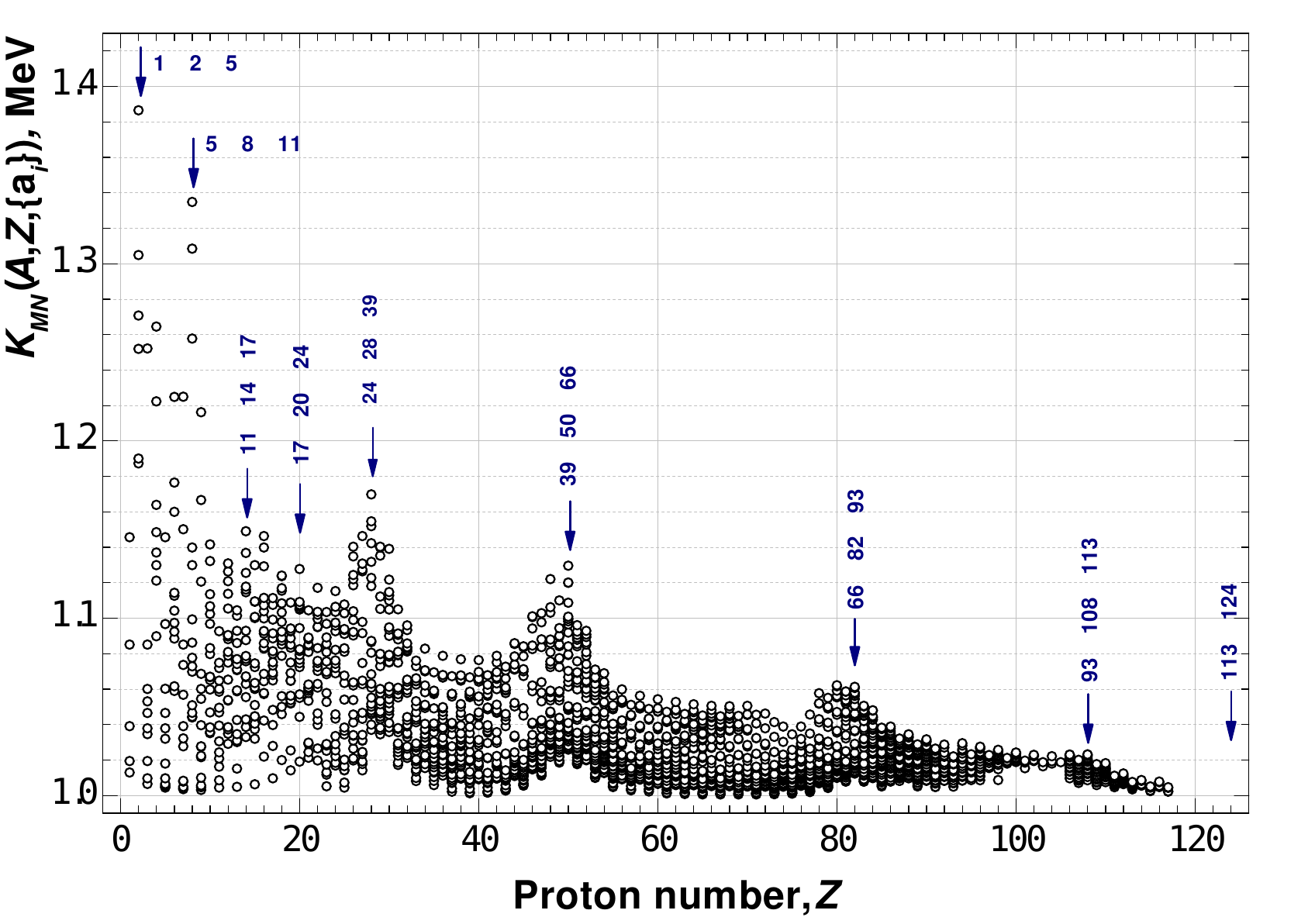}
}
\subfigure[]{
\includegraphics[scale=0.31]{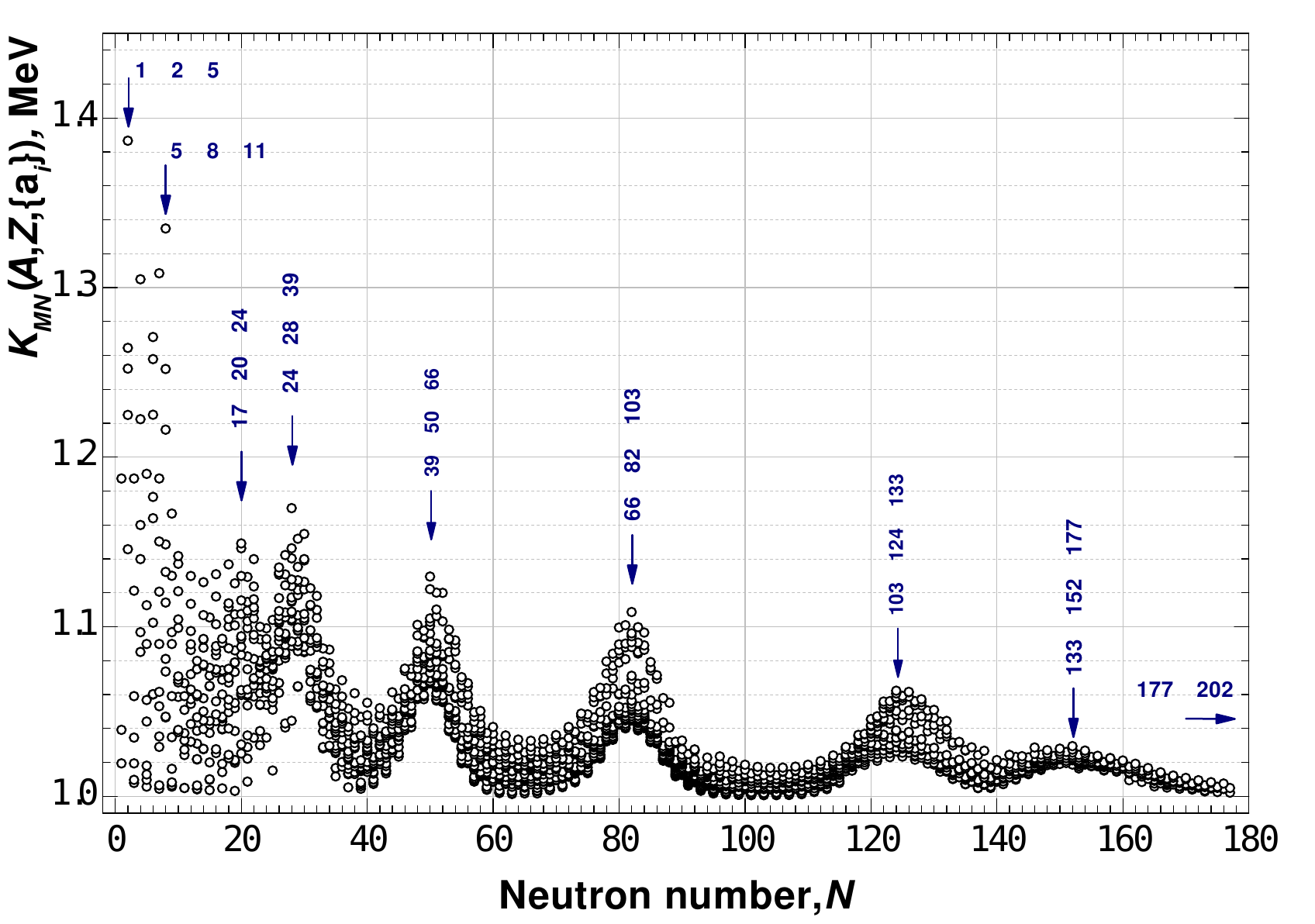}
}

\subfigure[]{
\includegraphics[scale=0.50]{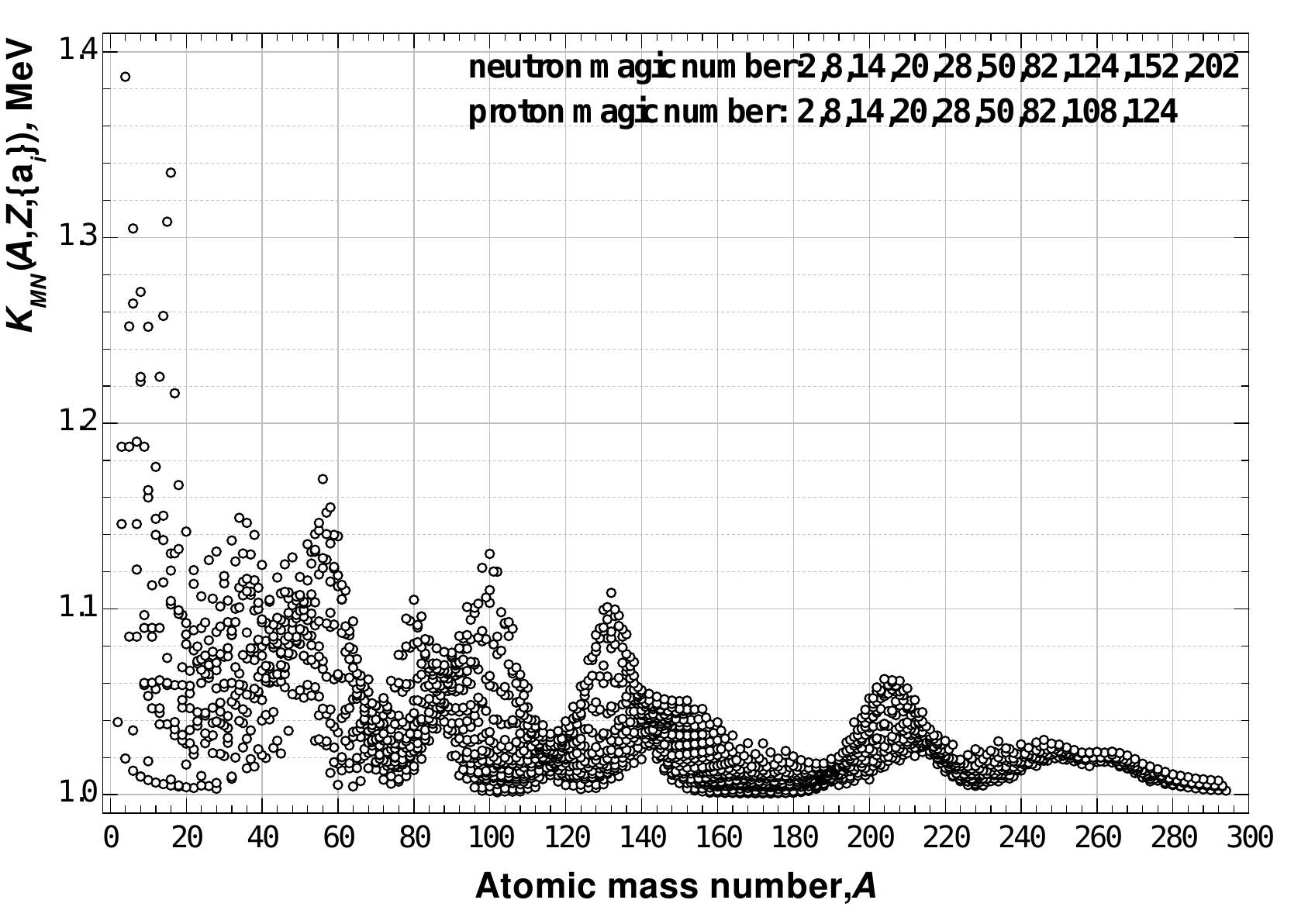}
}
\caption{
The shell correction energy function $K_{\rm{MN}}(A,Z, \{a_{i}\} )$ based on the magic numbers, see Eq.\eqref{eq:CorMN_func}, as function of the proton $Z$, neutron $N$ and atomic mass number $A$ respectively. The proton and neutron magic numbers displayed in respective pads within the limits of influences for each of the magic numbers. The magnitude of magicity of the different numbers slightly differs. A large shell correction energy corresponds to shell closure at certain nucleon numbers. 
The magic numbers are indicated with arrows,  the boundaries of each the discovered magic numbers on top of the arrows (see Table~\ref{tab:ZN_magnumbers} for more details). The traditional magic numbers ($Z=20,28,50,82$)  and  ($N=20,28,50,82$) are clearly seen close to the stable nuclei, while the peaks close to magic numbers ($Z=2,8$) and ($N=2,8$) are less clear.
}
\label{fig:CorMagicNumbers}
\end{figure*}

The amplitudes in Eq.\eqref{eq:func_CorGam} are written as:
\begin{equation}
\begin{split}
&{\mathcal{A}}_{Z}(\upsilon,\{a_{i}\}_{10})=w_{Z}+Q(\upsilon, \{a_{i}\}_{10} ),\\
&{\mathcal{A}}_{N}(\upsilon,\{a_{i}\}_{11})=w_{N}+Q(\upsilon, \{a_{i}\}_{11} ).\\
\end{split}
\label{eq:func_CorGam_Amplitude}
\end{equation}

The scale functions, $Q(\upsilon, \{a_{i}\}_{j})$ in Eq.\eqref{eq:func_CorGam}, which specify the half-width at half-maximum, are given by this relation:
\begin{equation}
\begin{split}
&Q(\upsilon,\{a_{i}\}_{j} )={\rm{exp}}\left(a_{i+15}-  \left(\sum_{k=1}^{4} c_{k}(\upsilon, \{a_{i}\}_{j} ) \right)^{2}  \right).\\
\end{split}
\label{eq:func_CorGam_Gam_ex}
\end{equation}
As one may note, the Eq.\eqref{eq:func_CorGam_Gam_ex} is similar to the $P(\upsilon, \{a_{i}\}_{j})$ function from Eq.\eqref{eq:func_CorPow}, but here we have an extra term, $a_{i+15}$. 
Due to this term the $j$ index is now mapping 15 parameters for each of the function in the expressions Eq.\eqref{eq:func_CorGam} and Eq.\eqref{eq:func_CorGam_Amplitude}:
\begin{equation}
\begin{split}
& \{a_{i}\}_{10}\rightarrow a_{136},a_{137},\dots,a_{150},\\
& \{a_{i}\}_{11}\rightarrow a_{151},a_{152},\dots,a_{165},\\
& \cdots\\
& \{a_{i}\}_{13}\rightarrow a_{181},a_{182},\dots,a_{195},
\end{split}
\label{eq:ParamMappring_j1013}
\end{equation}

Since experimental data has systematic uncertainties we apply robust procedures in order to estimate the weights of the explicit form of the Breit-Wigner distributions, which has been obtained with the help of the LCH-weights method \cite{Aleksandrov2004519, LCH:1999}. 

In Fig.\ref{fig:BetheWeizsackerCorMNA} we plot the distribution functions  of Eq.\eqref{eq:func_CorGam}. In the top pad, we show for comparison the spectrum of the shell correction energy, Eq.\eqref{eq:CorMN_func}, that reflects effect of the magicity of the given atomic mass number. From this figure, it is clear how the adopted Breit-Wigner functions correlates with the shell correction  function.

In the given formalization, see Eqs.\eqref{eq:BE_j} and \ref{eq:BWparametrization_Wigner_CorMN} the problem of finding the unknown structure functions in Eq.\eqref{eq:BW_coeff} becomes similar to the problem that was stressed in the Sec.\ref{Theory_and_Method}, namely the problem of allocating the exponents with unknown decrement $h_{i}$ and amplitudes ($\rm{a}_{0}$, $\rm{a}_{i}$) on the finite values of their sum, $f(t_{k})$, which are known approximately \cite{Alexandrov:19701285}.

If we do not know the exact number of the exponents, then this problem can be resolved by assuming that number $n$ in Eq.\eqref{eq:NonLinearSystem} is very big, bigger than the expected number of the unknown exponents.  However, to obtain the correct dependences of the decrements on the level of the $A$, $Z$ numbers in Eq.\eqref{eq:BW_coeff}, it is necessary to carry out extensive and time-consuming computations of the amplitude of the attenuating oscillations for subsequent calculations of the different values of the decrements.

In the developed iteration scheme, the problem of choosing the initial approximations ($a~priori$ knowledge) is in some sense solved, by using values from \cite{Rohlf:1994}, see the fifth row in Table~\ref{tab:BW_CoeffFits}, as initial values for structure functions. The choice of $p_{i}$ is arbitrary, and in present application we suppose, that initial values correspond to the liquid-drop model, hence we start from the power factors $p_{1}=1/3$, $p_{2}=4/3$, $p_{3}=2/3$ and $p_{4}=3/2$. Of course, in standard regularization $a~priori$ knowledge leads to restriction on the solution space.  However, this approximation was chosen not because we think that it is a good description of the binding energy, but rather because it is a necessary first step towards the required formalism. Indeed as will be shown further, this approximation is a useful simplification of the parameterized powers, see  Eq.\eqref{eq:BW_Powers_coeff}. The result power factors we plot in Fig.\ref{fig:PowerFactors_AZN}.

\noindent {\bf{Independent variables and free parameters}}\\
For next step forward, one may need to define the number and the form of the independent variables. Thus the next problem is to choose the independent variables of the functions $a_{vol}$, $a_{surf}$, $a_{comb}$, $a_{sym}$, $a_{Wigner}$, and $p_{j}$,  where $j=1,2,3,4$.  The hypothesis that they are linearly independent permits one to choose them as follow:
\begin{equation}
\upsilon_{1}=\frac{Z}{A}, ~\upsilon_{2}=\frac{N}{A}, ~\upsilon_{3}=\frac{N-Z}{A}, ~\upsilon_{4}=\frac{Z}{N+1},
\label{eq:IndependentVariable_14}
\end{equation}
 where $Z$ and $N$ are the numbers of protons and neutrons in nuclei, and $A=Z+N$. The next two the linearly independent variables are
\begin{equation}
\begin{split}
& \upsilon_{5}={\rm{log}}(A+1), ~\upsilon_{6}=\frac{1}{\upsilon_{5}},\\
\end{split}
\label{eq:IndependentVariable_56}
\end{equation}

We have also the corresponding four square and the four cubic terms of Eq.\eqref{eq:IndependentVariable_14} that used in Eq.\eqref{eq:func_CorGam_coeff}. In addition to those parameters three spin variables $\upsilon_{7}$, $\upsilon_{8}$, $\upsilon_{9}$ were implemented in order to estimate the impact of valence proton and neutron in the nuclei to the binding energy, as well as to obtain dependencies of the binding energy with respect to the odd-odd, even-even, odd-even nuclei. They are defined in the following way
\begin{equation}
\begin{split}
&\upsilon_{7}=
\begin{cases} 
0,~~\mbox{for odd}~A\\ 
1,~~\mbox{otherwise}~\\
\end{cases}\\
&\upsilon_{8}=
\begin{cases} 
0,~~\mbox{for odd}~Z\\ 
1,~~\mbox{otherwise}~\\
\end{cases}\\
&\upsilon_{9}=
\begin{cases} 
0,~~\mbox{for odd}~N\\ 
1,~~\mbox{otherwise}~\\
\end{cases}
\label{eq:AAA}
\end{split}
\end{equation}

Thus, totally we use 17 linearly independent variables.

Summing up all above, let us compute the number of unknown parameters in our parameterization. The parameters that we found in order to describe the whole set of the experimental data are:
\begin{equation}
\begin{split}
& n_{str}=5, ~~n_{p}=4, ~~n_{map}=14,\\
\end{split}
\label{eq:n_BWpowers}
\end{equation}
where $n_{str}$ is the number of the structures namely, volume, surface, Coulomb, asymmetry, Wigner and magic number terms, see Eq.\eqref{eq:BW_coeff}; $n_{p}$ is the number of the power factors that used in our parameterization, see Eq.\eqref{eq:BW_Powers_coeff}, while the total number of the structures and power factors defined as $n_{SP}=n_{str}+n_{p}$.  Thus only for the structure functions, Eq.\eqref{eq:BW_coeff},  in the BW mass formula we implement
\begin{equation}
n_{SP}+n_{SP}n_{map}=135,
\label{eq:BW_paramNr}
\end{equation}
where $n_{map}$ is the number of parameters that we use for each of the unknown functions in Eq.\eqref{eq:func_CorGam_coeff}.

The number of parameters for the function that determines the correction energy, Eq.\eqref{eq:CorMN_func}, with respect to the magic numbers namely, the Breit-Wigner function, is equal to 
\begin{equation}
n_{\rm{BrWign}}=2n_{\rm{width}}+2n_{\rm{amp}}=60,
\label{eq:nBW}
\end{equation}
where we use 15 parameters to define the widths, $n_{\rm{width}}$, and 15 parameters to define the amplitudes, $n_{\rm{amp}}$.

The number of unknown parameters in our mass function before the correction of magic numbers starts affecting the final solution defined as $n_{0}$ and written as: 
\begin{equation}
\begin{split}
& n_{0}=n_{SP}+n_{SP}n_{map}+n_{\rm{BrWign}} = 195.\\
\end{split}
\label{eq:n_BWpowers}
\end{equation}
Therefore, in total we have:
\begin{equation}
\begin{split}
& {\mathcal{N}}_{\rm{param}}=n_{0}+2n_{mn,Z}+2n_{mn,N}+8 = 241,
\end{split}
\label{eq:Ntotal}
\end{equation}
where the total number of the proton magic numbers is $n_{mn,Z}=9$, the neutron $\textendash$ $n_{mn,N}=10$. 
The factor of 2 in the front of the magic numbers appears due to fact, that one needs take into account the effect of the boundaries between two closest magic numbers with respect to the examined nuclei. ${\mathcal{N}}_{\rm{param}}$ is the total number of unknown free parameters that describes the semi-empirical mass formula in the inverse problem framework Eq.\eqref{eq:BWparametrization_Wigner_CorMN}. Determination of these parameters of the binding energies represents nothing but a solution of the inverse problem with approximate initial conditions.

Here we have to emphasize that many inverse problems include nuisance parameters, such as variance parameters, regularization parameters, application-specific tuning parameters, or degrees of freedom parameters, that are not of primary interest but can have significant influence on the estimation of primary parameters. Such kind of issues arise in a great variety of applications, including seismic inverse problems \cite{Tarantola:1984}, dynamic systems \cite{Fahrmeir:2010}, optimal experimental design \cite{Horesh:2011},  uncertainty quantification \cite{Flath:2011} and pharmacokinetic modeling \cite{Bradley:1996}. 
Nevertheless, one may note that this is quite a big number with respect to the 31 independent mass-related parameters used in the finite-range droplet model (FRDM) \cite{RevModPhys.75.1021}, or, for instance, 19 parameters used in the Hartree-Fock BCS-method, the HFBCS-1 mass formula, \cite{GORIELY2001311}, and in Hartree-Fock-Bogoliubov  HBF-1  mass formulas \cite{Samyn2002142}. However, this number is less than 477 parameters required in the mass relations used in \cite{RevModPhys.41.S1}, or by factor of three less than 928 parameters used in the mass predictions based on the Garvey-Kelson mass relations \cite{Janecke1988265}.
Indeed, we have a big number of parameters, because we do not know the exact number of exponents in this framework. However, in our case it is not an ultimate number, since most of these parameters are strongly correlated with each other, those in further computations, one can drop some of them off from the final solution in order to keep just uncorrelated ones. From our estimation,  the total number of unknown parameters can be reduced by a factor of $2$ or $4$. Unfortunately, the search for these kind of correlations between unknown parameters requires a lot of computation resources due to nonlinearity of the problem, which is currently not available for us. In light of this issue, we decide stick to this number in the current work. 

\begin{figure*}
	\centering
	\subfigure[]{
		\includegraphics[scale=0.31]{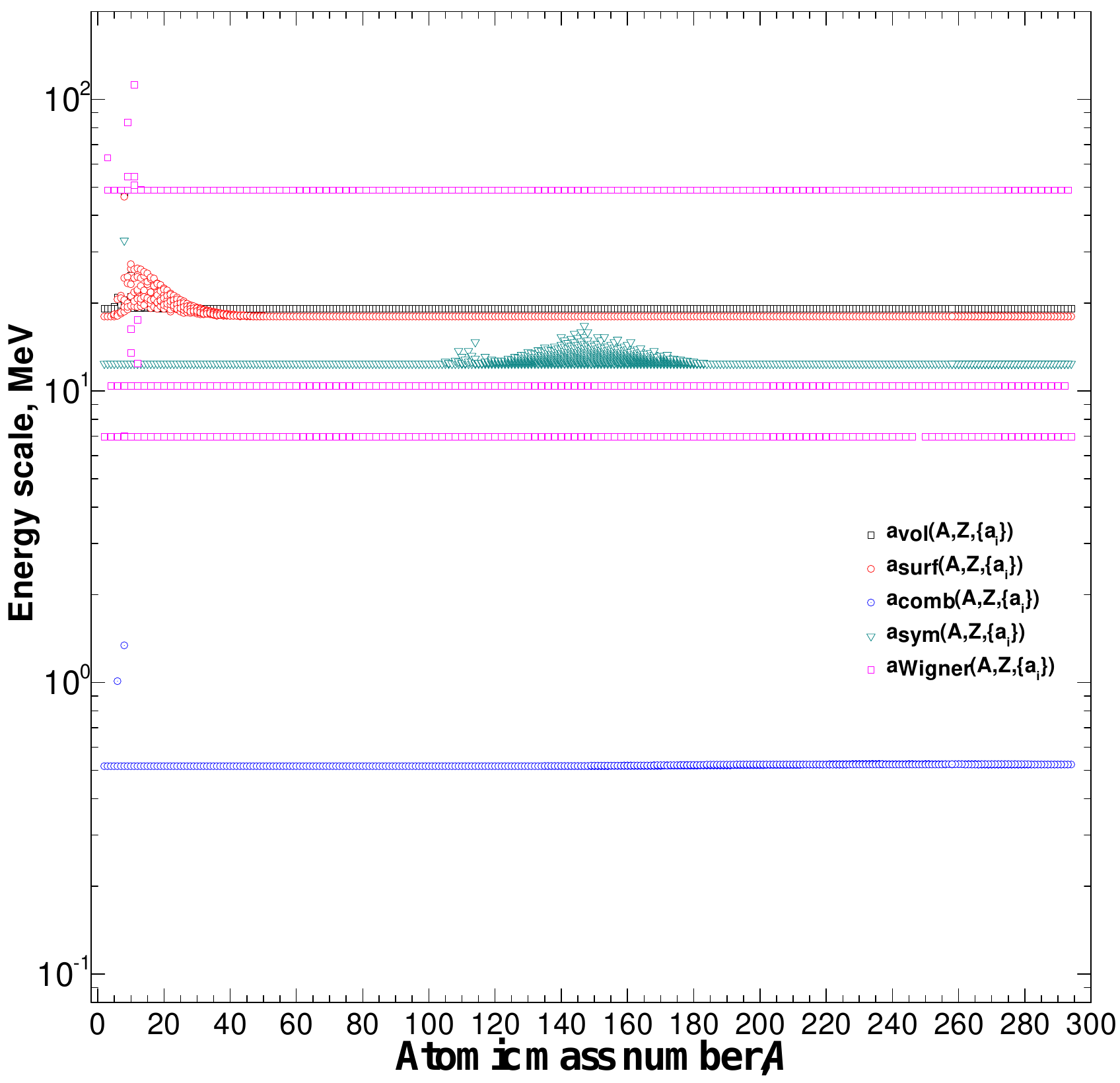}
	}
	\subfigure[]{
		\includegraphics[scale=0.31]{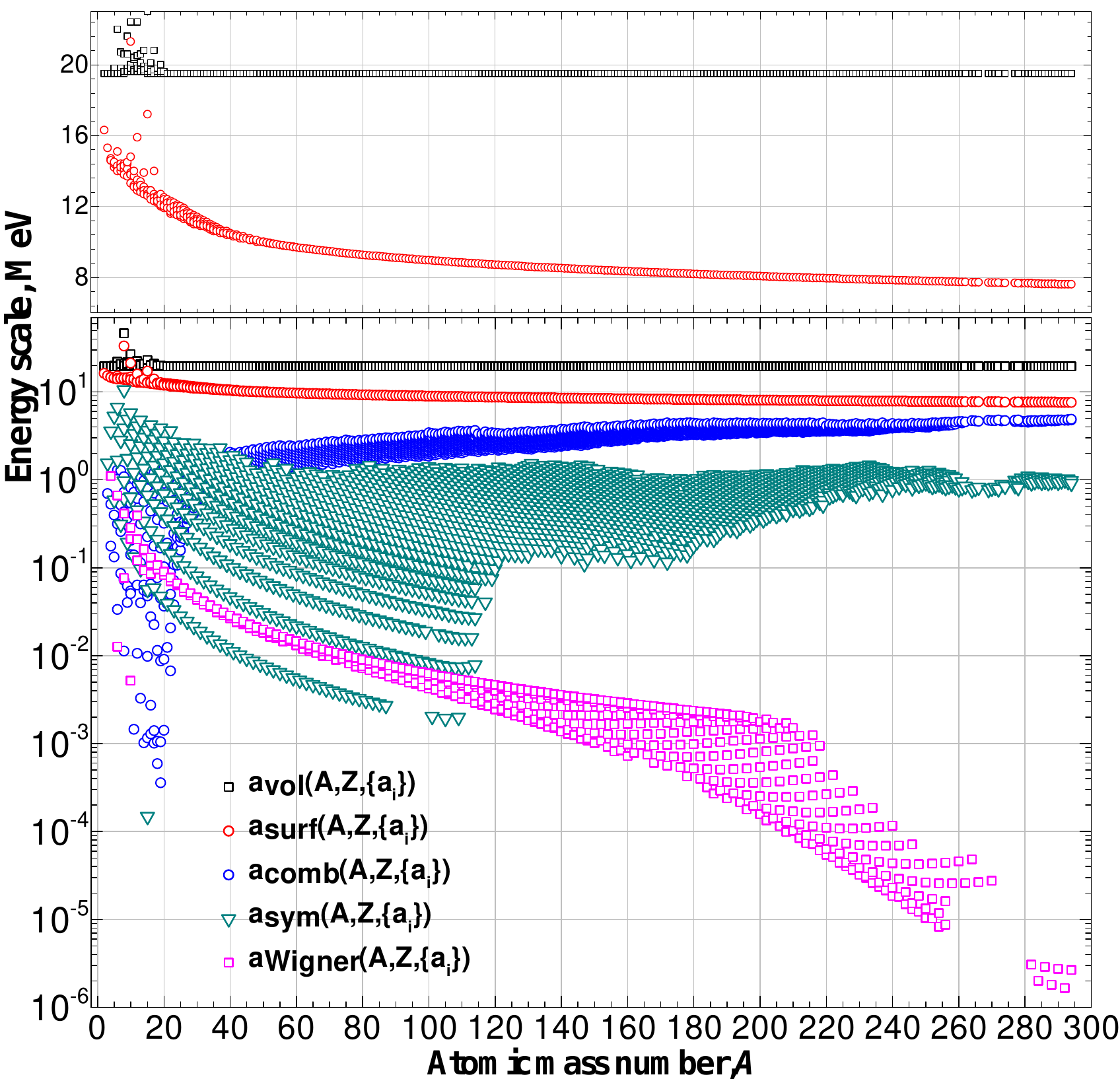}
	}
	
	\subfigure[]{
		\includegraphics[scale=0.31]{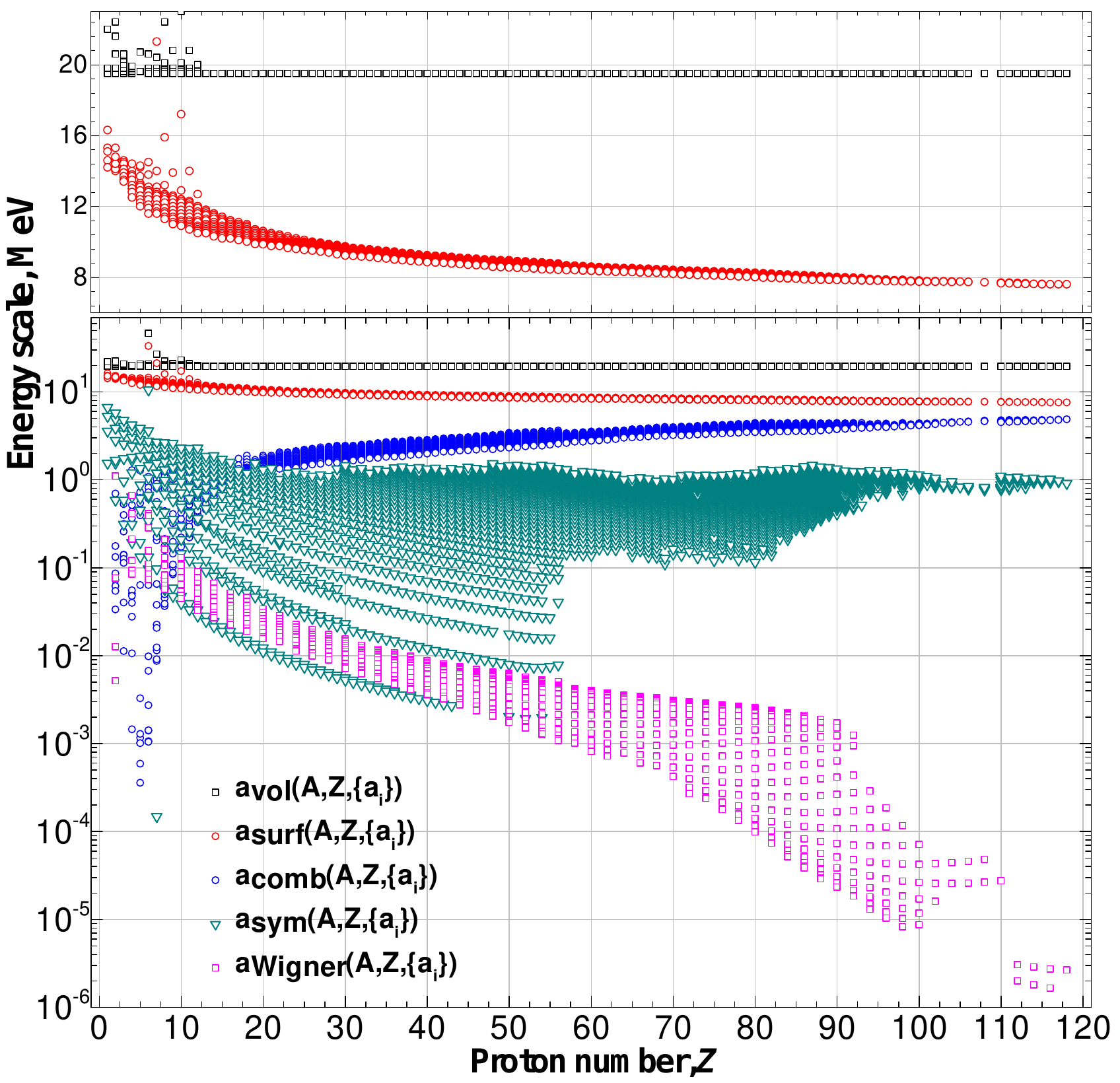}
	}
	\subfigure[]{
		\includegraphics[scale=0.31]{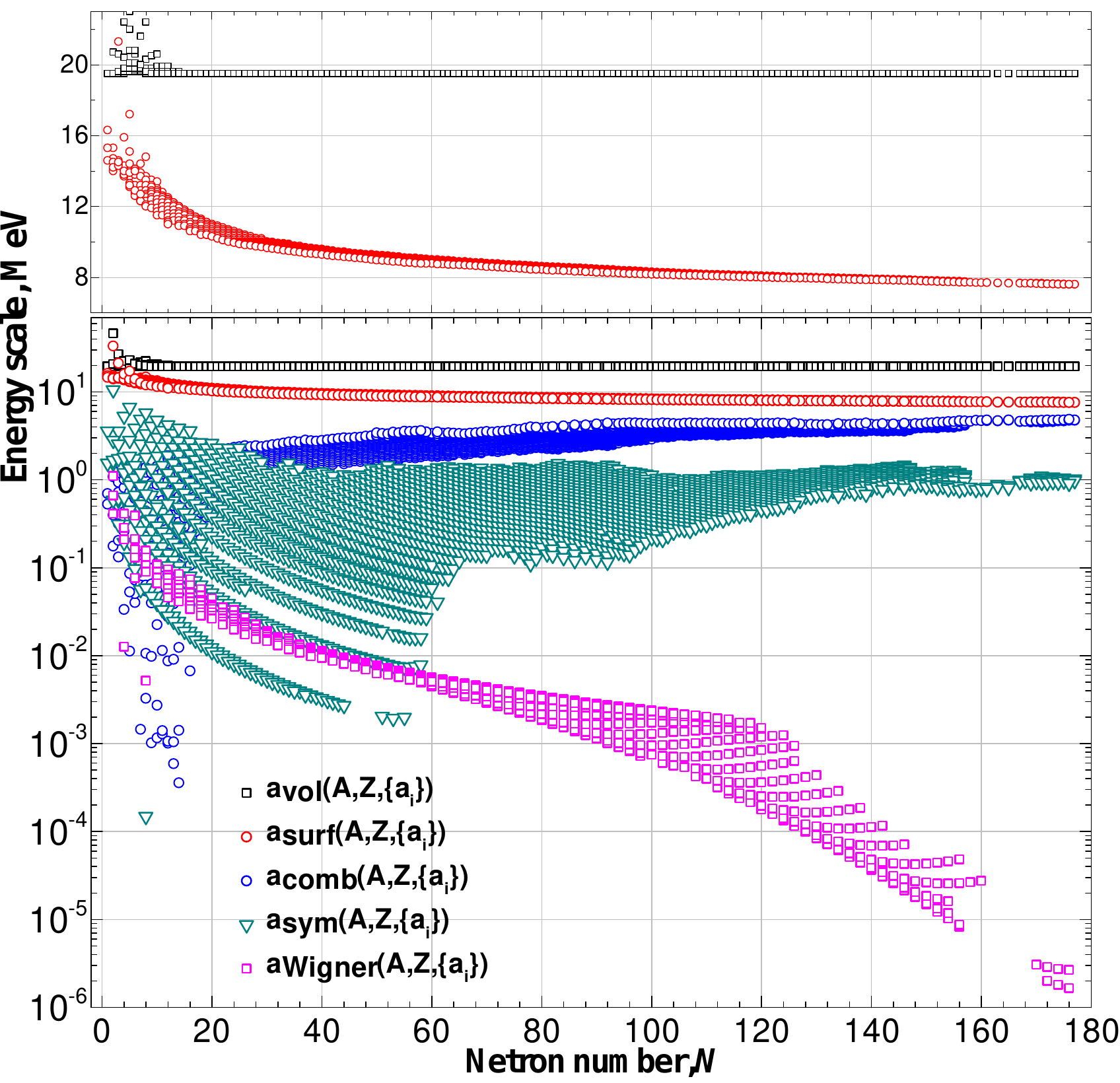}
	}
	\caption{Structure functions $\alpha_{vol}, \alpha_{surf}, \alpha_{comb}, \alpha_{sym}$ and $\alpha_{Wigner}$ shown with respect to the atomic mass numbers (a). The products of the structure functions with the corresponding factors from the BW formula, Eq.\eqref{eq:BWparametrization_Wigner_CorMN}, shown with respect to the proton, neutron and atomic mass numbers (b-d). Zoomed region in the bottom pad shows the $\alpha_{vol}$ and $\alpha_{surf}$ structure functions.
	}
	\label{fig:Structure_constants}
\end{figure*}

\section{Results}
\label{Results}

A major program in computational physics is to calculate nuclear properties from underlying realistic nuclear forces. A careful analysis of the AME2012 \cite{Ami2012_ChinPhysC, NUBASE2012_ChinPhysC} database using the dynamic auto-regularization method or, more precisely, the REGN program \cite{Alexandrov:19701285, Alexandrov:1973, Alexandrov:1982, Alexandrov:1983}, which is the constructive development of Tikhonov regularization method \cite{Tikhonov:1986, Tikhonov:1995, Tikhonov:1983}, for solving the nonlinear system of equations Eq.\eqref{eq:BWparametrization_Wigner_CorMN}, reveals phenomenological functional features of previously unknown parameters of the BW mass formula. 

The search and selection of the best solution has been done with the help of the conventional statistical methods, such as assessments of the $\chi^{2}$ test. The LCH-weighting procedure \cite{Aleksandrov2004519, LCH:1999} of the REGN program helps us to choose the better function out of two functions with the same $\chi^{2}$. 
The essential difference of the Alexandrov method \cite{Alexandrov:1973, Alexandrov:1982, Alexandrov:1983, Alexandrov:19701285} from other similar methods is extremely effective ideology regularization of inverse problem solution, which on each iteration step controls not only the actual decision, but, very importantly, uncertainty of the solution. At the same time, it should be noted that the transition from the mathematical theory of the auto-regularized iterative processes, which is based on meaningful L. Aleksandrov's theorems of convergence \cite{Alexandrov:1970}, to Fortran codes, for example, REGN-Dubna \cite{Alexandrov:1983}, FXY-Sofia-Dubna \cite{AleksandrovPriv1997} is a complicated, but technically clear work.

The presented parameterization of the binding energy allows us to solve the formulated inverse problem, see Eqs.\eqref{eq:BE_j} - \ref{eq:MExess_j}, with the help of the Aleksandrov's auto-regularized method. This solution provides us description of the 2564 nucleus masses and their binding energies starting from the atomic mass number, $A=2$, with relative error $-1.1924\times10^{-6}$ and $3.2197\times10^{-4}$ for atomic mass and binding energy respectively. Where the relative error for the binding energies, the atomic masses and etc. was computed as:
\begin{equation}
\bar{\epsilon}_{rel}=\frac{1}{N}\sum_{j=1}^{N=2564} \frac{E_{B,j}^{\rm{Expt}}(A, Z) -  E_{B,j}^{\rm{Th}}(A, Z, \{a_{i}\})}{E_{B,j}^{\rm{Expt}}(A, Z)}.
\label{eq:MeanAbsoluteError}
\end{equation}

Mean absolute error $\bar{\epsilon}_{abs}$ is $-0.0509$ and $7.8488\times10^{-4}$ for atomic mass and binding energy respectively. The maximum absolute deviation is less than 2.6 MeV for the atomic mass and less than 0.82 MeV for the binding energy. In addition, the various characteristics have been obtained, see Fig.\ref{fig:PowerFactors_AZN} and Fig.\ref{fig:Structure_constants}. Here we would like to note, that one might reach a deviation less than 1.9 MeV for atomic masses by exclusion the  Hydrogen atom from the consideration.
\begin{figure*}
	\centering
	\includegraphics[scale=0.84]{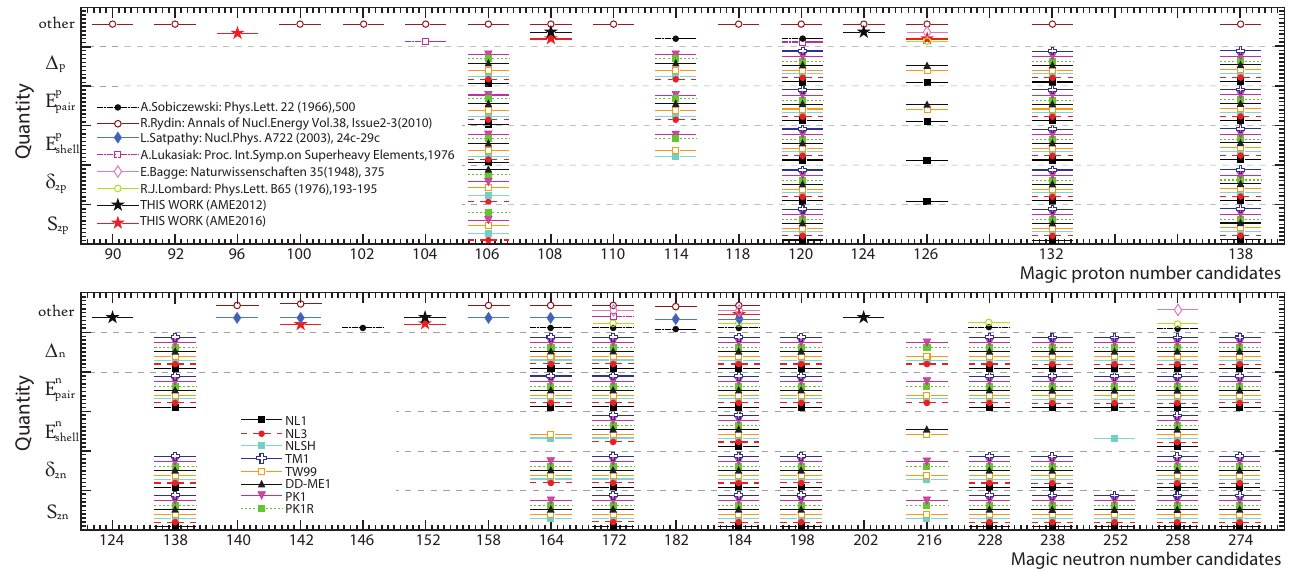}
	\caption{
		The possible magic proton numbers suggested by two-proton separation energies, two-proton gaps, shell correction energies, pairing energies and effective pairing gaps for the proton (top). The possible magic neutron number suggested by two-neutron separation energies, two-neutron gaps, shell correction energies, pairing energies and effective pairing gaps for the neutron (bottom). Magic numbers are shown for the relativistic continuum Hartree-Bogoliubov (RCHB) theory calculation with the mostly used effective interactions 
		NL1 \cite{PhysRevLett.57.2916}, NL3 \cite{PhysRevC.55.540}, NLSH \cite{PhysRevC.45.2514}, TM1 \cite{SUGAHARA1994557}, TW99 \cite{Typel1999331}, DD-ME1 \cite{PhysRevC.66.024306}, PK1 \cite{PhysRevC.69.034319} and PK1R \cite{PhysRevC.69.034319}. Data taken from \cite{Zhang2005106}. The additional magic number candidates correspond to predictions of these models \cite{Satpathy2003C24,  Sorlin2008602, Rydin2011238, Rydin20112356, EBagge:1948, Sobiczewski1966500, Rouben1972385, Lukasiak:1976, Lombard1976193,  ROUBEN19776,  Kanungo200258, Prelas20142467}, and shown in the horizontal bin $\textendash$ $other$. Predictions of the current work for the super-heavy region are shown with a black star. Note, that magic numbers are shown along the $x$-axis with a fixed arbitrary scale.
	}
	\label{fig:MagNumberList}
\end{figure*}

The model standard deviation computed using the M\"{o}ller and Nix formula:
\begin{equation}
\begin{split}
&\sigma_{\rm{mod}}^{2} =\\
& \frac{1}{\sum w_{j}} \sum_{j=1}^{2564} w_{j}\left[ \left(   M_{a.m.,j}^{\rm{Expt}}(A,Z) - M_{a.m.,j}^{\rm{Th}}(A,Z, \{a_{i}\})  \right)^{2} - (\sigma_{j}^{\rm{Expt}})^{2} \right],
\label{eq:MollerFit_quality}
\end{split}
\end{equation}
where $M_{a.m.,j}^{\rm{Th}}(A,Z, \{a_{i}\})$ is the model prediction for the data point $i$, $\sigma_{j}^{\rm{Expt}}=\sigma_{j}^{\rm{Expt}}(A,Z)$ is the experimental error with which each of the data points $M_{a.m.,j}^{\rm{Expt}}(A,Z)$ is measured, assuming that the inherent errors of the model follow a Gaussian distribution, see for details \cite{Moller1988213}. The same formula been used to compute the rms values for the binding energy. The results are shown in Table~\ref{tab:RMS_error}. 
Such big mean errors for the atomic mass are driven by the very light mass sector, which is not well described by this approach.

The mean model error is 
\begin{equation}
\begin{split}
& \bar{\epsilon}_{\rm{mod}} =\\
& \frac{1}{\sum w_{j}} \sum_{j=1}^{2564} w_{j}\left[ \left(   M_{a.m.,j}^{\rm{Expt}}(A,Z) - M_{a.m.,j}^{\rm{Th}}(A,Z, \{a_{i}\})  \right)  \right],
\label{eq:MollerFit_meanModelError}
\end{split}
\end{equation}
\begin{table}[ht]
\begin{center}
\begin{tabular}{llccc}
\hline \hline
\multicolumn{2}{c}{\small{2012 data (2564 nuclei)}}\\
\multicolumn{1}{c}{Proton magic numbers}
&\multicolumn{1}{c}{Neutron magic numbers}\\
\hline
$~~~~1 ~\leq~ 2  ~\textless~ 5$        &~~~~$1 ~\leq~ 2  ~\textless~ 5$\\
$~~~~5 ~\leq~ 8  ~\textless~ 11$       &~~~~$5 ~\leq~ 8  ~\textless~ 11$\\
$~~11 ~\leq~ 14  ~\textless~ 17$     &~~$11 ~\leq~ 14  ~\textless~ 17$\\
$~~17 ~\leq~ 20  ~\textless~ 24$     &~~$17 ~\leq~ 20  ~\textless~ 24$\\
$~~24 ~\leq~ 28  ~\textless~ 39$     &~~$24 ~\leq~ 28  ~\textless~ 39$\\
$~~39 ~\leq~ 50  ~\textless~ 66$     &~~$39 ~\leq~ 50  ~\textless~ 66$\\
$~~66 ~\leq~ 82  ~\textless~ 95$     &~~$66 ~\leq~ 82  ~\textless~ 103$\\
$~~95 ~\leq~ 108  ~\textless~ 116$   &$103 ~\leq~ 124  ~\textless~ 138$\\
$116 ~\leq~ 124  ~\textless~ {\rm{unknown}}$   &$138 ~\leq~ 152  ~\textless~ 177$\\
&$177 ~\leq~ 202  ~\textless~ {\rm{unknown}}$\\
\hline
\multicolumn{2}{c}{\small{2016 data (2496 nuclei)}}\\
\hline
$~~66 ~\leq~ 82  ~\textless~ 89$     &~~$66 ~\leq~ 82  ~\textless~ 104$\\
$~~89 ~\leq~ 96  ~\textless~ 102$   &$104 ~\leq~ 126  ~\textless~ 134$\\
$102 ~\leq~ 108  ~\textless~ 117$   &$134 ~\leq~ 142  ~\textless~ 147$\\
$117 ~\leq~ 126  ~\textless~ {\rm{unknown}}$  &$147 ~\leq~ 152  ~\textless~ 168$\\
                                                                                &$168 ~\leq~ 184  ~\textless~ {\rm{unknown}}$\\
\hline \hline
\end{tabular}
\caption{
The range of influence of proton and neutron magics numbers obtained in this work using AME2012 data. The bottom lines correspond to the magic numbers obtained on the base of the AME2016 using solution of the inverse problem for the AME2012 data. 
}
\label{tab:ZN_magnumbers}
\end{center}
\end{table}

The weight factor is 
\begin{equation}
w_{j}= \left( \left(\sigma_{j}^{\rm{Expt}}(A,Z)\right)^{2} + \left(\sigma_{j}^{\rm{Th}}(A,Z)\right)^{2}\right)^{-k}. \\
\label{eq:SigmaRMSweight_factor}
\end{equation}
The benefit of such weighting in the form given by Eq.\eqref{eq:MollerFit_quality} is that one may ask if $k = 2$ is the only possibility, or if other values of $k$ are possible. We shall not discuss this question in depth here but instead will comment that if all $(\sigma_{j}^{\rm{Expt}})^{2}$ are equal, then all values of $k$ yield the same equation. 
However, the more interesting case is how $k$ should be chosen if the $(\sigma_{j}^{\rm{Expt}})^{2}$ have different values, for details see \cite{Moller1988213}. To gain some additional insight into the properties of the applied method we compare the root-mean-square deviation of our calculated masses from the AME2012 for three different choices of $k$ factor, namely 1, 1.5 and 2. In Table~\ref{tab:RMS_error} we provide our $\sigma_{\rm{mod}}$ for the different $k$'s.

The mean modified error, Eq.\eqref{eq:MollerFit_meanModelError}, of our solution is of 1.232(1\%) $\textendash$ 1.231(5\%) MeV, the $\sigma_{\rm{mod}}$ of 1.11(1\%) $\textendash$ 1.565(5\%) MeV (assuming that the theoretical uncertainty varies between 1\% and 5\%, see Eq.\eqref{eq:MeanError}, and $k=2$) for the atomic mass. While for the binding energy, using the same assumption for band of the theoretical uncertainties, we obtain $\bar{\epsilon}_{\rm{mod}}$, of fit of 0.044(1\%) $\textendash$ -0.0057(5\%) MeV, the $\sigma_{\rm{mod}}$ of 0.209(1\%) $\textendash$ 0.223(5\%) MeV,  which can be compared with the latest fits for all modern mass formulas \cite{PhysRevLett.16.197, Janecke1988265, Liran1976431, Nayak1999213, Koura200047, PhysRevC.52.R23}, see Table~\ref{tab:RMS_error}. 

From this table one may note, that $\sigma$ is different for the binding energy and atomic mass. However, from Eq.\eqref{eq:AtomicMassFunc} it is clear that the binding energy and atomic mass are related with each other by a constant. Therefore, where did this deference come from? In fact, during our analysis we work with the binding energy per nucleon, as it presented in the AME data base. This is turn means, that if one will multiply both the experimental binding energies and the binding energies predicted by the inverse problem framework by the corresponding atomic mass number. Then will take the difference of these two energies. The observed discrepancy will gone. 

The mean error of the model is computed, assuming that the theoretical uncertainty is modeled through a multi-band set of 3\% and 5\%, in the following way:
\begin{equation}
\begin{split}
&\bar{\epsilon}_{\%}=\frac{1}{N}\sum_{j=1}^{N=2564}  E_{B,j}^{\rm{Th}}(A, Z, \{a_{i}\})\times s_{2}\\
\label{eq:MeanError}
\end{split}
\end{equation}

The hypothesis testing has been done with the help of formula (see Eq.6 in \cite{Ami2012_ChinPhysC}):
\begin{equation}
\chi^{2}=\sum_{j=1}^{N=2564}\left(  \frac{E_{B,j}^{\rm{Expt}}(A, Z)-E_{B,j}^{\rm{Th}}(A, Z, \{a_{i}\})}{\sigma_{j}(A, Z)}   \right)^{2},
\label{eq:Chi2_test}
\end{equation}
where
\begin{equation}
\sigma_{j}(A, Z)=s_{1} \sigma_{j}^{\rm{Expt}}(A, Z)+s_{2}E_{B}^{\rm{Expt}}(A, Z).
\label{eq:Sigma_treat}
\end{equation}
Here $\sigma_{j}^{\rm{Expt}}(A, Z)$ is the uncertainty of a nucleus as it has been reported in AME2012,  $s_{1}$ and $s_{2}$ are the nuisance parameters, where $s_{2}$ is the percentage of the given experimental value. Eq.\eqref{eq:Chi2_test} has been applied to all considered nuclei in this work. Table~\ref{tab:Chi2Sigma_treatment} illustrates the quality of the descriptions of the binding energy, the nuclear and atomic masses, and the mass excess assuming a different hypothesis for the nuisance parameters $s_{1}$ and $s_{2}$.  Note, that some masses of nuclei are measured with very high precision, which can be noticed from the mass excess column in AME2012, but due to artificial cutoff of the significant digits the uncertainties for these nuclei are given as zero uncertainty. Since we do not know the exact numbers, we treat uncertainties for these nuclei as 1\% of the given experimental value.

We used such analyses in order to be able to allocate questionable uncertainties in the 2012 Audi-Wapstra tables. In Table~\ref{tab:Chi2Sigma_treatment} we report the value of corrections that have been made.

In addition, one can define as above the normalized chi, $\chi_{n}$ (or `consistency factor' or Birge ratio): 
\begin{equation}
\chi_{n}=\sqrt{\frac{\chi^{2}}{N-{\mathcal{N}}_{\rm{param}}}}
\label{eq:NormalizedChi}
\end{equation}
where $N-{\mathcal{N}}_{\rm{param}}$ is the number of degrees of freedom.

\begin{table}[t]
\begin{center}
\begin{tabular}{lccccccc}
\hline \hline
\multicolumn{1}{c}{}
&\multicolumn{1}{c}{$s_{1}$}
&\multicolumn{1}{c}{$s_{2},\%$}
&\multicolumn{1}{c}{$\chi^{2}$}
&\multicolumn{1}{c}{$\chi_{n}$}\\
\hline
\multicolumn{1}{c}{\small{$E_{B}(A, Z, \{a_{i}\})$}}
&\multicolumn{1}{c}{\small{$1$}}
&\multicolumn{1}{c}{\small{$0.12\times10^{-1}$}}
&\multicolumn{1}{c}{\small{$0.230\times10^{4}$}}
&\multicolumn{1}{c}{\small{$0.996$}}\\
\multicolumn{1}{c}{\small{$M_{n.m.}(A, Z, \{a_{i}\})$}}
&\multicolumn{1}{c}{\small{$1$}}
&\multicolumn{1}{c}{\small{$0.337\times10^{-4}$}}
&\multicolumn{1}{c}{\small{$0.232\times10^{4}$}}
&\multicolumn{1}{c}{\small{$0.999$}}\\
\multicolumn{1}{c}{\small{$M_{a.m.}(A, Z, \{a_{i}\})$}}
&\multicolumn{1}{c}{\small{$1$}}
&\multicolumn{1}{c}{\small{$0.337\times10^{-4}$}}
&\multicolumn{1}{c}{\small{$0.232\times10^{4}$}}
&\multicolumn{1}{c}{\small{$0.999$}}\\
\multicolumn{1}{c}{\small{$\Delta m(A, Z, \{a_{i}\})$}}
&\multicolumn{1}{c}{\small{$1$}}
&\multicolumn{1}{c}{\small{$0.304$}}
&\multicolumn{1}{c}{\small{$0.331\times10^{4}$}}
&\multicolumn{1}{c}{\small{$0.979$}}\\
\hline
\multicolumn{1}{c}{\small{$E_{B}(A, Z, \{a_{i}\})$}}
&\multicolumn{1}{c}{\small{$1$}}
&\multicolumn{1}{c}{\small{$0$}}
&\multicolumn{1}{c}{\small{$0.139\times10^{13}$}}
&\multicolumn{1}{c}{\small{$0.245\times10^{5}$}}\\
\multicolumn{1}{c}{\small{$M_{n.m.}(A, Z, \{a_{i}\})$}}
&\multicolumn{1}{c}{\small{$1$}}
&\multicolumn{1}{c}{\small{$0$}}
&\multicolumn{1}{c}{\small{$0.202\times10^{15}$}}
&\multicolumn{1}{c}{\small{$0.294\times10^{6}$}}\\
\multicolumn{1}{c}{\small{$M_{a.m.}(A, Z, \{a_{i}\})$}}
&\multicolumn{1}{c}{\small{$1$}}
&\multicolumn{1}{c}{\small{$0$}}
&\multicolumn{1}{c}{\small{$0.252\times10^{15}$}}
&\multicolumn{1}{c}{\small{$0.330\times10^{6}$}}\\
\multicolumn{1}{c}{\small{$\Delta m(A, Z, \{a_{i}\})$}}
&\multicolumn{1}{c}{\small{$1$}}
&\multicolumn{1}{c}{\small{$0$}}
&\multicolumn{1}{c}{\small{$0.254\times10^{15}$}}
&\multicolumn{1}{c}{\small{$0.331\times10^{6}$}}\\
\hline \hline
\end{tabular}
\caption{
The calculated $\chi^{2}$  and $\chi_{n}$  for the simplest choice of $s_{1}$ and $s_{2}$ values, where $s_{2}$ is given in percent. Our case is the top subsection of the table, see Eq.\eqref{eq:Sigma_treat}, and the case without uncertainty turning is the bottom subsection.
}
\label{tab:Chi2Sigma_treatment}
\end{center}
\end{table}

\begin{table*}[ht]
\begin{center}
\begin{tabular}{l|cc|cc|cclc|cc|cccc}
\hline \hline
\multicolumn{1}{l|}{}
&\multicolumn{2}{c}{\small{1995 data}}
&\multicolumn{2}{c}{\small{2001 data}}
&\multicolumn{4}{c}{\small{2003 data}}
&\multicolumn{2}{c}{\small{2003 data}}
&\multicolumn{2}{c}{\small{2012 data}}\\
\multicolumn{1}{l|}{}
&\multicolumn{2}{c}{\small{(1768 nuclei)}}
&\multicolumn{2}{c}{\small{(2135 nuclei)}}
&\multicolumn{4}{c}{\small{(382 nuclei)}}
&\multicolumn{2}{c}{\small{(2149 nuclei)}}
&\multicolumn{2}{c}{\small{(2353 nuclei)}}\\
\cline{2-13}
\multicolumn{1}{l|}{Model}
&\multicolumn{1}{c}{$\sigma$}
&\multicolumn{1}{c|}{$\bar{\epsilon}$}
&\multicolumn{1}{c}{$\sigma$}
&\multicolumn{1}{c|}{$\bar{\epsilon}$}
&\multicolumn{1}{c}{$\sigma$}
&\multicolumn{1}{c}{$\bar{\epsilon}$}
&\multicolumn{1}{l}{$\sigma_{\rm{mod}}$}
&\multicolumn{1}{c|}{$\bar{\epsilon}_{\rm{mod}}$}
&\multicolumn{1}{c}{$\sigma$}
&\multicolumn{1}{c|}{$\sigma_{\rm{RBF}}$}
&\multicolumn{1}{c}{$\sigma$}
&\multicolumn{1}{c}{$\sigma_{\rm{RBF}}$}\\
\hline
\small{RMF \cite{Niu:2013hda}} 
&\footnotesize{}  
&\footnotesize{}  
&\footnotesize{}  
&\footnotesize{}  
&\footnotesize{}  
&\footnotesize{}  
&\footnotesize{}  
&\footnotesize{}  
&\footnotesize{}  
&\footnotesize{}  
&\footnotesize{2.217}  
&\footnotesize{0.488$^{\ast}$}\\  
\small{HFBCS-1 \cite{GORIELY2001311}} 
&\footnotesize{0.738} 
&\footnotesize{0.102} 
&\footnotesize{0.805} 
&\footnotesize{0.180} 
&\footnotesize{1.115}  
&\footnotesize{0.494}  
&\footnotesize{1.056}  
&\footnotesize{0.460}
&\footnotesize{}  
&\footnotesize{}  
&\footnotesize{}  
&\footnotesize{}\\  
\small{HFB-1 \cite{Samyn2002142}} 
&\footnotesize{0.772}  
&\footnotesize{0.026}  
&\footnotesize{0.764}  
&\footnotesize{-0.057} 
&\footnotesize{1.123}  
&\footnotesize{0.510}  
&\footnotesize{1.091}  
&\footnotesize{0.494} 
&\footnotesize{}  
&\footnotesize{}  
&\footnotesize{}  
&\footnotesize{}\\  
\small{HFB-2 \cite{PhysRevC.66.024326}} 
&\footnotesize{}  
&\footnotesize{} 
&\footnotesize{0.674}  
&\footnotesize{0.000}
&\footnotesize{0.769}  
&\footnotesize{0.377}  
&\footnotesize{0.724}  
&\footnotesize{0.356} 
&\footnotesize{}  
&\footnotesize{}  
&\footnotesize{}  
&\footnotesize{}\\  
\small{HFB-21 \cite{Goriely:2010bm, Niu:2013hda}} 
&\footnotesize{}  
&\footnotesize{}  
&\footnotesize{}  
&\footnotesize{}  
&\footnotesize{}  
&\footnotesize{}  
&\footnotesize{}  
&\footnotesize{}  
&\footnotesize{}  
&\footnotesize{}  
&\footnotesize{0.572}  
&\footnotesize{0.410$^{\ast}$} \\
\small{FRDM \cite{Moller1995185, MYERS1974186, Moller1988213,Niu:2013hda}} 
&\footnotesize{0.678}  
&\footnotesize{0.023}  
&\footnotesize{0.676}  
&\footnotesize{0.072}  
&\footnotesize{0.655}  
&\footnotesize{0.247}  
&\footnotesize{0.485}  
&\footnotesize{0.202} 
&\footnotesize{0.656}  
&\footnotesize{0.283$^{\ast}$}  
&\footnotesize{0.654}  
&\footnotesize{0.268$^{\ast}$}\\  
\small{TF-FRDM \cite{Myers1996141}} 
&\footnotesize{0.662}    
&\footnotesize{$-0.034$} 
&\footnotesize{0.655}    
&\footnotesize{$-0.036$} 
&\footnotesize{0.655}   
&\footnotesize{-0.085}  
&\footnotesize{0.511}   
&\footnotesize{-0.121} 
&\footnotesize{}  
&\footnotesize{}  
&\footnotesize{}  
&\footnotesize{}\\  
\small{DZ10 \cite{PhysRevC.52.R23, PhysRevC.59.R2347, Niu:2013hda}} 
&\footnotesize{0.375}  
&\footnotesize{$-0.010$} 
&\footnotesize{0.373}  
&\footnotesize{0.009}  
&\footnotesize{0.479}  
&\footnotesize{0.054}  
&\footnotesize{0.378}  
&\footnotesize{0.028} 
&\footnotesize{}  
&\footnotesize{}  
&\footnotesize{0.591}  
&\footnotesize{0.225$^{\ast}$} \\
\small{DZ31 \cite{PhysMexicRevS54.129, Niu:2013hda}} 
&\footnotesize{}  
&\footnotesize{}  
&\footnotesize{}  
&\footnotesize{}  
&\footnotesize{}  
&\footnotesize{}  
&\footnotesize{}  
&\footnotesize{}  
&\footnotesize{}  
&\footnotesize{}  
&\footnotesize{0.397}  
&\footnotesize{0.204$^{\ast}$}\\
\small{Koura et al. \cite{Koura200047}} 
&\footnotesize{0.680}  
&\footnotesize{0.012}  
&\footnotesize{0.682}  
&\footnotesize{0.053}  
&\footnotesize{0.755}  
&\footnotesize{0.200}  
&\footnotesize{0.676}  
&\footnotesize{0.163}  
&\footnotesize{}  
&\footnotesize{}  
&\footnotesize{}  
&\footnotesize{}\\  
\small{KTUY \cite{doi:10.1143/PTP.113.305, Niu:2013hda}} 
&\footnotesize{}  
&\footnotesize{}  
&\footnotesize{}  
&\footnotesize{}  
&\footnotesize{}  
&\footnotesize{}  
&\footnotesize{}  
&\footnotesize{}  
&\footnotesize{}  
&\footnotesize{}  
&\footnotesize{0.701}  
&\footnotesize{0.210$^{\ast}$}\\
\small{Nayak-Satpathy \cite{Nayak1999213}}
&\footnotesize{0.359}  
&\footnotesize{0.000}  
&\footnotesize{0.485}  
&\footnotesize{0.047}  
&\footnotesize{0.837}  
&\footnotesize{0.229}  
&\footnotesize{0.779}  
&\footnotesize{0.208} 
&\footnotesize{}  
&\footnotesize{}  
&\footnotesize{}  
&\footnotesize{}\\  
\small{GK \cite{PhysRevLett.16.197, RevModPhys.41.S1}\cite{Cheng:2014uha}} 
&\footnotesize{0.163}  
&\footnotesize{$-0.010$} 
&\footnotesize{}    
&\footnotesize{}   
&\footnotesize{0.717}  
&\footnotesize{0.127}  
&\footnotesize{0.653}  
&\footnotesize{0.096} 
&\footnotesize{}  
&\footnotesize{}  
&\footnotesize{0.0677 -- 0.1665}  
&\footnotesize{}\\ 
\small{J\"{a}necke-Masson \cite{Janecke1988265}} 
&\footnotesize{0.247}  
&\footnotesize{$-0.010$} 
&\footnotesize{0.319}  
&\footnotesize{0.010}  
&\footnotesize{0.540}  
&\footnotesize{0.070}  
&\footnotesize{0.451}  
&\footnotesize{0.071} 
&\footnotesize{}  
&\footnotesize{}  
&\footnotesize{}  
&\footnotesize{}\\  
\small{Liran-Zeldes \cite{Liran1976431}} 
&\footnotesize{0.276}  
&\footnotesize{$-0.005$} 
&\footnotesize{0.586}  
&\footnotesize{$-0.036$}
&\footnotesize{0.722}  
&\footnotesize{-0.226}  
&\footnotesize{0.554}  
&\footnotesize{-0.253} 
&\footnotesize{}  
&\footnotesize{}  
&\footnotesize{}  
&\footnotesize{}\\  
\small{ETFSI-2 \cite{doi:10.1063/1.1361389}} 
&\footnotesize{}  
&\footnotesize{}  
&\footnotesize{}  
&\footnotesize{}  
&\footnotesize{}  
&\footnotesize{}  
&\footnotesize{}  
&\footnotesize{}  
&\footnotesize{}  
&\footnotesize{}  
&\footnotesize{0.719}  
&\footnotesize{0.360$^{\ast}$}\\
\small{WS3 \cite{Wang:2011hxa,Niu:2013hda}} 
&\footnotesize{}  
&\footnotesize{}  
&\footnotesize{}  
&\footnotesize{}  
&\footnotesize{}  
&\footnotesize{}  
&\footnotesize{}  
&\footnotesize{}  
&\footnotesize{0.336}  
&\footnotesize{0.223$^{\ast}$}  
&\footnotesize{0.335}  
&\footnotesize{0.207$^{\ast}$}\\ 
\hline
\multicolumn{1}{l}{}
&\multicolumn{2}{c}{\small{}}
&\multicolumn{2}{c}{\small{}}
&\multicolumn{4}{c}{\small{2003 (371 nuclei $(N,Z \geq 28)$ )}}
&\multicolumn{2}{c}{\small{}}\\
\hline
\small{LDM \cite{Mendoza-Temis:2008ztv, Morales:2010zza}} 
&\footnotesize{}  
&\footnotesize{}  
&\footnotesize{}  
&\footnotesize{}  
&\footnotesize{1.9307}  
&\footnotesize{}  
&\footnotesize{0.8763$^{\ast\ast}$} 
&\footnotesize{}  
&\footnotesize{}\\  
\small{LDMM \cite{Mendoza-Temis:2008ztv, Morales:2010zza}} 
&\footnotesize{}  
&\footnotesize{}  
&\footnotesize{}  
&\footnotesize{}  
&\footnotesize{0.9955}  
&\footnotesize{}  
&\footnotesize{0.3718$^{\ast\ast}$} 
&\footnotesize{}  
&\footnotesize{}\\  
\small{DZ10 \cite{PhysRevC.52.R23, PhysRevC.59.R2347, Morales:2010zza}}
&\footnotesize{}  
&\footnotesize{}  
&\footnotesize{}  
&\footnotesize{}  
&\footnotesize{0.3348}  
&\footnotesize{}  
&\footnotesize{0.2727$^{\ast\ast}$} 
&\footnotesize{}  
&\footnotesize{}\\  
\hline\hline
\end{tabular}
\begin{tabular}{l|cccccccccc}
\multicolumn{1}{l|}{}
&\multicolumn{4}{c}{\small{2012 data (2436 nuclei $(N,Z \geq 1)$)}}\\
\cline{2-5}
\multicolumn{1}{l|}{\bf{This work}}
&\multicolumn{1}{c}{$\sigma$}
&\multicolumn{1}{c}{$\bar{\epsilon}$}
&\multicolumn{1}{c}{$\sigma_{\rm{mod}}$}
&\multicolumn{1}{c}{$\bar{\epsilon}_{\rm{mod}}$}\\
\hline
\footnotesize{} 
&\footnotesize{} 
&\footnotesize{}
&\scriptsize{(for $k=2$)}\footnotesize{~0.197 -- 0.194}  
&\footnotesize{0.085 -- (-0.033)}\\  
\small{$E_{B}(A,Z,\{ a_{i}\})$} 
&\footnotesize{0.0280}  
&\footnotesize{$6.173\times10^{-4}$}
&\scriptsize{(for $k=1.5$)}\footnotesize{~0.824 -- 0.474}   
&\footnotesize{1.85 -- (-0.195)}\\ 
\small{} 
&\footnotesize{}  
&\footnotesize{}
&\scriptsize{(for $k=1$)}\footnotesize{~3.064 -- 0.867}   
&\footnotesize{20.458 -- (-0.654)}\\  
\small{} 
&\footnotesize{}  
&\footnotesize{}
&\scriptsize{(for $k=2$)}\footnotesize{~1.257 -- 1.1257}  
&\footnotesize{0.822 -- (-0.821)}\\
\small{$M_{a.m.}(A,Z,\{ a_{i}\})$} 
&\footnotesize{0.5487}  
&\footnotesize{0.001}
&\scriptsize{(for $k=1.5$)}\footnotesize{~0.232 -- 0.104}  
&\footnotesize{(28.073  -- 5.614)$\times 10^{-3}$}\\ 
\footnotesize{} 
&\footnotesize{}  
&\footnotesize{}
&\scriptsize{(for $k=1$)}\footnotesize{~0.033 -- 0.007}  
&\footnotesize{(57.607 -- 2.304)$\times 10^{-5}$}\\
\hline
\multicolumn{1}{l|}{}
&\multicolumn{4}{c}{\small{2012 data (2564 nuclei $(N,Z \geq 1)$)}}\\
\hline
\footnotesize{} 
&\footnotesize{} 
&\footnotesize{}
&\scriptsize{(for $k=2$)}\footnotesize{~0.209 -- 0.223}  
&\footnotesize{0.044 -- (-0.057)}\\  
\small{$E_{B}(A,Z,\{ a_{i}\})$} 
&\footnotesize{0.0312}  
&\footnotesize{$7.849\times10^{-4}$}
&\scriptsize{(for $k=1.5$)}\footnotesize{~0.805 -- 0.534}   
&\footnotesize{0.650 -- (-0.330)}\\ 
\small{} 
&\footnotesize{}  
&\footnotesize{}
&\scriptsize{(for $k=1$)}\footnotesize{~2.946 -- 0.970}   
&\footnotesize{8.701 -- (-1.089)}\\  
\small{} 
&\footnotesize{}  
&\footnotesize{}
&\scriptsize{(for $k=2$)}\footnotesize{~1.11 -- 1.565}  
&\footnotesize{1.232 -- 1.231}\\
\small{$M_{a.m.}(A,Z,\{ a_{i}\})$} 
&\footnotesize{0.5508}  
&\footnotesize{0.051}
&\scriptsize{(for $k=1.5$)}\footnotesize{~0.205 -- 0.129}  
&\footnotesize{(42.157 -- 8.422)$\times 10^{-3}$}\\ 
\footnotesize{} 
&\footnotesize{}  
&\footnotesize{}
&\scriptsize{(for $k=1$)}\footnotesize{~0.0294 -- 0.0083}  
&\footnotesize{(86.531 -- 3.4549)$\times 10^{-5}$}\\
\hline
\multicolumn{1}{l|}{}
&\multicolumn{4}{c}{\small{2016 data (2496 nuclei $(N,Z \geq 1)$)}}\\
\hline
\footnotesize{} 
&\footnotesize{}  
&\footnotesize{}
&\scriptsize{(for $k=2$)}\footnotesize{~0.191 -- 0.176}   
&\footnotesize{0.084 -- (-0.02)}\\  
\small{$E_{B}(A,Z,\{ a_{i}\})$} 
&\footnotesize{0.028}  
&\footnotesize{$6.519\times10^{-4}$}
&\scriptsize{(for $k=1.5$)}\footnotesize{~0.799 -- 0.426}   
&\footnotesize{1.469 -- (-0.117)}\\  
\small{} 
&\footnotesize{}  
&\footnotesize{}
&\scriptsize{(for $k=1$)}\footnotesize{~2.977 -- 0.777}   
&\footnotesize{20.38 -- (-0.389)}\\  
\small{} 
&\footnotesize{}  
&\footnotesize{}
&\scriptsize{(for $k=2$)}\footnotesize{~1.264 --1.264}   
&\footnotesize{0.8291 -- 0.8291}\\  
\small{$M_{a.m.}(A,Z,\{ a_{i}\})$} 
&\footnotesize{0.5573}  
&\footnotesize{0.0011}
&\scriptsize{(for $k=1.5$)}\footnotesize{~0.233 -- 0.104}   
&\footnotesize{(28.254 -- 5.651)$\times 10^{-3}$}\\  
\footnotesize{} 
&\footnotesize{}  
&\footnotesize{}
&\scriptsize{(for $k=1$)}\footnotesize{~0.033 -- 0.007}   
&\footnotesize{(57.29 -- 2.292)$\times 10^{-5}$}\\  
\hline \hline
\end{tabular}
\caption{
The simplest rms error ($\sigma$) and mean error ($\bar{\epsilon}$) of fits given by various mass formulas to the masses of the AME2001 \cite{Wapstra2001129} evaluations. 
These numbers taken from \cite{RevModPhys.75.1021} with the typo corrections for the 1995 data. Since, the original ($\sigma$) for the
HFBCS-1 \cite{GORIELY2001311}, HFB-1 \cite{Samyn2002142}, DZ10 \cite{PhysRevC.52.R23, PhysRevC.59.R2347}, Koura \cite{Koura200047}, Garvey \cite{PhysRevLett.16.197, RevModPhys.41.S1}
and Liran-Zeldes \cite{Liran1976431} is different with respect to the numbers, that were shown by Lunney in \cite{RevModPhys.75.1021}
(see Table 1 in \cite{RevModPhys.75.1021} for details). 
Mass formula HFB-2 was fitted to the 2001 compilation, but all others were fitted to the 1995 or earlier evaluations \cite{AUDI19931, Audi1995409}. 
The FRDM stays for the finite-range droplet model \cite{Moller1995185}, ETFSI is the extended Thomas-Fermi plus Strutinsky integral model \cite{doi:10.1063/1.1361389},
KTUY is the Koura-Tachibana-Uno-Yamada model \cite{doi:10.1143/PTP.113.305}, WS3 is the latest version of the Weizc\"{a}cker-Skyrme model \cite{Wang:2011hxa}, 
DZ10 and DZ31 corresponds to the Duflo-Zuker model with 10 and 31 free parameters respectively \cite{PhysRevC.52.R23, PhysRevC.59.R2347, PhysMexicRevS54.129}, LDM is the liquid drop model, LDMM is an improved LDM by inclusion of schematic microscopic shell corrections \cite{Mendoza-Temis:2008ztv}. 
$(\ast)$ denotes the rms values obtained as result of the improvement of the given mass model prediction by application of the radial basis function (RBF) approach corrections to the considered model \cite{Niu:2013hda}. 
$(\ast\ast)$ denotes the rms values obtained as result of the improvement of the given mass model prediction with the help of the CLEAN method \cite{1980AA89.377C, 1980AA89.377C, doi:10.1146/annurev.aa.24.090186.001015} to the considered model \cite{Morales:2010zza}.
All errors are in MeV, except the  mean relative error, which is dimensionless. The weighted rms error ($\sigma_{\rm{mod}}$), mean error ($\bar{\epsilon}_{\rm{mod}}$) were computed with the help of the M\"{o}ller-Nix formula, see Eqs.\eqref{eq:MollerFit_quality} and \eqref{eq:MollerFit_meanModelError}. 
Results are shown assuming that the theoretical uncertainty varies between 1\% and 5\% for the atomic mass. 
}
\label{tab:RMS_error}
\end{center}
\end{table*}

As shown in Fig.\ref{fig:Structure_constants}, the products of the structure functions with the corresponding terms in the BW formula are not independent of each other but rather strongly correlated, see Eq.\eqref{eq:BWparametrization_Wigner_CorMN}. From this figure, it is clear how the effect of the binding energy saturation reflected on each of the structure functions. The mean value of the structure functions are shown in Table.\ref{tab:BW_CoeffFits}.

The mean value of the power factors in the parameterized BW mass formula do not deviate too much from the initial approximation, while $p_{1}$ decreases by a factor of two with respect to the initial approximation, $p_{1}$ = 1/3. The mean values are $\langle p_{1}(A,Z,\{a_{i}\}) \rangle$ = 0.164398, $\langle p_{2}(A,Z,\{a_{i}\}) \rangle$ = 1.311083, $\langle p_{3}(A,Z,\{a_{i}\}) \rangle$ = 1.88221, $\langle p_{4}(A,Z,\{a_{i}\}) \rangle$ = 1.847739. The dependencies of the power factors with respect to the atomic mass number, $Z$ and $N$ are shown in Fig.\ref{fig:PowerFactors_AZN}.

The correction function Eq.\eqref{eq:CorMN_func} in our work plays an important role in the parameterization procedure. Since one has to take into account the fact that there is an especially high binding energy for nuclei with double shell closures, that is, where the neutron and proton shells are both closed. The shell model does a good job of explaining the stability of nuclei with neutron and proton shell closures at 2, 8, 20, 28, 50, and 82, and when the neutron shell closes at 126. For instance, Helium-4, which has a double shell closure with 2 neutrons and 2 protons, is an example of a highly stable element. But where the closed shells are located in the heavy element region is less certain. Various models predict that it should occur at $Z=114,120$, or 126 for the protons and $N=150, 164, 172$ and 184 for the neutrons \cite{Bender200142, Zhang2005106, Satpathy2003C24, Sorlin2008602, Rydin2011238, Rydin20112356}. The defined correction function in Eq.\eqref{eq:CorMN_func} provides an indicator about the deviation in the level structure of nuclei away from closed shells. In addition, it includes the pairing effects, which can also provide reliable information about the magic number. Fig.\ref{fig:CorMagicNumbers} shows the correction energies, see Eq.\eqref{eq:CorMN_func}, as a function of mass numbers $Z, N$  and $A$. The peaks on Fig.\ref{fig:CorMagicNumbers} represent the possible magic proton and neutron numbers. A peak is enhanced if a magic proton number exists, the boundaries of the discovered magic numbers are shown in Table~\ref{tab:ZN_magnumbers}. Therefore the analysis of the latent regularities allows us to rediscover the already known set of the magic numbers ($2, 8, 20, 28, 50, 82$), and obtain the additional magic numbers for proton and neutrons. Also as a consequence of our parameterized solution, we obtain the total number of magic numbers. We have compared our magic number candidates with results of previous works, this result is illustrated in Fig.\ref{fig:MagNumberList}.

The obtained results and validation of our approach on the big set of the experimental data allows us to probe the outer boundary of the island of enhanced nuclear stability.  
Therefore, we make a prediction for the binding energy, nuclear and atomic mass, and mass excess of the recently discovery nuclei at the DGFRS separator \cite{Oganessian201562}, see Table~\ref{tab:Utyonkov_data}. 
In addition, we performed the blind analysis of the 2496 nuclei with $N,Z\geq 1$ from the AME2016 \cite{Ami2016_ChinPhysC} data using solution obtained on the base of the AME2012 data. Results of this analysis are shown in Table~\ref{tab:RMS_error}, new set of magic numbers is shown in Table \ref{tab:ZN_magnumbers}.

\begin{figure*}
\centering
\includegraphics[scale=0.25]{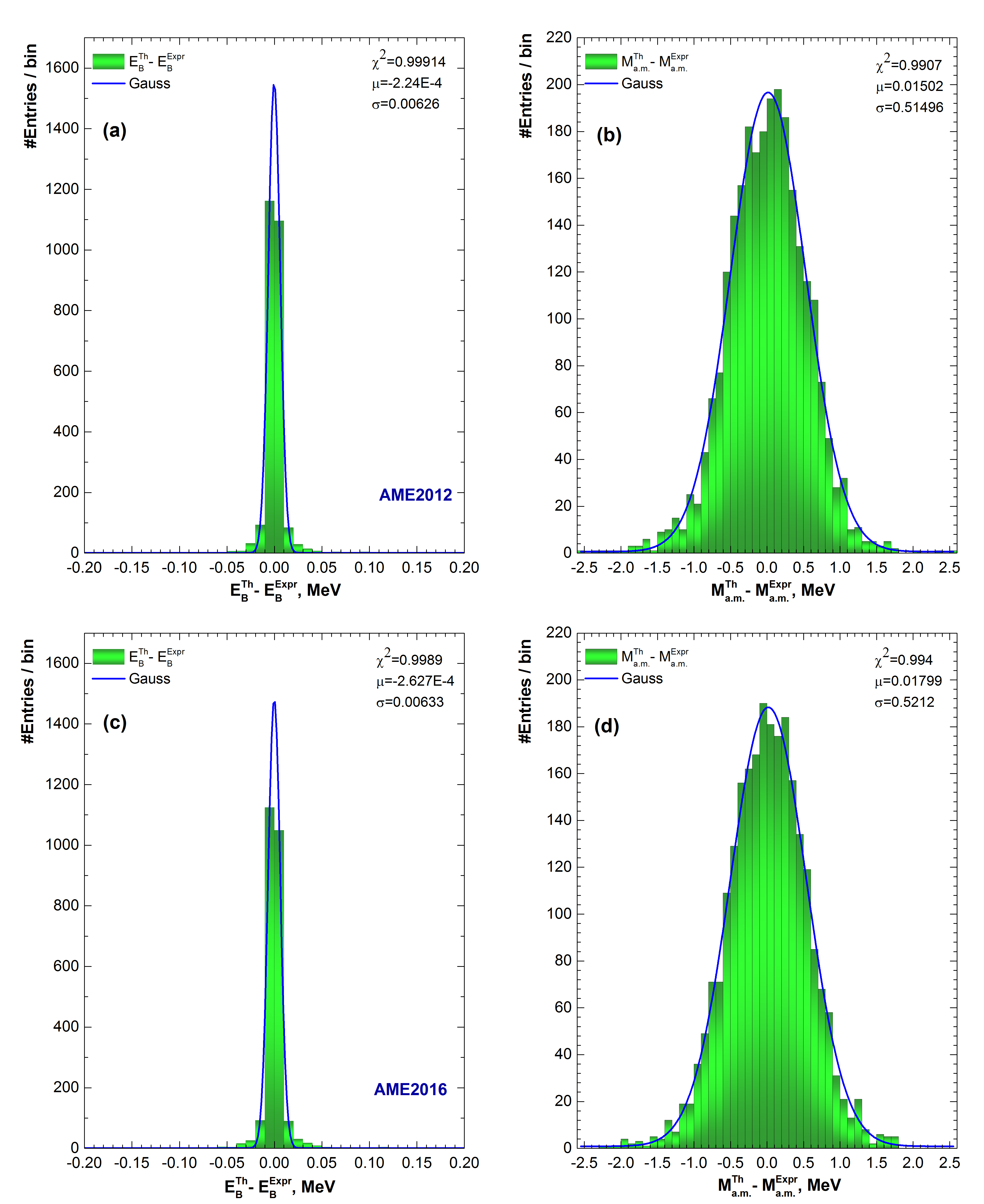}
\includegraphics[scale=0.285]{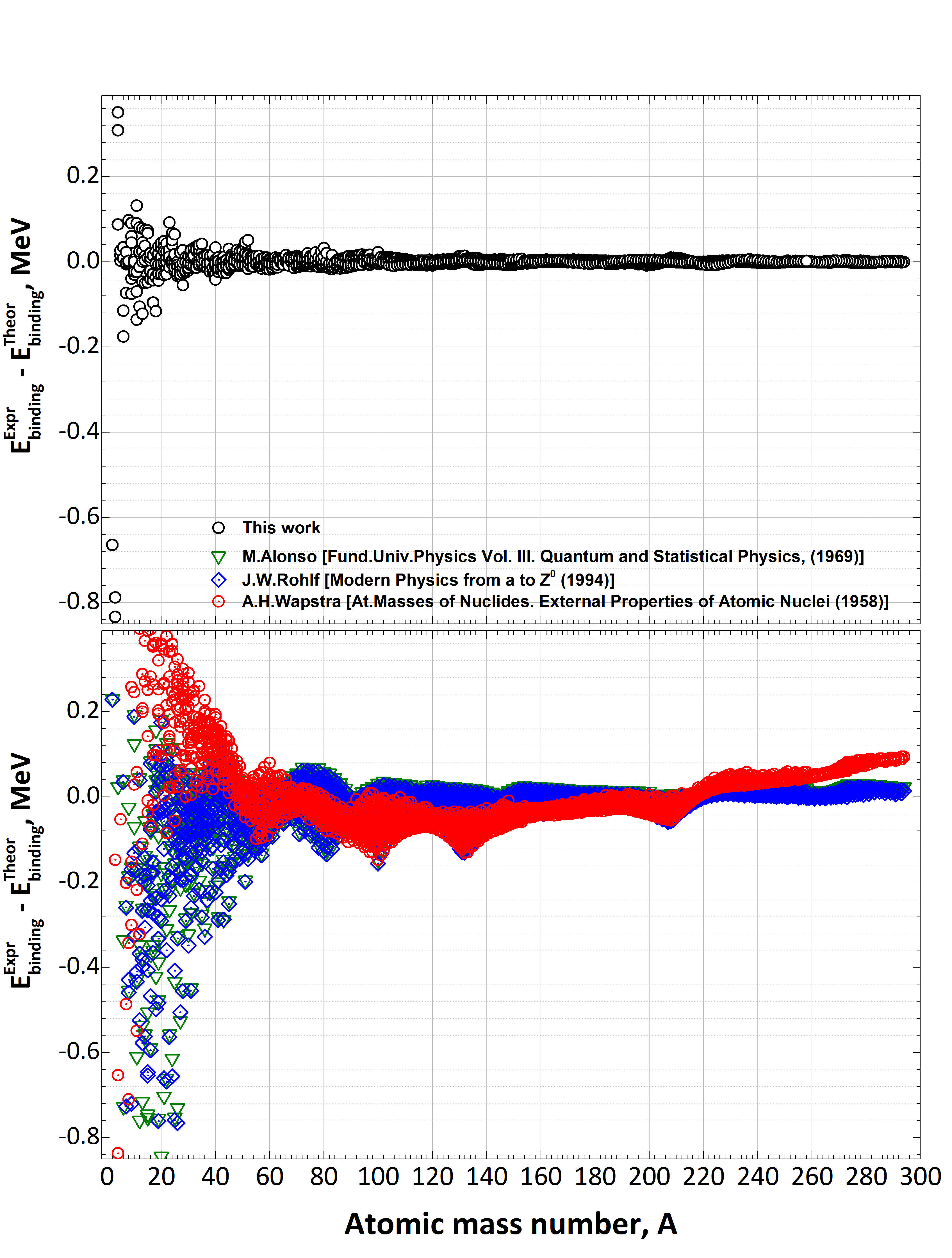}
\caption{
(a)(b) -- Residuals of the experimental data and result of theoretical predictions of the generalized BW mass formula fitted with the help of the Gaussian distribution function using the AME2012 (2564 nuclei), (c)(d) -- results of the blind analysis of the AME2016 (2496 nuclei) data using solution obtained on the base of the AME2012 data. Result of the binding energies fit is shown on the left panel, result for the atomic masses fits -- center panel.
Right panel -- deviations of the different tunes of the BW mass formula, namely, Wapstra\cite{Wapstra:1958}, Rohlf\cite{Rohlf:1994}, Alonso \cite{Alonso1969} with the experimental data, results of this work shown on top pad.
}
\label{fig:GaussFit_BindingEnergyA}
\end{figure*}

\begin{figure}
\centering
\includegraphics[scale=0.31]{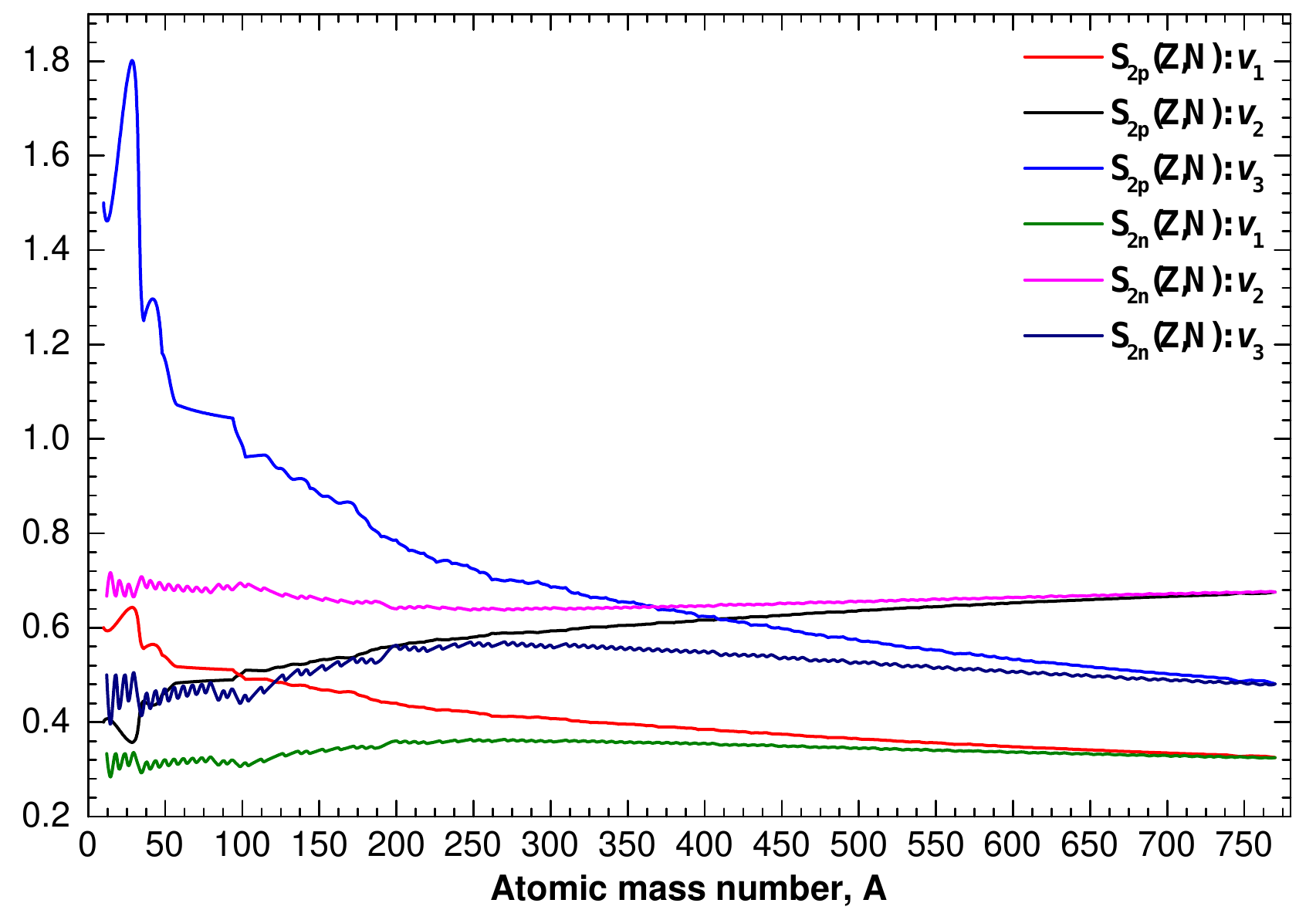}
\caption{
The behavior of variables $\upsilon_{1}$, $\upsilon_{2}$, $\upsilon_{3}$, see Eq.\eqref{eq:IndependentVariable_14}, which are the discovered arguments of drip-lines of $S_{2p}(Z,N)$  and $S_{2n}(Z,N)$ presented in Fig.\ref{fig:DripLines_Predictions}.
}
\label{fig:FitParam_Asimptotic}
\end{figure}

The nuclear landscape, illustrated in Fig.\ref{fig:DripLines_Predictions}, is visualized by a plot of all known nuclei (green squares) and the valley of stability (black squares), with neutron number on the $x$-axis and proton number on the $y$- axis. The point where some of the nucleons, protons or neutrons, will be completely unbound is known as a nuclear drip line, because it is as if the extra nucleons drip right off the nucleus. To find where these drip lines are on the nuclear landscape, the separation energy is needed. In finding both the proton and neutron drip lines, even-even nuclei were used. Since only even-even nuclei were considered, two neutron ($S_{2n}$) and two proton ($S_{2p}$) separation energies were used. These energies required to remove two neutrons and two protons
\begin{equation}
S_{2n}(N,Z)=(Z+N)E_{B}(N, Z, \{a_{i}\}) - (Z+(N-2))E_{B}(N-2, Z, \{a_{i}\}),
\label{eq:2NeutronSeparationEnergy}
\end{equation}
in the case of neutron shells, and the two-proton separation energy
\begin{equation}
S_{2p}(N,Z)=(Z+N)E_{B}(N, Z, \{a_{i}\}) - ((Z-2)+N)E_{B}(N, Z-2, \{a_{i}\}),
\label{eq:2ProtonSeparationEnergy}
\end{equation}
in the case of proton shells. The two-neutron and two proton drip lines are reached when $S_{2n}\approx 0$ and $S_{2p}\approx 0$, respectively. In the real math we compute the separation energies until they do change the sign to minus, then we took the previous even-even nuclei for which the values of  $S_{2n}$ and $S_{2p}$ are very close to zero as our drip-line limit. Since by convention, our values for the binding energy are per nucleon, that is, we put the additional factors in front of the binding energy. The result drip lines are shown in Fig.\ref{fig:DripLines_Predictions}. We observe that the drip lines reach some asymptotic limit, which occurs due to the asymptotic behavior of the variables $\upsilon_{1}$, $\upsilon_{2}$, $\upsilon_{3}$, see Fig.\ref{fig:FitParam_Asimptotic}, which were introduced in Eq.\eqref{eq:IndependentVariable_14}, and
represent the discovered arguments of the separation energies.

\section{Summary}
\label{Summary}

Summarizing the above results, we may say that based on the obtained solution, the signatures of the traditional magic numbers 2, 8, 20, 28, 50, 82, which are clearly seen to be close to the stable nuclei, there is indication for some other shell closures at $Z = 14, 108, 124$  and $N = 14, 124, 152, 202$, that are supported by the correction energies, see Eq.\eqref{eq:CorMN_func}. However, as one may note, the peaks for these magic numbers looks more smeared than those from the general set. The function Eq.\eqref{eq:CorMN_func} used for the  correction energy, representing the overall behavior of the single-particle spectra, may be a good candidate to identify the magicity of nuclei. As seen from Fig.\ref{fig:MagNumberList} the obtained magic proton number candidates have been proposed in the modified Lucas' geometrical packing scheme \cite{Rydin2011238}, while the magic neutron number candidates are new and do not match any models. 
Blind analysis of the AME2016 dataset with the same solution gives us a slightly different spectra of magic numbers. With a tail of $Z = 96,108,126$  and $N = 126, 142, 152, 184$, see Table~\ref{tab:ZN_magnumbers}.

We found that results of the verification of the Bethe-Weizs\"{a}cker mass formula in the inverse problem framework greatly improve the agreement between the experimental masses and the calculated ones, see Fig.\ref{fig:GaussFit_BindingEnergyA}, and thus predicts the drip-lines more accurately than calculated earlier \cite{PhysRevC.65.037301, ADHIKARI:2004, BASU:2004} with the modified Bethe-Weizs\"{a}cker mass formula alone. This demonstration show us, that the implementation of  inverse problem method with the different mass models can become a very effective tool for providing mass predictions in regions far from known nuclear masses. 

We adopt a procedure in which we compute all structure functions from scratch, which allows us to provide some predictions regarding the magic numbers. A correlation between the surface, Coulomb, asymmetry and Wigner terms has been found which allows us to avoid the introduction of different radius constants for the nuclear and the spin-orbit potentials. 

The volume and surface symmetry terms are shown, see Fig.\ref{fig:Structure_constants},  to be largely independent of the other terms in the formula, while the Coulomb diffuseness correction $Z(Z-1)/A^{p_{1}}$ or the charge exchange correction $(N-Z)^{2}/A^{p_{3}}$, asymmetry and Wigner correction terms are of critical importance in determining the nucleon binding energy and play the main role to improve the accuracy of the mass formula. The Wigner term and the curvature energy can also be used separately for the same purpose. The interplay between different terms is found to be important.

We were able to calculate the borders of the nuclear landscape (drip lines) and show their limit, Fig.\ref{fig:DripLines_Predictions_limit}. This finding together with the asymptotic behavior, see Fig.\ref{fig:FitParam_Asimptotic}, and accurate predictions of the binding energies of all known isotopes allows us to obtain quite precise predictions for the location of the proton and neutron drip lines, and claim the exact number of bound nuclei in the nuclear landscape. We would like to comment that this result is not in a conflict with respect to the problem of the critical charge in QED, for interested reader we refer to \cite{Popov2001}, where one may find all necessary details. We are of course aware that the super-heavy isotopes in Fig.\ref{fig:DripLines_Predictions_limit} cannot be produced at present and may be even at future facilities and that, so far, only theoretical studies may be carried out in such region of the nuclear chart.

The concept of ill-posed problems and the associated regularization theories seem to provide a satisfactory framework to solve nuclear physics problems. This new perspective may be used for testing the applications of the liquid model in different areas, for forecasting the mass values away from the valley of stability, for calculation of kinetic and total energy of nuclear proton, alpha, cluster decays and spontaneous fission, for preliminary research of island stability problem and the possibility of creating new super-heavy elements in the future.

\vspace{1cm}
{\bf{ACKNOWLEDGMENTS:}}
We are grateful to {\bf{Lubomir Aleksandrov}} for the provided REGN program. Strachimir Cht. Mavrodiev would like to thank {\bf{Alexey Sissakian}} and {\bf{Lubomir Aleksandrov}} for many years of a very constructive collaboration and friendship. Unfortunately, they passed away, but they made enormous contributions to the REGN development and application of it for solving different inverse problems, especially for the discovering latent regularities. 

This work was supported by the Chinese Academy of Sciences President's International Fellowship Initiative under Grant No. 2016PM043.



\vspace{2cm}
{\bf{REFERENCES:}}

\bibliographystyle{elsarticle-num}

\appendix
\onecolumn

\newpage
\section{Binding energy in the limit of high atomic mass}
\label{Binding energy in the limit of high atomic mass}
\begin{figure*}[hb]
\centering
\subfigure[]{
\includegraphics[scale=0.3]{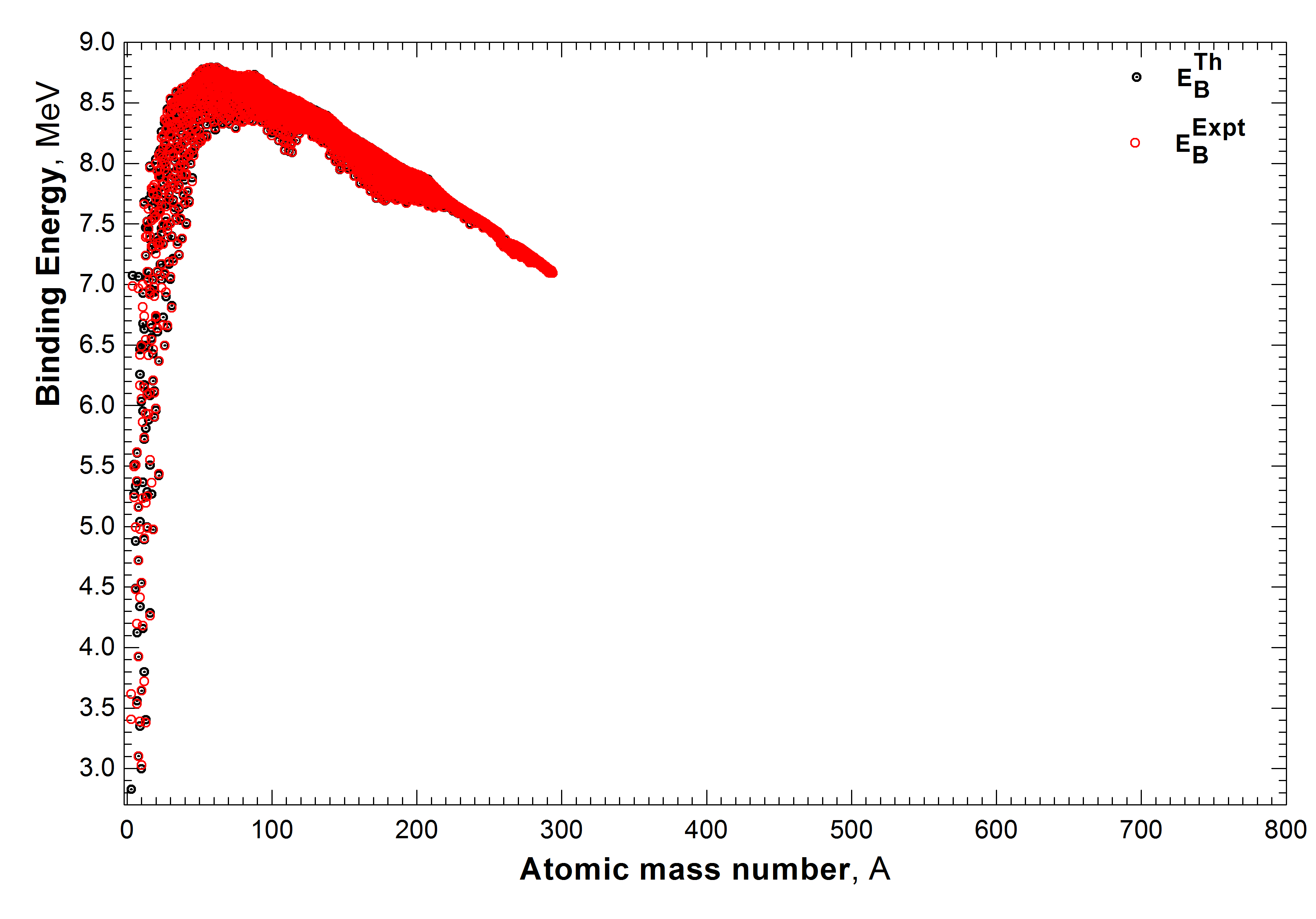}
}
\subfigure[]{
\includegraphics[scale=0.3]{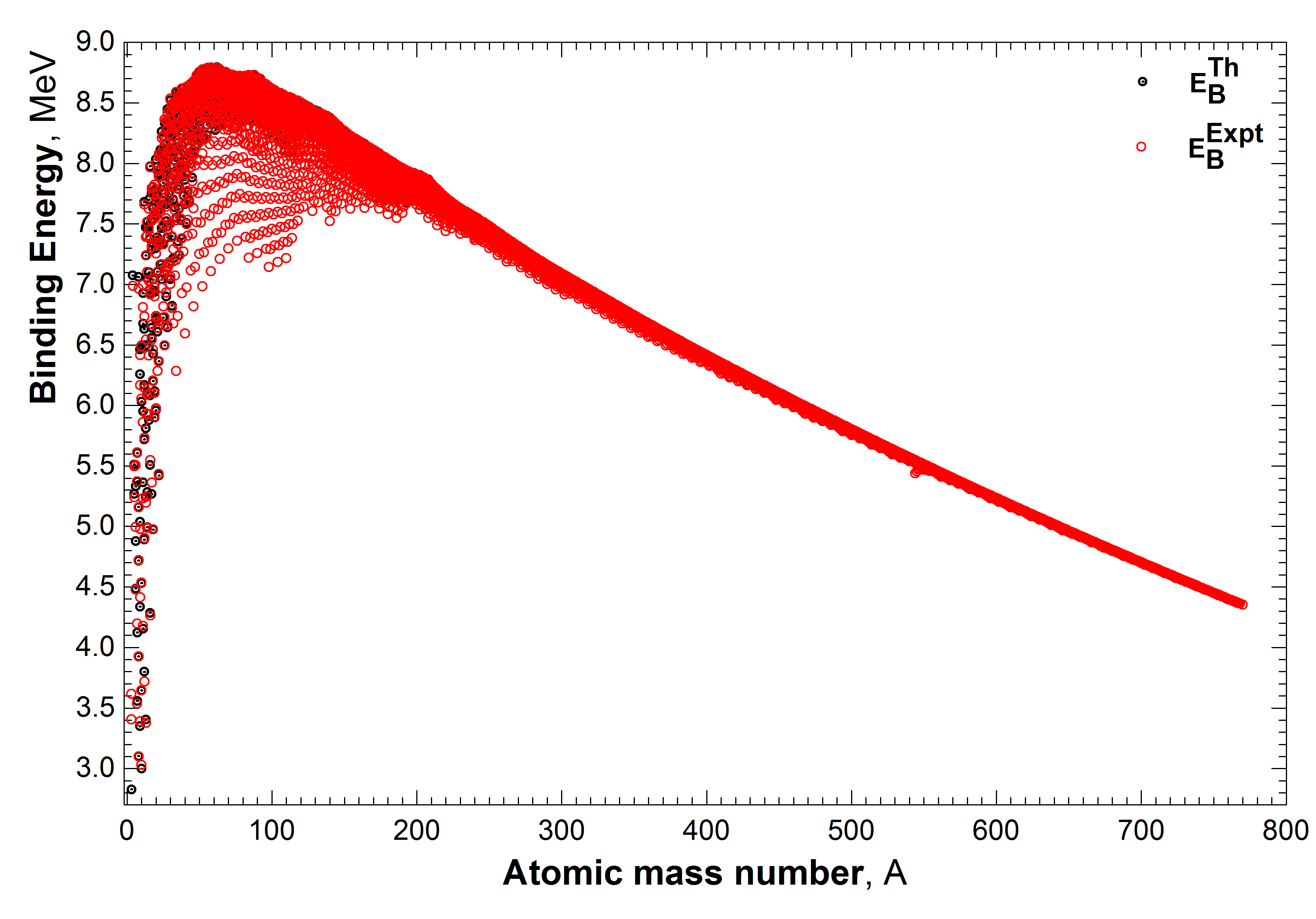}
}

\subfigure[]{
\includegraphics[scale=0.3]{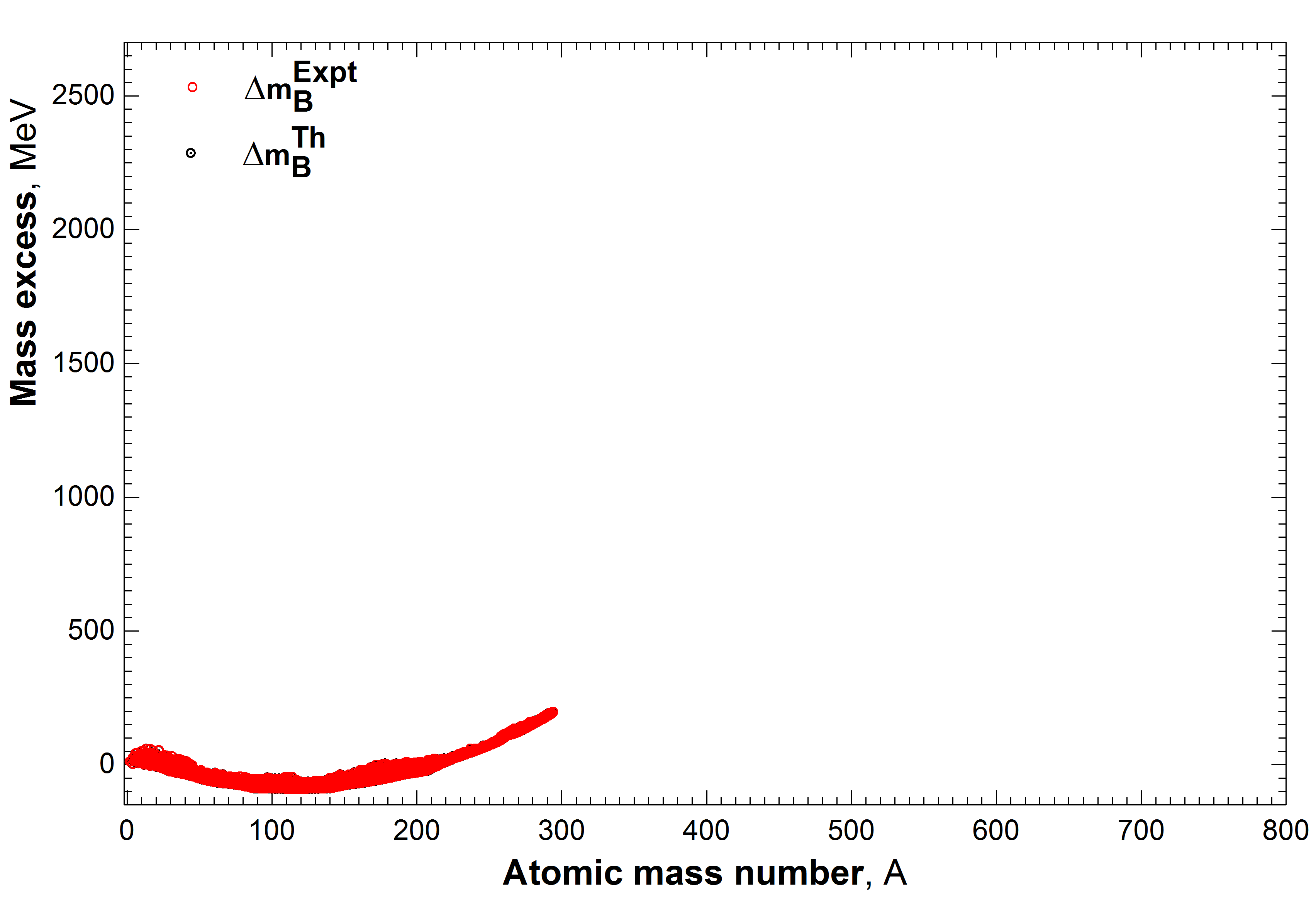}
}
\subfigure[]{
\includegraphics[scale=0.3]{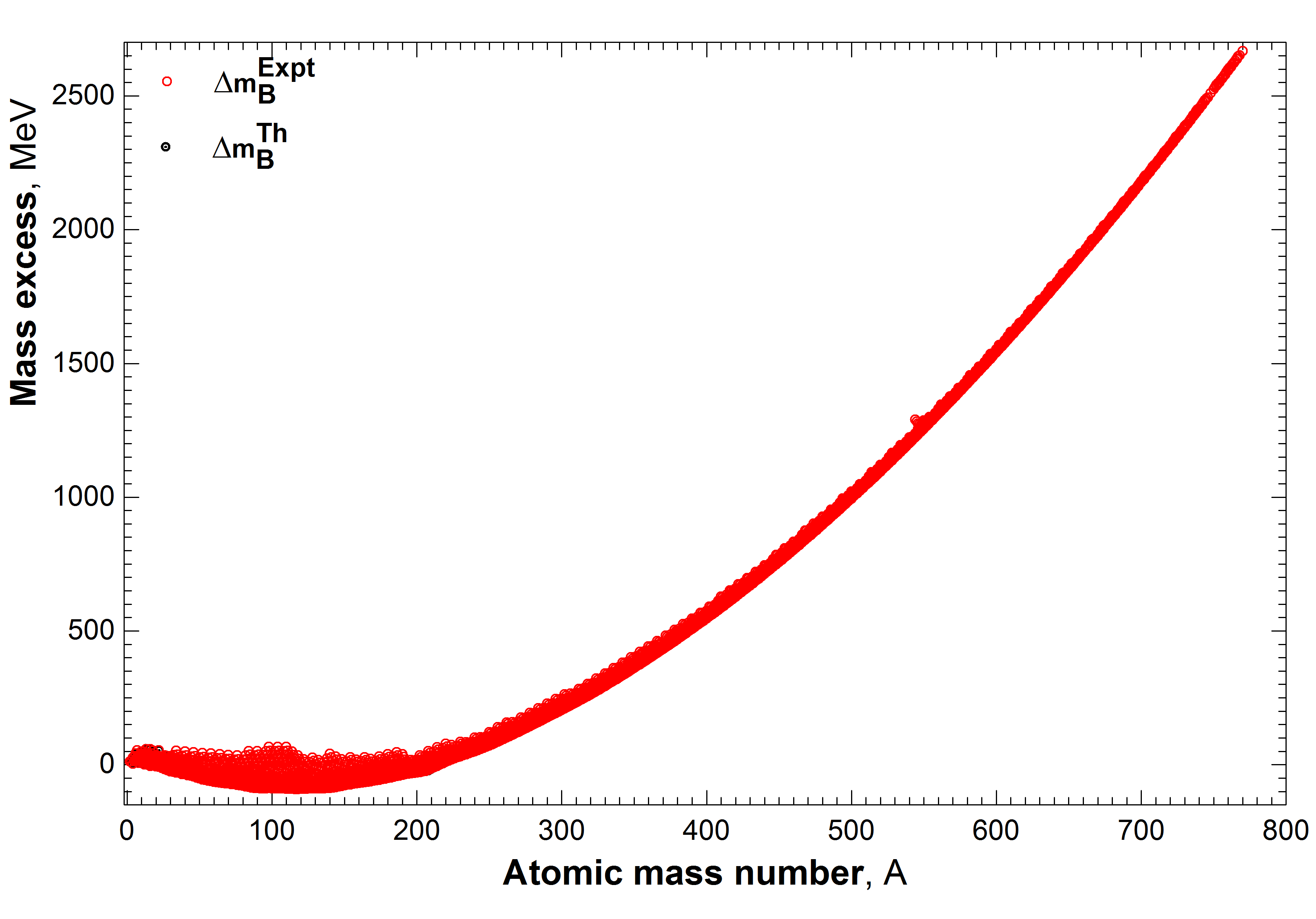}
}
\caption{
Current observation of binding energy and mass excess -- left panels, the theoretical prediction coming from the inverse problem approach -- right panels.
}
\label{fig:BinEn_vs_A_inTheLimit900A}
\end{figure*}

\newpage
\begin{landscape}
\section{Resulting table}
\label{Resulting table}

\begin{center}
\LTcapwidth=10.2in
\begin{longtable}{ccccccrr|crr|crr|crr}
\hline \hline
\multicolumn{1}{c}{$\boldsymbol{\#}$}
&\multicolumn{1}{c}{\scriptsize{\bf{El}}}
&\multicolumn{1}{c}{\scriptsize{$A$}}
&\multicolumn{1}{c}{\scriptsize{$Z$}}
&\multicolumn{1}{c}{\scriptsize{$N$}}
&\multicolumn{1}{c}{\scriptsize{$E_{B}^{Expt}(A,Z)$,}}
&\multicolumn{1}{c}{\scriptsize{$E_{B}^{Th}(A,Z)$,}}
&\multicolumn{1}{c|}{\scriptsize{Residual,}}
&\multicolumn{1}{c}{\scriptsize{$M_{n.m.}^{Expt}(A,Z)$,}}
&\multicolumn{1}{c}{\scriptsize{$M_{n.m.}^{Th}(A,Z,a_{i})$,}}
&\multicolumn{1}{c|}{\scriptsize{Residual,}}
&\multicolumn{1}{c}{\scriptsize{$M_{a.m.}^{Expt}(A,Z)$,}}
&\multicolumn{1}{c}{\scriptsize{$M_{a.m.}^{Th}(A,Z,a_{i})$,}}
&\multicolumn{1}{c|}{\scriptsize{Residual,}}
&\multicolumn{1}{c}{\scriptsize{$\Delta m^{Expt}(A,Z)$,}}
&\multicolumn{1}{c}{\scriptsize{$\Delta m^{Th}(A,Z,a_{i})$,}}
&\multicolumn{1}{c}{\scriptsize{Residual,}}\\
\multicolumn{1}{c}{}
&\multicolumn{1}{c}{}
&\multicolumn{1}{c}{}
&\multicolumn{1}{c}{}
&\multicolumn{1}{c}{}
&\multicolumn{1}{c}{~\scriptsize{\rm{\bf{MeV}}}~}  
&\multicolumn{1}{c}{~\scriptsize{\rm{\bf{MeV}}}~}  
&\multicolumn{1}{c|}{~\scriptsize{\rm{\bf{MeV}}}~}  
&\multicolumn{1}{c}{~\scriptsize{\rm{\bf{MeV}}}~}  
&\multicolumn{1}{c}{~\scriptsize{\rm{\bf{MeV}}}~}  
&\multicolumn{1}{c|}{~\scriptsize{\rm{\bf{MeV}}}~}  
&\multicolumn{1}{c}{~\scriptsize{\rm{\bf{MeV}}}~}  
&\multicolumn{1}{c}{~\scriptsize{\rm{\bf{MeV}}}~}  
&\multicolumn{1}{c|}{~\scriptsize{\rm{\bf{MeV}}}~}  
&\multicolumn{1}{c}{~\scriptsize{\rm{\bf{MeV}}}~}  
&\multicolumn{1}{c}{~\scriptsize{\rm{\bf{MeV}}}~}  
&\multicolumn{1}{c}{~\scriptsize{\rm{\bf{MeV}}}~}\\
\hline
\endfirsthead
\hline \hline
\multicolumn{1}{c}{$\boldsymbol{\#}$}
&\multicolumn{1}{c}{\scriptsize{\bf{El}}}
&\multicolumn{1}{c}{\scriptsize{$A$}}
&\multicolumn{1}{c}{\scriptsize{$Z$}}
&\multicolumn{1}{c}{\scriptsize{$N$}}
&\multicolumn{1}{c}{\scriptsize{$E_{B}^{Expt}(A,Z)$,}}
&\multicolumn{1}{c}{\scriptsize{$E_{B}^{Th}(A,Z,a_{i})$,}}
&\multicolumn{1}{c|}{\scriptsize{Residual,}}
&\multicolumn{1}{c}{\scriptsize{$M_{n.m.}^{Expt}(A,Z)$,}}
&\multicolumn{1}{c}{\scriptsize{$M_{n.m.}^{Th}(A,Z,a_{i})$,}}
&\multicolumn{1}{c|}{\scriptsize{Residual,}}
&\multicolumn{1}{c}{\scriptsize{$M_{a.m.}^{Expt}(A,Z)$,}}
&\multicolumn{1}{c}{\scriptsize{$M_{a.m.}^{Th}(A,Z,a_{i})$,}}
&\multicolumn{1}{c|}{\scriptsize{Residual,}}
&\multicolumn{1}{c}{\scriptsize{$\Delta m^{Expt}(A,Z)$,}}
&\multicolumn{1}{c}{\scriptsize{$\Delta m^{Th}(A,Z,a_{i})$,}}
&\multicolumn{1}{c}{\scriptsize{Residual,}}\\
\multicolumn{1}{c}{}
&\multicolumn{1}{c}{}
&\multicolumn{1}{c}{}
&\multicolumn{1}{c}{}
&\multicolumn{1}{c}{}
&\multicolumn{1}{c}{~\scriptsize{\rm{\bf{MeV}}}~}  
&\multicolumn{1}{c}{~\scriptsize{\rm{\bf{MeV}}}~}  
&\multicolumn{1}{c|}{~\scriptsize{\rm{\bf{MeV}}}~}  
&\multicolumn{1}{c}{~\scriptsize{\rm{\bf{MeV}}}~}  
&\multicolumn{1}{c}{~\scriptsize{\rm{\bf{MeV}}}~}  
&\multicolumn{1}{c|}{~\scriptsize{\rm{\bf{MeV}}}~}  
&\multicolumn{1}{c}{~\scriptsize{\rm{\bf{MeV}}}~}  
&\multicolumn{1}{c}{~\scriptsize{\rm{\bf{MeV}}}~}  
&\multicolumn{1}{c|}{~\scriptsize{\rm{\bf{MeV}}}~}  
&\multicolumn{1}{c}{~\scriptsize{\rm{\bf{MeV}}}~}  
&\multicolumn{1}{c}{~\scriptsize{\rm{\bf{MeV}}}~}  
&\multicolumn{1}{c}{~\scriptsize{\rm{\bf{MeV}}}~}\\
\hline
\endhead
\scriptsize{3}  &\scriptsize{H}  &\scriptsize{2}   &\scriptsize{1}   &\scriptsize{1}   &\scriptsize{1.112283E+00}   &\scriptsize{1.816214E+00}   &\scriptsize{-7.0393E-01}   &\scriptsize{1.875613E+03}   &\scriptsize{1.874204E+03}   &\scriptsize{1.41E+00}   &\scriptsize{1.87612375E+03}   &\scriptsize{1.874717E+03}   &\scriptsize{1.4065E+00}   &\scriptsize{1.313572E+01}   &\scriptsize{1.172713E+01}   &\scriptsize{1.4086E+00}\\
\scriptsize{4}  &\scriptsize{H}  &\scriptsize{3}   &\scriptsize{1}   &\scriptsize{2}   &\scriptsize{2.827266E+00}   &\scriptsize{3.635907E+00}   &\scriptsize{-8.0864E-01}   &\scriptsize{2.808921E+03}   &\scriptsize{2.806494E+03}   &\scriptsize{2.43E+00}   &\scriptsize{2.80943185E+03}   &\scriptsize{2.807004E+03}   &\scriptsize{2.4276E+00}   &\scriptsize{1.494981E+01}   &\scriptsize{1.252316E+01}   &\scriptsize{2.4266E+00}\\
\scriptsize{5}  &\scriptsize{He}  &\scriptsize{3}   &\scriptsize{2}   &\scriptsize{1}   &\scriptsize{2.572681E+00}   &\scriptsize{3.425578E+00}   &\scriptsize{-8.5290E-01}   &\scriptsize{2.808391E+03}   &\scriptsize{2.805831E+03}   &\scriptsize{2.56E+00}   &\scriptsize{2.80941325E+03}   &\scriptsize{2.806852E+03}   &\scriptsize{2.5612E+00}   &\scriptsize{1.493122E+01}   &\scriptsize{1.237107E+01}   &\scriptsize{2.5601E+00}\\
\scriptsize{6}  &\scriptsize{H}  &\scriptsize{4}   &\scriptsize{1}   &\scriptsize{3}   &\scriptsize{1.720450E+00}   &\scriptsize{1.391116E+00}   &\scriptsize{3.2933E-01}   &\scriptsize{3.750086E+03}   &\scriptsize{3.751403E+03}   &\scriptsize{-1.32E+00}   &\scriptsize{3.75059718E+03}   &\scriptsize{3.751914E+03}   &\scriptsize{-1.3166E+00}   &\scriptsize{2.462112E+01}   &\scriptsize{2.593774E+01}   &\scriptsize{-1.3166E+00}\\
\end{longtable}
\end{center}
\end{landscape}

\section{Predicted binding energies for super-heave nuclei}
\label{Predicted binding energies for super heave nuclei}

\begin{center}
\LTcapwidth=6.4in
\begin{longtable}{lccccccccc}
\hline \hline
\multicolumn{1}{l}{$\#$}
&\multicolumn{1}{c}{\small{Element}}
&\multicolumn{1}{c}{\small{$A$}}
&\multicolumn{1}{c}{\small{$Z$}}
&\multicolumn{1}{c}{\small{$N$}}
&\multicolumn{1}{c}{\small{$N-Z$}}
&\multicolumn{1}{c}{\small{$E_{B}(A,Z,a_{i})$,MeV}}
&\multicolumn{1}{c}{\small{$M_{\rm{n.m.}}(A,Z,a_{i})$,MeV}}
&\multicolumn{1}{c}{\small{$M_{\rm{a.m.}}(A,Z,a_{i})$,MeV}}
&\multicolumn{1}{c}{\small{$\Delta m(A,Z,a_{i})$,MeV}}\\
\hline
\endfirsthead
\hline \hline
\multicolumn{1}{l}{$\#$}
&\multicolumn{1}{c}{\small{Element}}
&\multicolumn{1}{c}{\small{$A$}}
&\multicolumn{1}{c}{\small{$Z$}}
&\multicolumn{1}{c}{\small{$N$}}
&\multicolumn{1}{c}{\small{$N-Z$}}
&\multicolumn{1}{c}{\small{$E_{B}(A,Z,a_{i})$,MeV}}
&\multicolumn{1}{c}{\small{$M_{\rm{n.m.}}(A,Z,a_{i})$,MeV}}
&\multicolumn{1}{c}{\small{$M_{\rm{a.m.}}(A,Z,a_{i})$,MeV}}
&\multicolumn{1}{c}{\small{$\Delta m(A,Z,a_{i})$,MeV}}\\
\hline
\endhead
\scriptsize{1} &\scriptsize{Rf} &\scriptsize{266} &\scriptsize{104} &\scriptsize{162} &\scriptsize{58} &\scriptsize{7.347272E+00} &\scriptsize{2.478344E+05} &\scriptsize{2.478886E+05} &\scriptsize{1.111566E+02}\\
\scriptsize{2} &\scriptsize{Rf} &\scriptsize{267} &\scriptsize{104} &\scriptsize{163} &\scriptsize{59} &\scriptsize{7.337124E+00} &\scriptsize{2.487693E+05} &\scriptsize{2.488235E+05} &\scriptsize{1.145900E+02}\\
\scriptsize{3} &\scriptsize{Rf} &\scriptsize{265} &\scriptsize{104} &\scriptsize{161} &\scriptsize{57} &\scriptsize{7.351197E+00} &\scriptsize{2.469011E+05} &\scriptsize{2.469553E+05} &\scriptsize{1.093922E+02}\\
\scriptsize{4} &\scriptsize{Db} &\scriptsize{270} &\scriptsize{105} &\scriptsize{165} &\scriptsize{60} &\scriptsize{7.309896E+00} &\scriptsize{2.515720E+05} &\scriptsize{2.516268E+05} &\scriptsize{1.233610E+02}\\
\scriptsize{5} &\scriptsize{Db} &\scriptsize{268} &\scriptsize{105} &\scriptsize{163} &\scriptsize{58} &\scriptsize{7.324159E+00} &\scriptsize{2.497037E+05} &\scriptsize{2.497584E+05} &\scriptsize{1.180157E+02}\\
\scriptsize{6} &\scriptsize{Db} &\scriptsize{267} &\scriptsize{105} &\scriptsize{162} &\scriptsize{57} &\scriptsize{7.332720E+00} &\scriptsize{2.487692E+05} &\scriptsize{2.488239E+05} &\scriptsize{1.149828E+02}\\
\scriptsize{7} &\scriptsize{Sg} &\scriptsize{271} &\scriptsize{106} &\scriptsize{165} &\scriptsize{59} &\scriptsize{7.303195E+00} &\scriptsize{2.525048E+05} &\scriptsize{2.525600E+05} &\scriptsize{1.251554E+02}\\
\scriptsize{8} &\scriptsize{Sg} &\scriptsize{269} &\scriptsize{106} &\scriptsize{163} &\scriptsize{57} &\scriptsize{7.315578E+00} &\scriptsize{2.506369E+05} &\scriptsize{2.506922E+05} &\scriptsize{1.202882E+02}\\
\end{longtable}
\end{center}

\end{document}